\documentclass[12pt,preprint]{aastex}

\usepackage{graphicx}
\usepackage{url}
\usepackage{amsmath}
\usepackage{epstopdf}
\usepackage{mathrsfs}
\usepackage{paralist}

\shorttitle{Spin-orbit alignment of exoplanet systems}
\shortauthors{Campante et al.}

\begin{document}

\title{Spin-orbit alignment of exoplanet systems: ensemble analysis using asteroseismology}

\author{T.~L.~Campante\altaffilmark{1,2}}
\email{campante@bison.ph.bham.ac.uk}
\and
\author{
M.~N.~Lund\altaffilmark{1,2},
J.~S.~Kuszlewicz\altaffilmark{1,2},
G.~R.~Davies\altaffilmark{1,2},
W.~J.~Chaplin\altaffilmark{1,2},
S.~Albrecht\altaffilmark{2},
J.~N.~Winn\altaffilmark{3,4},
T.~R.~Bedding\altaffilmark{5,2},
O.~Benomar\altaffilmark{6,7},
D.~Bossini\altaffilmark{1,2},
R.~Handberg\altaffilmark{2,1},
A.~R.~G.~Santos\altaffilmark{8,1},
V.~Van Eylen\altaffilmark{2,4},
S.~Basu\altaffilmark{9},
J.~Christensen-Dalsgaard\altaffilmark{2},
Y.~P.~Elsworth\altaffilmark{1,2},
S.~Hekker\altaffilmark{10,2},
T.~Hirano\altaffilmark{11},
D.~Huber\altaffilmark{5,12},
C.~Karoff\altaffilmark{13,2},
H.~Kjeldsen\altaffilmark{2},
M.~S.~Lundkvist\altaffilmark{2},
T.~S.~H.~North\altaffilmark{1,2},
V.~Silva Aguirre\altaffilmark{2},
D.~Stello\altaffilmark{5,2},
T.~R.~White\altaffilmark{2,14}
}

\altaffiltext{1}{School of Physics and Astronomy, University of Birmingham, Edgbaston, Birmingham, B15 2TT, UK}
   
\altaffiltext{2}{Stellar Astrophysics Centre (SAC), Department of Physics and Astronomy, Aarhus University, Ny Munkegade 120, DK-8000 Aarhus C, Denmark}

\altaffiltext{3}{Physics Department, Massachusetts Institute of Technology, 77 Massachusetts Avenue, Cambridge, MA 02139, USA}

\altaffiltext{4}{MIT Kavli Institute for Astrophysics \& Space Research, 70 Vassar Street, Cambridge, MA 02139, USA}   

\altaffiltext{5}{Sydney Institute for Astronomy, School of Physics, University of Sydney, Sydney, Australia}

\altaffiltext{6}{Department of Astronomy, The University of Tokyo, School of Science, 7-3-1 Hongo, Bunkyo-ku, Tokyo 113-0033, Japan}

\altaffiltext{7}{NYUAD Institute, Center for Space Science, New York University Abu Dhabi, PO Box 129188, Abu Dhabi, UAE}

\altaffiltext{8}{Instituto de Astrof\'isica e Ci\^encias do Espa\c co, Universidade do Porto, Rua das Estrelas, 4150-762 Porto, Portugal}

\altaffiltext{9}{Department of Astronomy, Yale University, New Haven, CT 06520, USA}

\altaffiltext{10}{Max Planck Institute for Solar System Research, Justus-von-Liebig-Weg 3, 37077 G\"ottingen, Germany}

\altaffiltext{11}{Department of Earth and Planetary Sciences, Tokyo Institute of Technology, 2-12-1 Ookayama, Meguro-ku, Tokyo 152-8551, Japan}

\altaffiltext{12}{SETI Institute, 189 Bernardo Avenue \#100, Mountain View, CA 94043, USA}

\altaffiltext{13}{Department of Geoscience, Aarhus University, H{\o}egh-Guldbergs Gade 2, DK-8000 Aarhus C, Denmark}

\altaffiltext{14}{Institut fur Astrophysik, Georg-August-Universit\"at G\"ottingen, Friedrich-Hund-Platz 1, 37077 G\"ottingen, Germany}

\begin{abstract}
The angle $\psi$ between a planet's orbital axis and the spin axis of its parent star is an important diagnostic of planet formation, migration, and tidal evolution. We seek empirical constraints on $\psi$ by measuring the stellar inclination $i_{\rm s}$ via asteroseismology for an ensemble of 25 solar-type hosts observed with NASA's {\it Kepler} satellite. Our results for $i_{\rm s}$ are consistent with alignment at the 2-$\sigma$ level for all stars in the sample, meaning that the system surrounding the red-giant star Kepler-56 remains as the only unambiguous misaligned multiple-planet system detected to date. The availability of a measurement of the projected spin-orbit angle $\lambda$ for two of the systems allows us to estimate $\psi$. We find that the orbit of the hot-Jupiter HAT-P-7b is likely to be retrograde ($\psi\!=\!116.4\degr^{+30.2}_{-14.7}$), whereas that of Kepler-25c seems to be well aligned with the stellar spin axis ($\psi\!=\!12.6\degr^{+6.7}_{-11.0}$). While the latter result is in apparent contradiction with a statement made previously in the literature that the multi-transiting system Kepler-25 is misaligned, we show that the results are consistent, given the large associated uncertainties. Finally, we perform a hierarchical Bayesian analysis based on the asteroseismic sample in order to recover the underlying distribution of $\psi$. The ensemble analysis suggests that the directions of the stellar spin and planetary orbital axes are correlated, as conveyed by a tendency of the host stars to display large inclination values.
\end{abstract}


\section{Introduction}\label{sec:intro}
The angle $\psi$ between the planetary orbital axis and the stellar spin axis (the true obliquity or spin-orbit angle) is a fundamental geometric property of planetary systems. Furthermore, it has been recognized as an important diagnostic of theories of planet formation, migration, and tidal evolution. For all these reasons, seeking empirical constraints on $\psi$ is a matter of the utmost importance.

In the case of an exoplanet found through the radial-velocity (RV) method, no information is available about the degree of spin-orbit alignment. For transiting exoplanets, on the other hand, the Rossiter--McLaughlin (RM) effect has now been widely exploited \citep[e.g.,][]{Queloz00,Winn05,Winn09,HAT-P-11b,WinnCircumbinary,XO3,Triaud10,Hirano11,Albrecht12,Albrecht13}. This technique is sensitive to the angle $\lambda$ between the sky-projected orbital and spin axes (the projected spin-orbit angle). At the time of writing there are 87 planets with published measurements of the RM effect (see the online compilation\footnote{\url{http://www.physics.mcmaster.ca/~rheller/}} by R.~Heller), of which 36 ($\sim\!40\,\%$) show substantial misalignments according to at least one publication \citep[see also fig.~7 of][]{Xue14}. Other techniques that allow obliquity measurements of transiting systems include the analysis of planetary transits over starspots \citep[e.g.,][]{Kepler17b,Nutzman11,WASP4,SONature}, Doppler tomography \citep[e.g.,][]{Cameron10,Gandolfi12,Albrecht13,Johnson14}, the analysis of the effect of gravity darkening on the transit light curve \citep[e.g.,][]{Barnes09,Barnes11,Ahlers14}, and the analysis of the photometric amplitude distribution of stellar rotation \citep{Mazeh15}.

Most obliquity measurements to date have been for systems harboring hot Jupiters, owing to the fact that the RM effect scales as the planet-to-star area ratio and to the increased opportunities for obtaining follow-up spectroscopic observations due to the frequent transit events. Empirical evidence has been found that the obliquities of hot-Jupiter systems are affected by tidal evolution (\citealt{Schlaufman,Winn10,Morton11,Triaud11,Albrecht12}; although see \citealt{Mazeh15} for evidence against this theory): systems expected to undergo strong planet-star tidal interactions are preferentially found to have low obliquities, while systems with weaker tidal interactions display a wide range of obliquities that, besides well-aligned planets, also include planets in polar or even retrograde orbits. This suggests that the orbital plane has changed relative to the plane of the protoplanetary disk by the time hot Jupiters are formed and before tides have had any impact on shaping the system, which presumably happens due to the same mechanism responsible for their migration.

The above discussion assumes that the protoplanetary disk is coplanar with the stellar equator. The possibility remains, however, that primordial star-disk misalignments are ubiquitous, meaning that large obliquities could be a generic feature of planetary systems and not specific to hot-Jupiter formation. This hypothesis may in principle be tested by measuring the obliquities of systems with multiple transiting planets, since the planetary orbits in these systems are nearly coplanar and presumably trace the plane of the protoplanetary disk \citep{Lissauer11,Fabrycky14}. Accordingly, if multi-transiting systems tend to have low obliquities, then the high obliquities observed for hot-Jupiter systems are likely to be associated with planet migration. If, instead, the obliquity distribution of multi-transiting systems is similar to that of hot-Jupiter systems, then this would indicate more general processes of stellar and planet formation: processes such as chaotic star formation \citep[e.g.,][]{Bate10,Thies11,Fielding15}, magnetic star-disk interactions \citep[e.g.,][]{Lai11}, torques due to internal gravity waves \citep[e.g.,][]{Rogers12}, or torques due to neighboring stars \citep[e.g.,][]{Batygin12}.

In order to study the dynamical histories of planetary systems across a comprehensive range of architectures and in a variety of environments, it is imperative to extend obliquity measurements to systems with smaller planets, longer-period planets, and multiple planets \citep[note that, according to the current state of knowledge, hot Jupiters rarely have nearby planetary companions and may thus occur preferentially as single planets;][]{Steffen12}. For long-period planets, however, the opportunities to observe transits occur less frequently, which limits the possibility of obtaining follow-up observations or identifying the transit geometry from starspot crossings. Furthermore, application of the RM technique becomes increasingly more challenging when dealing with multiple-planet systems, since these systems also tend to involve smaller planets \citep[e.g.,][]{Latham11}.

An alternative technique for measuring the obliquities of planetary systems, one that does not depend on the signal-to-noise ratio (S/N) of the transit data and hence on planet size, makes use of a combination of the photometric stellar rotation period, $P_{\rm rot}$, and the spectroscopically-determined projected rotational velocity, $v\sin i_{\rm s}$, and stellar radius, $R_{\rm s}$, to determine the sine of the angle $i_{\rm s}$ between the stellar spin axis and the line of sight (the stellar inclination angle). This technique evolved from the technique originally developed by \citet{Herbst86} and \citet{Hendry93}, having been revisited more recently by a number of authors \citep{JJ10,Schlaufman}, including a series of applications \citep[e.g.,][]{Hirano12,Hirano14,Walkowicz13,Morton14} to transiting systems observed with the NASA {\it Kepler} mission \citep{Kepler,Koch10}.

Finally, asteroseismology provides an independent means of directly determining $i_{\rm s}$. The asteroseismic estimation of $i_{\rm s}$ rests on our ability to resolve and extract signatures of rotation in the power spectra of non-radial modes of oscillation \citep[e.g.,][]{Gizon03,Ballot06,Ballot08,Benomar09,Campante11}. The asteroseismic technique requires bright targets and long-duration time series to attain the desired S/N and frequency resolution. The applicability of this technique depends entirely on the stellar properties and not on the planetary or orbital parameters, which is an advantage when measuring obliquities of systems with small and/or long-period planets. Following its application to host stars with single, non-transiting large planets discovered using the RV method \citep{Wright11,Gizon13}, the asteroseismic technique has been applied to several {\it Kepler} Sun-like hosts \citep{Chaplin13,Benomar14,Lund14,VanEylen14}. In addition, \citet{Huber13} used asteroseismology to measure a large obliquity for Kepler-56, a red giant hosting two transiting coplanar planets, thus showing that spin-orbit misalignments are not confined to hot-Jupiter systems. Another instance of an asteroseismic obliquity measurement of an evolved host is that of Kepler-432 \citep{KOI-1299}. Recently, the stellar inclination angles of the solar analogs 16 Cyg A and B (the B component hosts a Jovian planet) were determined using asteroseismology \citep{Davies16Cyg}.

Here we present the first analysis of an ensemble of asteroseismic obliquity measurements obtained for solar-type stars with transiting planets. The asteroseismic sample consists of 25 {\it Kepler} planet-candidate host stars (also designated as {\it Kepler} Objects of Interest or KOIs), of which 20 are confirmed hosts. The host stars are distributed along the main sequence and subgiant branch, and all exhibit solar-like oscillations. The rest of the paper is organized as follows. In Sect.~\ref{sec:geometry} we present a recap of the spin-orbit geometry. A detailed asteroseismic analysis of the individual planetary-system hosts follows in Sect.~\ref{sec:astero}. In Sect.~\ref{sec:stat} we place statistical constraints on the spin-orbit alignment. Finally, a discussion of the results and conclusions make up Sect.~\ref{sec:conclusions}.

\section{Spin-orbit geometry}\label{sec:geometry}
\begin{figure}[!t]
\centering
\includegraphics*[scale=0.8]{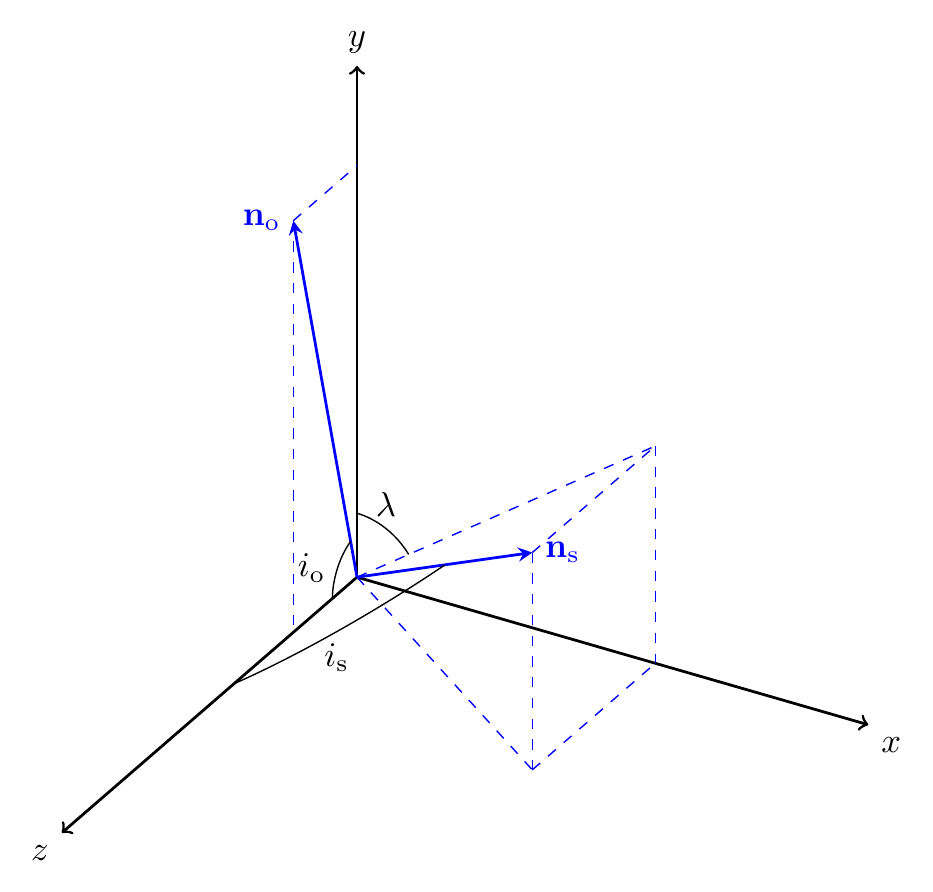}
\includegraphics*[scale=0.8]{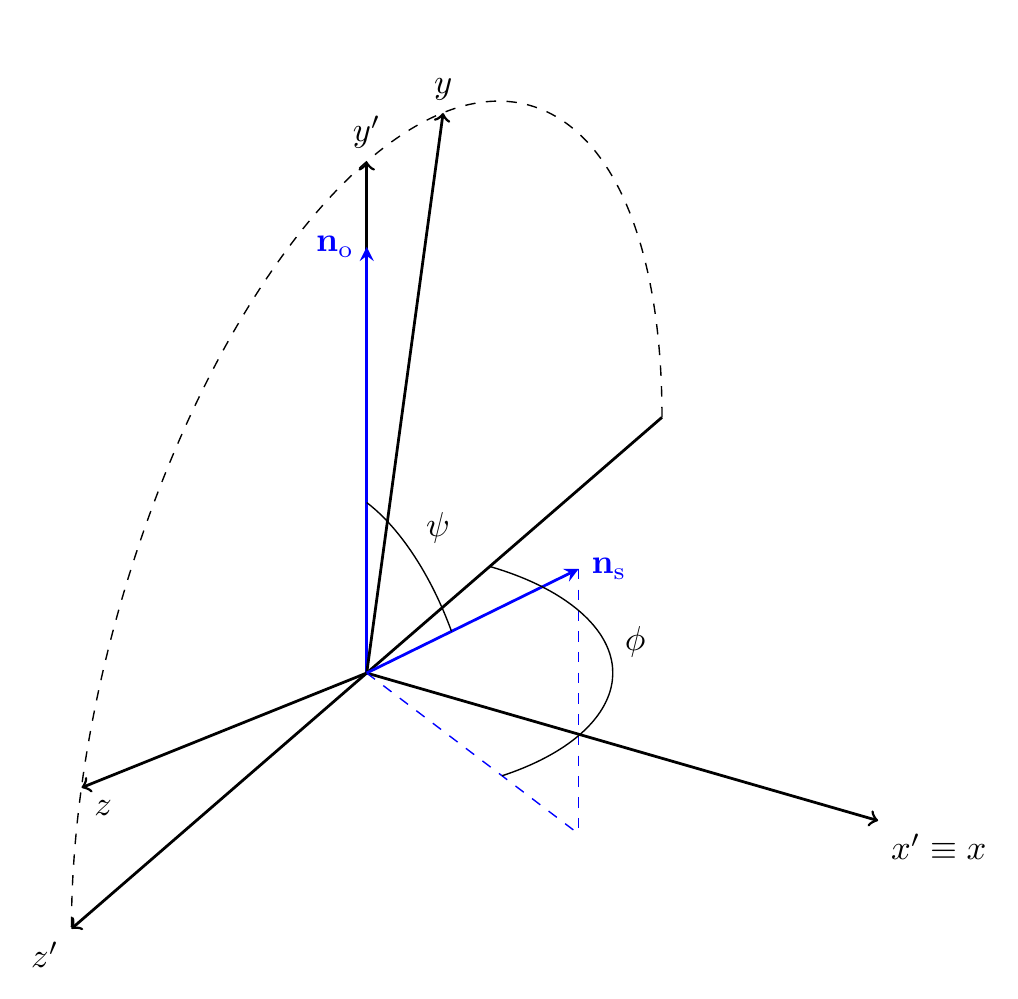}
\caption{\small Spin-orbit geometry. Left panel: Observer-oriented coordinate system. Here the $z$ axis points toward the observer, the $x$ axis points along the line of nodes, the $y$ axis completes a right-handed triad, and the $xy$ plane is the plane of the sky. Right panel: Orbit-oriented coordinate system (obtained from the observer-oriented system by a counterclockwise rotation of $\pi/2-i_{\rm o}$ about the $x'\!\equiv\!x$ axis). Here the $y'$ axis is the planetary orbital axis and the $x'z'$ plane is the orbital plane. The unit vectors $\mathbf{n}_{\rm o}$ and $\mathbf{n}_{\rm s}$ respectively denote the orbital and stellar rotation angular momentum unit vectors. All depicted angles are introduced in the text.\label{fig:geometry}}
\end{figure}

Figure \ref{fig:geometry} shows an observer-oriented coordinate system (left panel) and an orbit-oriented coordinate system (right panel). In the former, the orbital angular momentum unit vector, $\mathbf{n}_{\rm o}$, lies on the $yz$ plane and is solely determined by the angle $i_{\rm o}$ between the planetary orbital axis and the line of sight (the orbital inclination angle). The angle $i_{\rm o}$ can be measured for a transiting planet via transit photometry \citep[e.g.,][]{Charbonneau00,Henry00}, in which case one necessarily has $\sin i_{\rm o}\!\approx\!1$ (i.e., an edge-on orbit). Determination of the stellar rotation angular momentum unit vector, $\mathbf{n}_{\rm s}$, requires a polar and an azimuthal angle, respectively $i_{\rm s}$ and $\lambda$. To avoid degeneracies, we restrict $i_{\rm o}$ and $i_{\rm s}$ to the interval $[0,\pi/2]$, while $\lambda$ is allowed to vary in the interval $[-\pi,\pi]$. In the orbit-oriented coordinate system, $\mathbf{n}_{\rm s}$ is determined by the polar and azimuthal angles $\psi$ and $\phi$. The spin-orbit angle $\psi$ is the angle between $\mathbf{n}_{\rm o}$ and $\mathbf{n}_{\rm s}$, taking values in the interval $[0,\pi]$. Values of $\psi$ lower (greater) than $\pi/2$ correspond to prograde (retrograde) orbits. The specific cases of a parallel, an antiparallel, and a polar orbit then correspond to $\psi\!=\!0$, $\psi\!=\!\pi$, and $\psi\!=\!\pi/2$, respectively. The azimuthal angle $\phi$ is allowed to vary in the range $[-\pi,\pi]$ and takes the value $\pi$ along the line of sight.

The various angles are related according to the following equations \citep[for a derivation see, e.g.,][]{Fabrycky09}:
\begin{subequations}
\begin{gather}
\sin i_{\rm s}\sin\lambda = \sin\psi\sin\phi \, , \label{eq:geom1}\\ 
\cos\psi = \sin i_{\rm s}\cos\lambda\sin i_{\rm o} + \cos i_{\rm s}\cos i_{\rm o} \, , \label{eq:geom2}\\
\sin\psi\cos\phi = \sin i_{\rm s}\cos\lambda\cos i_{\rm o} - \cos i_{\rm s}\sin i_{\rm o} \, . \label{eq:geom3}
\end{gather}
\end{subequations}
Equation \eqref{eq:geom2} is of particular interest, as it allows determination of the spin-orbit angle $\psi$ provided that measurements of $i_{\rm o}$, $i_{\rm s}$, and $\lambda$ are available. A joint analysis of asteroseismology, the transit light curve, and the RM effect made it possible to determine $\psi$ for the hot-Jupiter system HAT-P-7 \citep[Kepler-2;][]{Benomar14,Lund14} and the multi-transiting system Kepler-25 \citep{Benomar14}. Both these systems are revisited in this work. 

\begin{figure}[!t]
\centering
\includegraphics[width=0.80\linewidth]{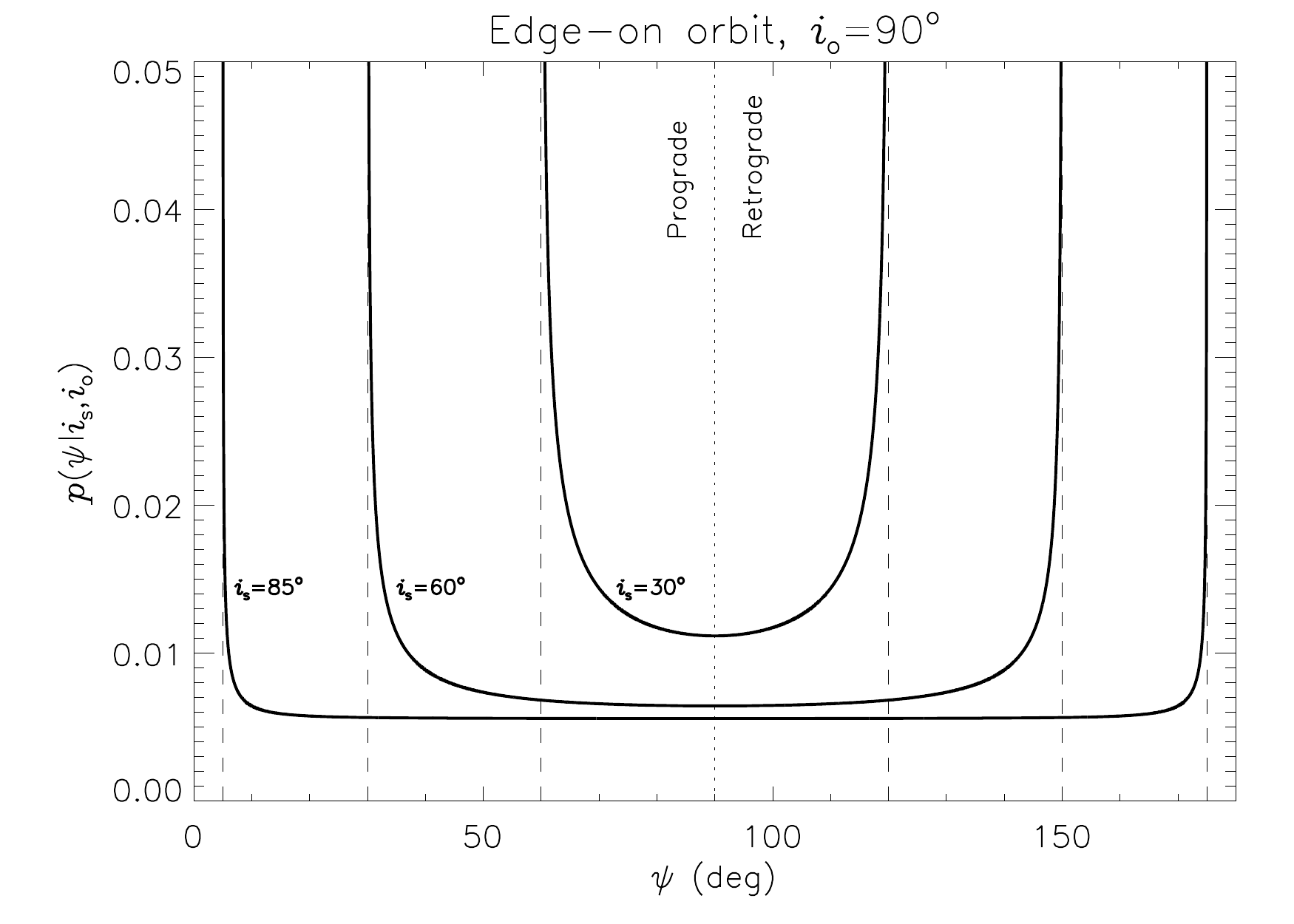}
\caption{\small Posterior probability distribution for the spin-orbit angle $\psi$ conditioned on $i_{\rm s}$ and $i_{\rm o}$, $p(\psi|i_{\rm s},i_{\rm o})$. We have assumed an edge-on orbit (i.e., $i_{\rm o}\!=\!90\degr$), having selected three error-free values for $i_{\rm s}$ ($i_{\rm s}\!=\!30\degr$, $i_{\rm s}\!=\!60\degr$, and $i_{\rm s}\!=\!85\degr$). The vertical dashed lines are placed at the asymptotes $\psi\!=\!|i_{\rm o}-i_{\rm s}|$ and $\psi\!=\!i_{\rm o}+i_{\rm s}$. The vertical dotted line at $\psi\!=\!\pi/2$ marks the transition between a prograde and a retrograde orbit.\label{fig:analyticalpost}}
\end{figure}

For an individual transiting system, we would still expect to place mild constraints on $\psi$ even when lacking a measurement of $\lambda$. In Fig.~\ref{fig:analyticalpost} we show the posterior probability distribution (after normalization) for $\psi$ conditioned on $i_{\rm s}$ and $i_{\rm o}$, $p(\psi|i_{\rm s},i_{\rm o})$ (see Appendix \ref{append:psipost} for a derivation of the analytical expression). We have assumed an edge-on orbit (i.e., $i_{\rm o}\!=\!90\degr$), having selected three error-free values for the stellar inclination angle ($i_{\rm s}\!=\!30\degr$, $i_{\rm s}\!=\!60\degr$, and $i_{\rm s}\!=\!85\degr$). The main conclusions to be drawn from this simple exercise follow immediately from an inspection of Fig.~\ref{fig:analyticalpost}: For a transiting system, a small value of $i_{\rm s}$ implies a spin-orbit misalignment. The converse is not true, since a large value of $i_{\rm s}$ is consistent with, but does not necessarily imply, a spin-orbit alignment. The interpretation of the spin-orbit alignment in terms of the measured $i_{\rm s}$ can thus be ambiguous. This provides one of the main motivations for this work, namely that, in order to draw general inferences about spin-orbit alignment, a statistical analysis of an ensemble of such measurements is needed.

\section{Asteroseismic analysis}\label{sec:astero}

\subsection{Sample characterization}\label{sec:sample}
Our asteroseismic sample consists of 25 solar-type KOIs, of which 20 are confirmed hosts and thus have been assigned a Kepler ID (see Table \ref{tb:sample} for a complete list). At the time of writing, all planets in the systems awaiting confirmation have been designated as `CANDIDATE' in the cumulative table of the NASA Exoplanet Archive\footnote{\url{http://exoplanetarchive.ipac.caltech.edu/}} \citep{Akeson13}. Fundamental properties (e.g., surface gravity, radius, mass, and age) from a detailed asteroseismic analysis are available for all the KOIs in the sample \citep{VSA15}.

\begin{figure}[!t]
\centering
\includegraphics[width=0.70\linewidth]{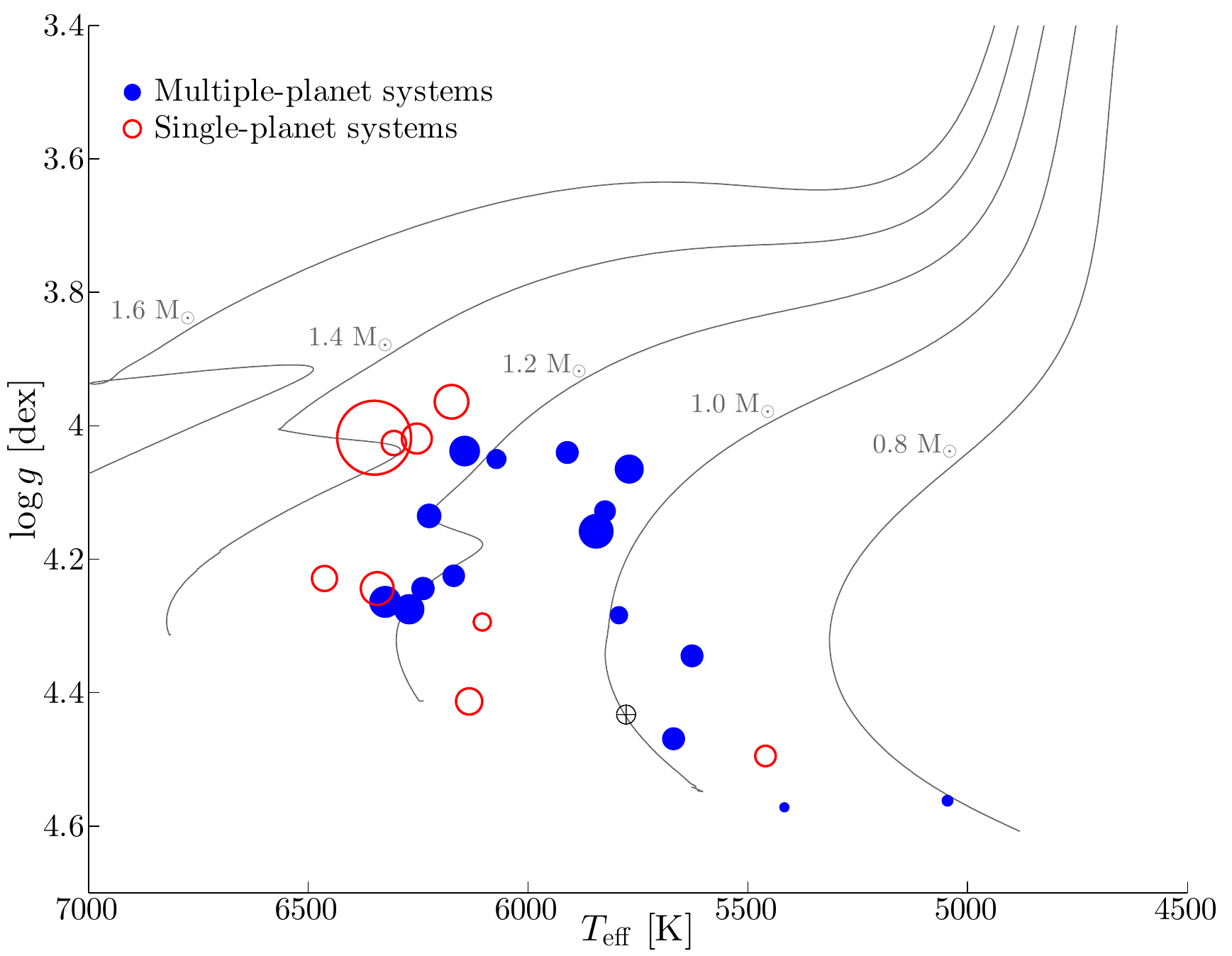}
\caption{\small Surface gravity vs.~effective temperature for the KOIs in the asteroseismic sample. Filled blue circles represent multiple-planet systems, while open red circles represent single-planet systems. Symbol size scales linearly with planetary size (for multiple-planet systems, the smallest planet is considered). For reference, a hypothetical solar twin harboring an Earth-size planet is represented by `$\earth$'. Solar-calibrated evolutionary tracks spanning the mass range $0.8$--$1.6\,{\rm M}_\sun$ (in steps of $0.2\,{\rm M}_\sun$) are shown as continuous lines. These tracks have been computed using the Modules for Experiments in Stellar Astrophysics \citep[MESA;][]{Paxton11,Paxton13} evolution code.\label{fig:HR}}
\end{figure}

A systematic study of {\it Kepler} planet-candidate hosts using asteroseismology was presented by \citet{HuberKOIs}, in which fundamental properties were determined for 66 host stars based on their average asteroseismic parameters. This raised the number of {\it Kepler} hosts with asteroseismic solutions to nearly 80 stars. Whether or not a given host star is included in the present sample was determined by our ability to resolve and extract signatures of rotation in the oscillation spectrum, which required relatively bright targets ({\it Kepler}-band magnitude $m_{\rm Kep}\!\la\!12$) and multi-quarter observations. The intrinsic stellar properties have also played a crucial role in this regard, since it is well known that the signatures of rotation tend to be substantially blended in the power spectra of main-sequence solar-like oscillators hotter than about $6400\:{\rm K}$ \citep[e.g.,][]{Appourchaux12}. Figure \ref{fig:HR} displays the sample on a $\log g$ vs.~$T_{\rm eff}$ diagram. They are predominantly positioned along the main sequence and range in spectral type from early K to late F (i.e., $5000\:{\rm K}\!\la\!T_{\rm eff}\!\la\!6500\:{\rm K}$). A number of stars in the sample seem to have evolved slightly beyond the main-sequence phase, one example being the bright subgiant Kepler-21 \citep{Kepler-21}. There is also varying level of evidence of mixed \citep[e.g.,][]{Osaki,Aizenman} quadrupole modes in the power spectra of Kepler-36, Kepler-100, Kepler-128, and Kepler-129, an indication that these stars may have already left the main sequence. The fact that central hydrogen has been depleted in models of these stars \citep[][]{VSA15} supports this scenario.

The sample contains 16 multiple-planet systems, of which all except Kepler-93 \citep[][]{Kepler-93} and Kepler-410 A \citep[][]{VanEylen14} are also multi-transiting systems\footnote{Although being a single-transiting system, transit-timing variations (TTVs) suggest the presence of at least one additional (non-transiting) planet in the Kepler-410 A system.}. Moreover, a non-transiting planet was revealed by RV measurements orbiting beyond the two transiting planets in the Kepler-25 \citep{Marcy14} and Kepler-68 \citep{Kepler-68} systems. Most of the multi-transiting systems in our sample have had the eccentricities of their planets measured from transit photometry \citep{EylenAlbrecht}, which revealed a tendency toward low eccentricities that are consistent with nearly circular orbits. Of the remaining 9 single-planet systems, only one \citep[HAT-P-7;][]{HAT-P-7b} is a hot-Jupiter system. From Table \ref{tb:sample}, we also see a clear prevalence of systems that contain small planets (i.e., $R_{\rm p}\!\leq\!4\,R_\earth$) and long-period planets (i.e., $P_{\rm o}\!>\!10\:{\rm d}$).

\subsection{Data preparation}\label{sec:data}
Raw pixel data \citep{Jenkins10} were downloaded from the {\it Kepler} Asteroseismic Science Operations Center\footnote{\url{http://kasoc.phys.au.dk}} (KASOC) database and subsequently run through the homonymous filter \citep[][]{KASOCfilter}. The KASOC filter has been specifically designed to automatically carry out the preparation of {\it Kepler} photometric time series of planet-candidate hosts for asteroseismic analysis. The time series used in this work were acquired in short-cadence mode ($\Delta t\!\sim\!58.85\:{\rm s}$) to allow investigation of solar-like oscillations in main-sequence stars, whose dominant periods are typically several minutes. A weighted least-squares sine-wave-fitting method was then used to compute rms-scaled power spectra of the time series for further analysis \citep[][]{Kjeldsen92,Frandsen95}.

\subsection{Estimation of the stellar inclination angle}\label{sec:incl}

\subsubsection{Principle}\label{sec:inclsub1}
Solar-like oscillations are predominantly acoustic global standing waves. Commonly called p modes, owing to the fact that pressure plays the role of the restoring force, these modes are intrinsically damped and stochastically excited by near-surface convection \citep[for a review see, e.g.,][]{JCDReview,CunhaReview,ChaplinReview}. The oscillation modes are characterized by the radial order $n$, the spherical degree $l$, and the azimuthal order $m$. Radial modes have $l\!=\!0$, whereas non-radial modes have $l\!>\!0$. Values of $m$ range from $-l$ to $l$, meaning that there are $2l+1$ azimuthal components for a given multiplet of degree $l$. Observed oscillation modes are typically high-order modes of low spherical degree, with the associated power spectrum showing a pattern of peaks with near-regular frequency separations \citep{Vandakurov,Tassoul}.

The asteroseismic estimation of $i_{\rm s}$ rests on our ability to resolve and extract signatures of rotation in the power spectra of non-radial modes of oscillation. Rotation lifts the degeneracy in the frequencies, $\nu_{nl}$, of non-radial modes and introduces a dependence of the oscillation frequencies on $m$, with prograde (retrograde) modes (with $m\!>\!0$ and $m\!<\!0$, respectively) having frequencies slightly higher (lower) than the axisymmetric mode ($m\!=\!0$) in the observer's frame of reference. For the fairly modest values of the stellar angular velocity $\Omega$ that are typical of solar-like oscillators, the effect of rotation can be treated following a perturbative analysis \citep[e.g.,][]{Reese06} and the star is generally assumed to rotate as a solid body (i.e., $\Omega\!=\!{\rm const.}$). In the limit of solid-body rotation, the frequency $\nu_{nlm}$ of a mode, as observed in an inertial frame, can be expressed to first order as \citep{Ledoux51}:
\begin{equation}
\label{eq:Ledoux}
\nu_{nlm}=\nu_{nl0} + m\frac{\Omega}{2\pi} (1-C_{nl}) \approx \nu_{nl0} + m\nu_{\rm s} \, .
\end{equation}
The kinematic splitting $m\Omega/(2\pi)$ is corrected for the effect of the Coriolis force through the dimensionless Ledoux constant, $C_{nl}$. For high-order p modes, $C_{nl}\!\ll\!1$, with the rotational splitting being dominated by advection and given approximately by the angular velocity, i.e., $\nu_{\rm s}\!\approx\!\Omega/(2\pi)$ \citep[e.g.,][]{LundDiff,Davies16Cyg}. 

To a second order of approximation, centrifugal effects that disrupt the equilibrium structure of the star can be taken into account through an additional frequency perturbation \citep[e.g.,][]{Ballot10}. This perturbation scales as the ratio of the centrifugal to the gravitational forces at the stellar surface, i.e., $\Omega_{\rm surf}^2R_{\rm s}^3/(GM_{\rm s})$, where $G$ is the gravitational constant. We made use of the available values of $P_{\rm rot}$ in Table \ref{tb:sample} to compute the ratios of the surface angular velocity to the Keplerian break-up rotation rate, i.e., $\Omega_{\rm surf}/\Omega_{\rm K}\equiv2\pi/(P_{\rm rot}\sqrt{GM_{\rm s}/R_{\rm s}^3})$. We obtained $(\Omega_{\rm surf}/\Omega_{\rm K})^2\!\la\!9\!\times\!10^{-4}$ for the stars in the asteroseismic sample and have thus decided to neglect second-order effects in the present analysis.

Assuming energy equipartition between multiplet components with different azimuthal order\footnote{The predicted power asymmetries for stars in the asteroseismic sample are of the order of $1\,\%$ \citep[for an $l\!=\!1$ mode at the frequency of maximum oscillation amplitude $\nu_{\rm max}$;][]{Belkacem09}, which are negligible for our analysis.}, the dependence of mode power on $m$ is given by \citep{Dziembowski77,Dziembowski85,Gizon03}:
\begin{equation}
\label{eq:relpower}
\mathscr{E}_{l m}(i_{\rm s}) = \frac{(l-|m|)!}{(l+|m|)!} \left[P_l^{|m|}(\cos i_{\rm s})\right]^2 \, ,
\end{equation}
where $P_l^m(x)$ are the associated Legendre functions and the sum over $m$ of $\mathscr{E}_{l m}(i_{\rm s})$ has been normalized to unity. Measurement of the relative power of the azimuthal components in a non-radial multiplet thus provides a direct estimate of the stellar inclination angle. The above formalism further relies on the assumption that contributions to the observed intensity across the visible stellar disk depend only on the angular distance from the disk center, which is valid for photometric observations. According to Eq.~(\ref{eq:relpower}), when the stellar spin axis points toward the observer (pole-on configuration), only the axisymmetric mode is visible and no inference can thus be made about rotation. When the spin axis lies on the plane of the sky (edge-on configuration), as is approximately the case of the Sun, observations are essentially sensitive only to modes with even $|l-m|$. Given sufficient frequency resolution and S/N, it will be the intrinsic ratio $\nu_{\rm s}/\Gamma$ (where $\Gamma$ is the full width at half maximum, or linewidth, of the mode profile of each azimuthal component; see Eq.~\ref{eq:specmodel}) which determines whether it is possible to resolve the azimuthal components \citep{Ballot06,Procyon}. Moreover, dipole ($l\!=\!1$) modes are approximately three times more prominent in the power spectra of intensity observations than quadrupole ($l\!=\!2$) modes of similar frequency, and consequently it is the former modes that ultimately determine our ability to constrain $i_{\rm s}$.

\subsubsection{Results}\label{sec:inclsub2}

\begin{figure}[!ht]
\centering
\includegraphics[width=0.8\linewidth]{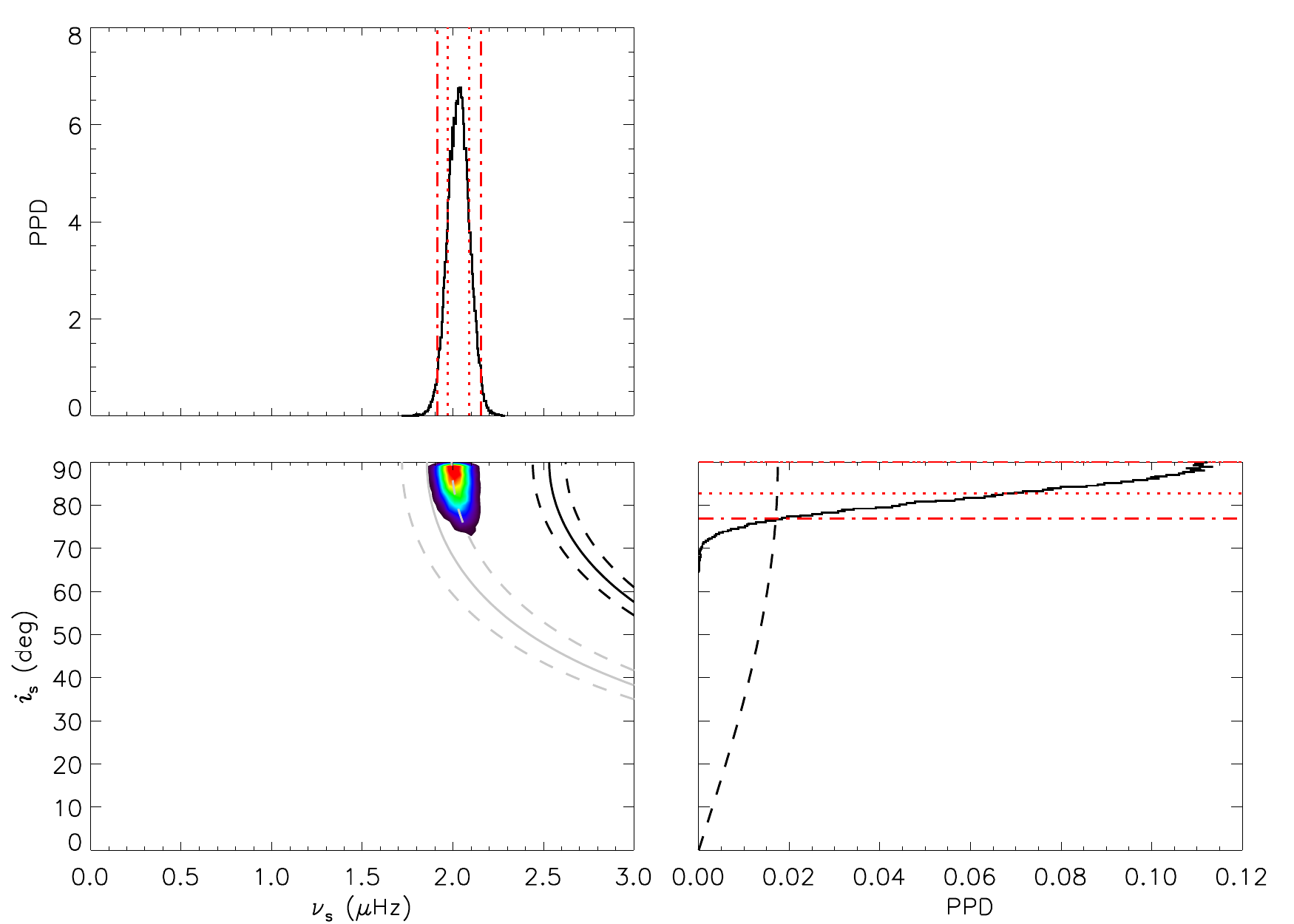}
\caption{\small Asteroseismic results on KIC~8866102 (Kepler-410 A, KOI-42). Bottom left panel: Joint PPD of $i_{\rm s}$ and $\nu_{\rm s}$ based on $80{,}000$ MCMC thinned (uniformly subsampled) samples. Also shown are two different estimates of the projected splitting (solid lines) with their associated 1-$\sigma$ envelopes (dashed lines): the black lines are based on the $v\sin i_{\rm s}$ from \citet{HuberKOIs} given in Table \ref{tb:sample}, whereas the gray lines are based on the value provided in \citet{Molenda13}. $v\sin i_{\rm s}$ values (and their literature sources) are listed in Table \ref{tb:sample} for all stars in the asteroseismic sample. Top left and bottom right panels: Marginalized PPDs. These are shown as histograms, with the number of bins defined according to Scott's normal reference rule \citep[][]{ScottBook}. Dotted and dot-dashed lines respectively enclose the $68.3\,\%$ and $95.4\,\%$ HPD credible regions \citep[e.g.,][chap.~3]{GregoryBook}. For reference, the dashed curve in the bottom right panel represents the (uninformative) isotropic prior on $i_{\rm s}$ adopted in the asteroseismic analysis.\label{fig:ppd008866102}}
\end{figure}

A detailed fitting of the modes of oscillation was conducted to extract signatures of rotation from the power spectra of stars in the asteroseismic sample (see Appendix \ref{append:specfit} for a description of the method; in Appendix \ref{append:artificial} we test the robustness of the returned uncertainties on $i_{\rm s}$). This allowed us to map the joint posterior probability distribution (PPD) of $i_{\rm s}$ and $\nu_{\rm s}$. An example is shown in Fig.~\ref{fig:ppd008866102} (bottom left panel) for Kepler-410 A, together with the corresponding marginalized PPDs (top left and bottom right panels). Also shown in the bottom left panel is an estimate of the projected splitting (black solid line), given by\footnote{$v\sin i_{\rm s}$ values (and their literature sources) are listed in Table \ref{tb:sample} for all stars in the asteroseismic sample.} $\nu_{\rm s}\sin i_{\rm s}\!\approx\!v\sin i_{\rm s}/(2\pi R_{\rm s})$. This assumes that the internal rotation rates probed by the asteroseismic splitting are similar to the surface rate of rotation. For a typical $1.2$-${\rm M}_\sun$ star in our sample (such as Kepler-410 A), however, the asteroseismic splitting is nearly equally sensitive to rotation in the radiative and convective zones \citep{LundDiff,Benomar15}, and the assumption that the surface and asteroseismic rotation periods are similar should thus be regarded as approximate. We also note that different authors very often disagree on the value of the measured $v\sin i_{\rm s}$. The bottom left panel of Fig.~\ref{fig:ppd008866102} displays such an example, where besides the projected splitting based on the $v\sin i_{\rm s}$ value from \citet{HuberKOIs} given in Table \ref{tb:sample} (black solid line), we also use the other available $v\sin i_{\rm s}$ value for this star from \citet{Molenda13} (gray solid line). Accurate and precise $v\sin i_{\rm s}$ measurements from spectroscopy are difficult to obtain for solar-type stars, since the rotational line broadening is often comparable to the effects of instrumental broadening and macroturbulence ($v_{\rm mac}$). In a recent work, \citet{Doyle14} used accurate $v\sin i_{\rm s}$ values from asteroseismology to break the degeneracy between the spectroscopic $v\sin i_{\rm s}$ and $v_{\rm mac}$ in spectral line profiles, thus obtaining a calibration for stellar macroturbulence. The marginalized PPDs are presented as histograms, with dotted and dot-dashed lines respectively enclosing the $68.3\,\%$ and $95.4\,\%$ highest posterior density (HPD) credible regions. Our results for the stellar inclination angle of Kepler-410 A are in very good agreement with those obtained by \citet{VanEylen14}. The stellar inclination is tightly constrained and its posterior distribution indicates an edge-on configuration. This was to be expected from inspection of the $l\!=\!1$ mode profiles in Fig.~\ref{fig:peakbag}.

\clearpage

\clearpage

\begin{figure}[!t]
\centering
\includegraphics[width=0.8\linewidth]{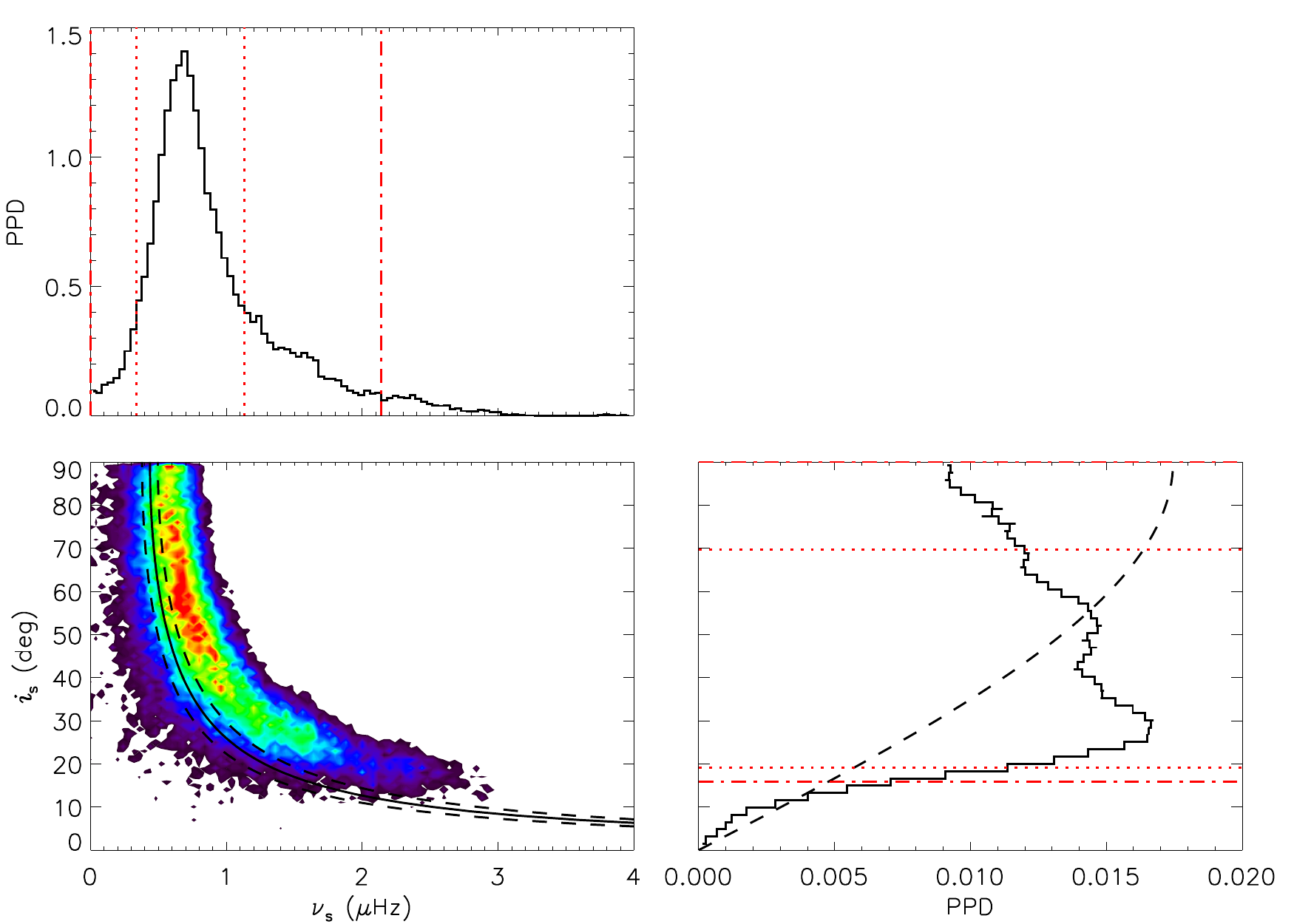}
\caption{\small Asteroseismic results on KIC~10666592 (HAT-P-7, Kepler-2, KOI-2). Similar to Fig.~\ref{fig:ppd008866102}.\label{fig:ppd010666592}}
\end{figure}

\begin{figure}[!b]
\centering
\includegraphics[width=0.6\linewidth]{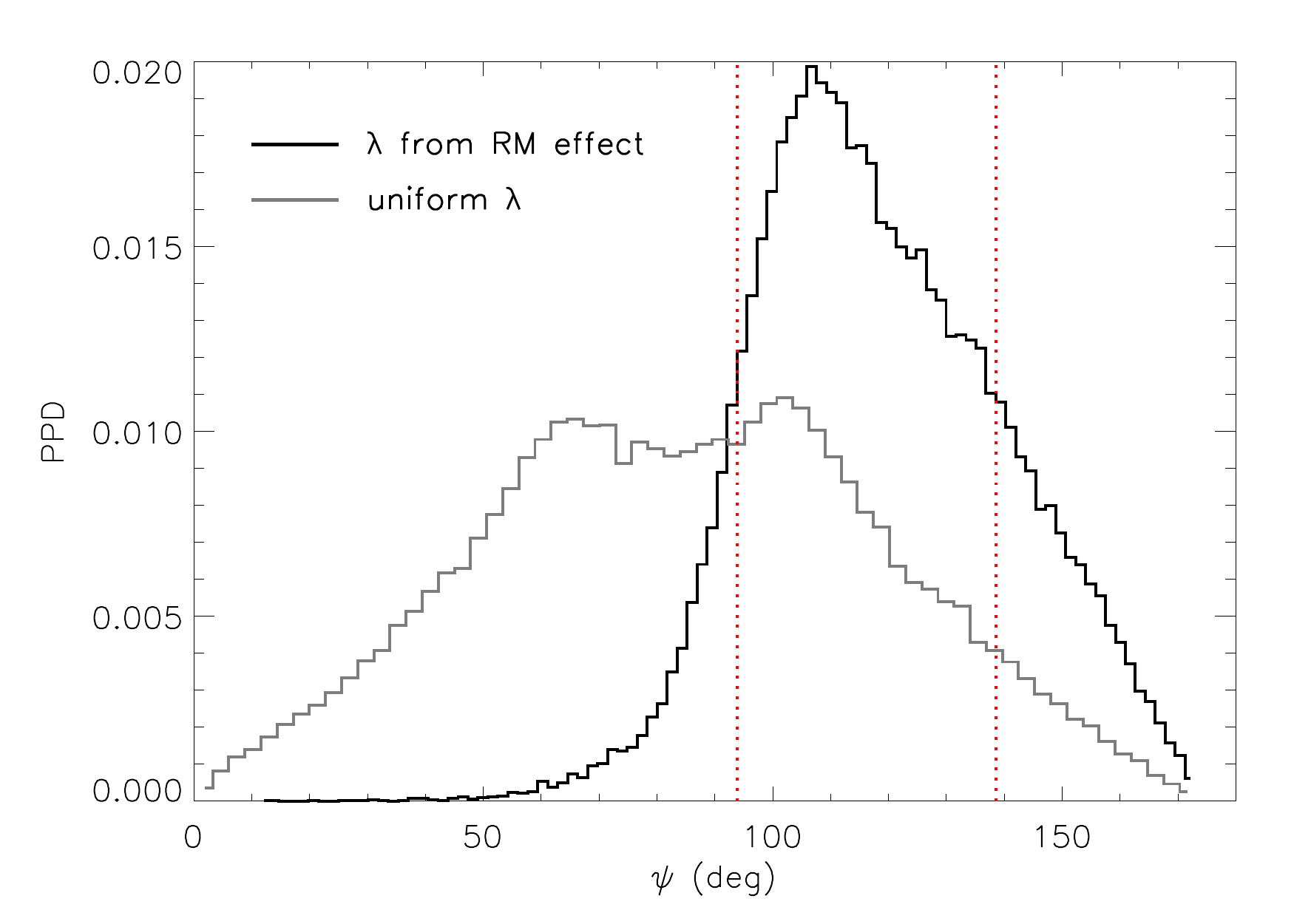}
\caption{\small Posterior probability distribution of the spin-orbit angle $\psi$ of KIC~10666592 (HAT-P-7, Kepler-2, KOI-2). The black histogram was obtained by means of Monte Carlo simulations using the PPD of $i_{\rm s}$ from our analysis, and assuming normal and uniform distributions, respectively, for $\lambda$ and $i_{\rm o}$ around their adopted literature values. Dotted lines enclose its associated $68.3\,\%$ HPD credible region. The gray histogram was obtained by assuming an isotropic distribution in $\lambda$.\label{fig:obl010666592}}
\end{figure}


\begin{figure}[!t]
\centering
\includegraphics[width=0.8\linewidth]{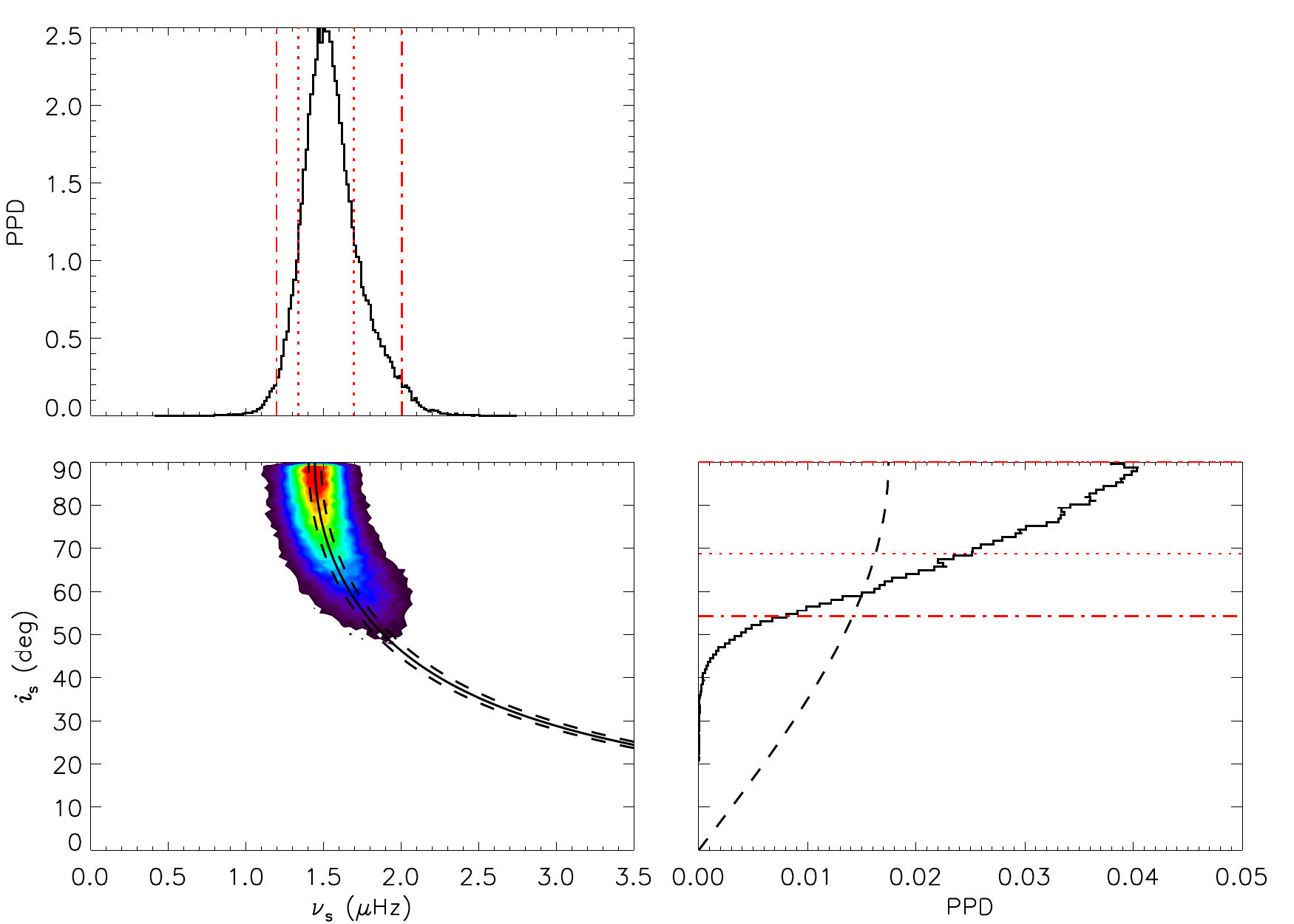}
\caption{\small Asteroseismic results on KIC~4349452 (Kepler-25, KOI-244). Similar to Fig.~\ref{fig:ppd008866102}.\label{fig:ppd004349452}}
\end{figure}

\begin{figure}[!b]
\centering
\includegraphics[width=0.6\linewidth]{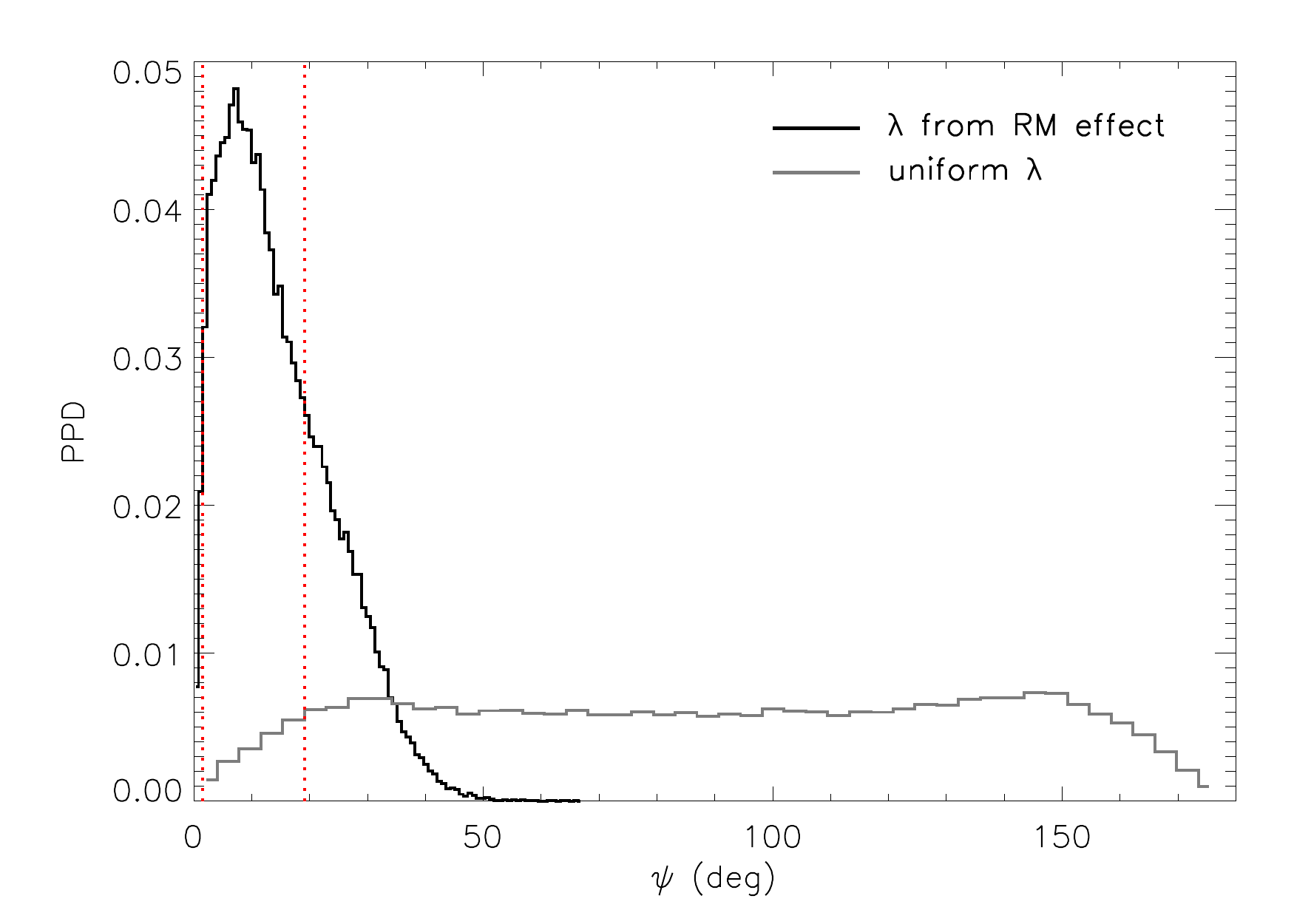}
\caption{\small Posterior probability distribution of the spin-orbit angle $\psi$ of KIC~4349452 (Kepler-25, KOI-244). Similar to Fig.~\ref{fig:obl010666592}.\label{fig:obl004349452}}
\end{figure}

\clearpage

Next, we present the results for the two targets in the asteroseismic sample that have an available $\lambda$ measurement in the literature, namely, HAT-P-7 and Kepler-25 (see Figs.~\ref{fig:ppd010666592}--\ref{fig:obl004349452}; see also Table \ref{tb:obliquities}). Of particular interest are Figs.~\ref{fig:obl010666592} and \ref{fig:obl004349452}, showing the derived PPDs of the spin-orbit angle $\psi$ (black histograms). These were sampled by means of Monte Carlo simulations (via Eq.~\ref{eq:geom2}) using the PPD of $i_{\rm s}$ from our analysis, and assuming normal and uniform distributions, respectively, for $\lambda$ and $i_{\rm o}$ around their adopted literature values (see Table \ref{tb:obliquities}). Also shown are the PPDs obtained by instead assuming an isotropic distribution in $\lambda$ (gray histograms). We find that the orbit of the hot-Jupiter HAT-P-7b is likely to be retrograde, while that of Kepler-25c is well aligned with the stellar spin axis. Our findings for Kepler-25 are not in agreement with the statement made by \citet{Benomar14} that this system is mildly misaligned and this point will be further discussed in Sect.~\ref{sec:conclusions}.

Asteroseismic results for the remainder of the stars in the sample are shown in Figs.~\ref{fig:ppd003425851}--\ref{fig:ppd011904151}. Even though $i_{\rm s}$ is independent of the rotational splitting $\nu_{\rm s}$, their measured values are highly correlated, as can be seen from the joint PPDs. Consequently, even when a constrained solution for $i_{\rm s}$ and $\nu_{\rm s}$ cannot be obtained, as is often the case in the present analysis, it is still possible to estimate the projected splitting $\nu_{\rm s}\sin i_{\rm s}$. Table \ref{tb:results} reports the $68.3\,\%$ and $95.4\,\%$ HPD credible regions for the stellar inclination angle $i_{\rm s}$ for all the stars in the asteroseismic sample. Note that the PPD of $i_{\rm s}$ is often too asymmetric to be adequately summarized by a single estimate and we have thus refrained from tabulating any measures of central tendency, such as the median or the mode of the distributions. We stress that Bayesian inference does not provide point estimates of parameters. Instead the Bayesian solution to the problem of parameter estimation is the full posterior probability distribution\footnote{Full posterior probability distributions are made available upon request.}. The several statistical analyses conducted in the next section are based on the full posteriors. Our results for $i_{\rm s}$ suggest that there are five other systems in the asteroseismic sample besides HAT-P-7 that could potentially be misaligned, namely, Kepler-50, Kepler-93, Kepler-145, Kepler-409, and KOI-280. And while this interpretation is certainly valid when considering the $68.3\,\%$ HPD credible regions, we note that our results are consistent with alignment at the 2-$\sigma$ level for all systems in the sample.

\begin{figure}[!t]
\centering
\includegraphics[width=0.75\linewidth]{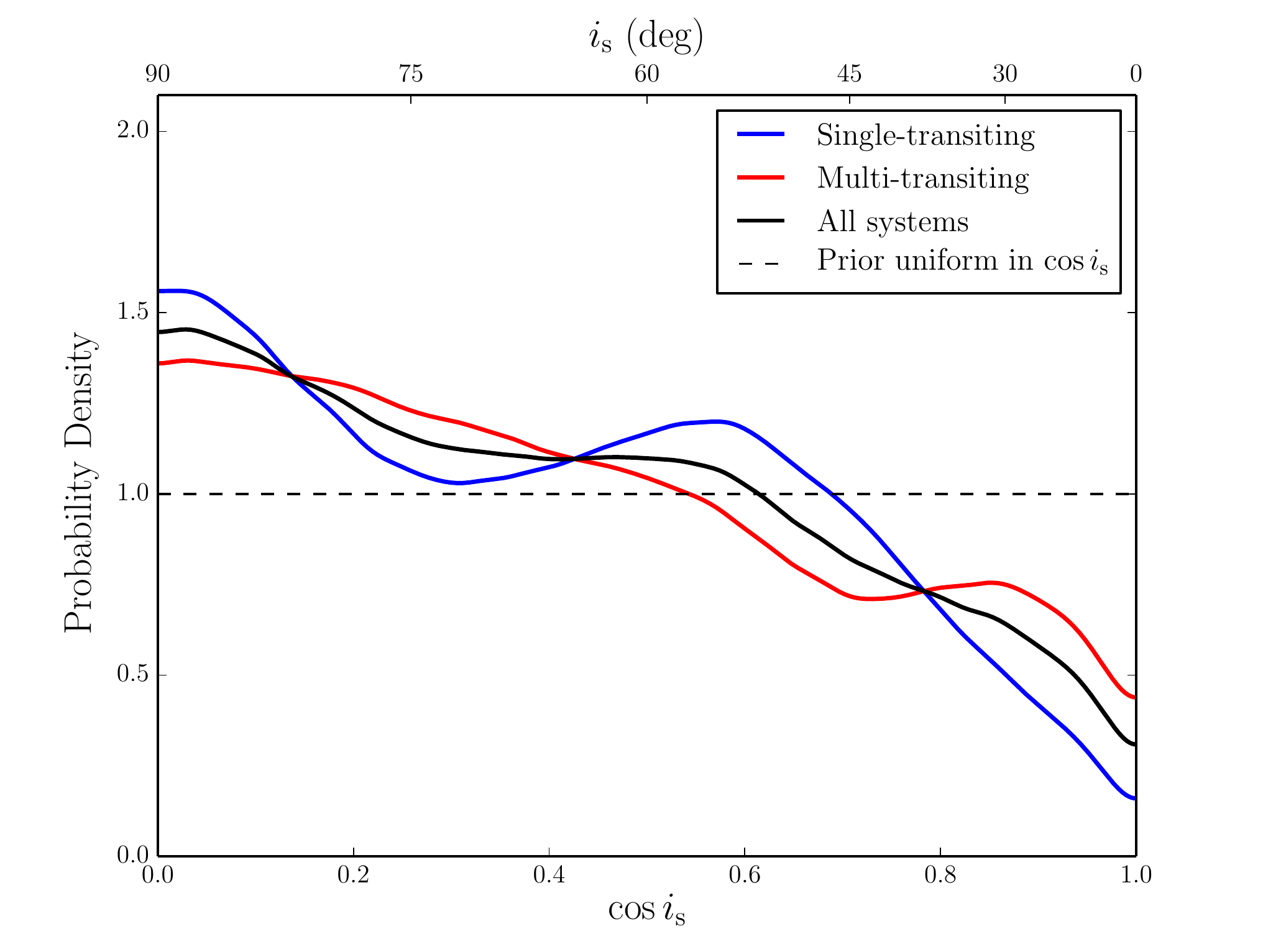}
\caption{\small Concatenated posterior probability distributions of $\cos i_{\rm s}$ (kernel density estimates) for stars in the asteroseismic sample. The overall concatenated posterior is depicted in black, whereas the concatenated posteriors for single- and multi-transiting systems are depicted in blue and red, respectively. For reference, the horizontal dashed line represents the (uninformative) isotropic prior on $i_{\rm s}$ (or, equivalently, the uniform prior on $\cos i_{\rm s}$) adopted in the asteroseismic analysis.
\label{fig:averageposterior}}
\end{figure}

\section{Statistical constraints on spin-orbit alignment}\label{sec:stat}
\subsection{Asteroseismic sample}\label{sec:statastero}
Figure \ref{fig:averageposterior} shows the concatenated posterior probability distributions of $\cos i_{\rm s}$, made by concatenating the individual posteriors and normalizing. This is shown for all stars in the asteroseismic sample (25 systems), as well as for the subsamples of single- (11) and multi-transiting (14) systems. The isotropic distribution is represented by a horizontal dashed line. We may have naively expected the concatenated posterior(s) not to exhibit the amount of structure seen in Fig.~\ref{fig:averageposterior}. This is a consequence of the small size of the asteroseismic sample, with a few individual posteriors eventually dominating at specific $i_{\rm s}$ ranges.

The stars in the asteroseismic sample appear to preferentially display large values of $i_{\rm s}$ (or, equivalently, small values of $\cos i_{\rm s}$), this being more accentuated than if the $i_{\rm s}$ had been drawn from an isotropic distribution. This departure from isotropy suggests that the directions of the stellar spin ($\mathbf{n}_{\rm s}$) and planetary orbital ($\mathbf{n}_{\rm o}$) axes are correlated. Given samplings of the posterior probability distributions of $\cos i_{\rm s}$ for the $N\!=\!25$ stars in the asteroseismic sample, $\{\cos i_{\rm s}\}_{n=1}^N$, we would like to infer the distribution of the spin-orbit angle $\psi$. As a first step, we need to devise a parameterized model for the distribution function of $\psi$, $f_{\boldsymbol{\alpha}}(\psi)$, as well as a prior on the model parameters $\boldsymbol{\alpha}$. We could then, in principle, use the observed data to constrain these parameters by means of a hierarchical Bayesian analysis (see Appendix \ref{append:hierarchical} for details). We note that the concatenated posteriors in Fig.~\ref{fig:averageposterior} are only shown here for comparison to fig.~9 of \citet{Hirano14} and fig.~5 of \citet{Morton14}. We do not use these concatenated posteriors directly for inference. Instead, we conduct a forward modeling of the observed data by defining a model distribution $f_{\boldsymbol{\alpha}}(\psi)$ and finding the model parameters $\alpha$ that best explain those data. Since we are modeling the distribution prior to observation, we are effectively performing a deconvolution that works on heteroscedastic data (i.e., with changing variance).

\begin{figure}[!t]
\centering
\includegraphics*[scale=0.68]{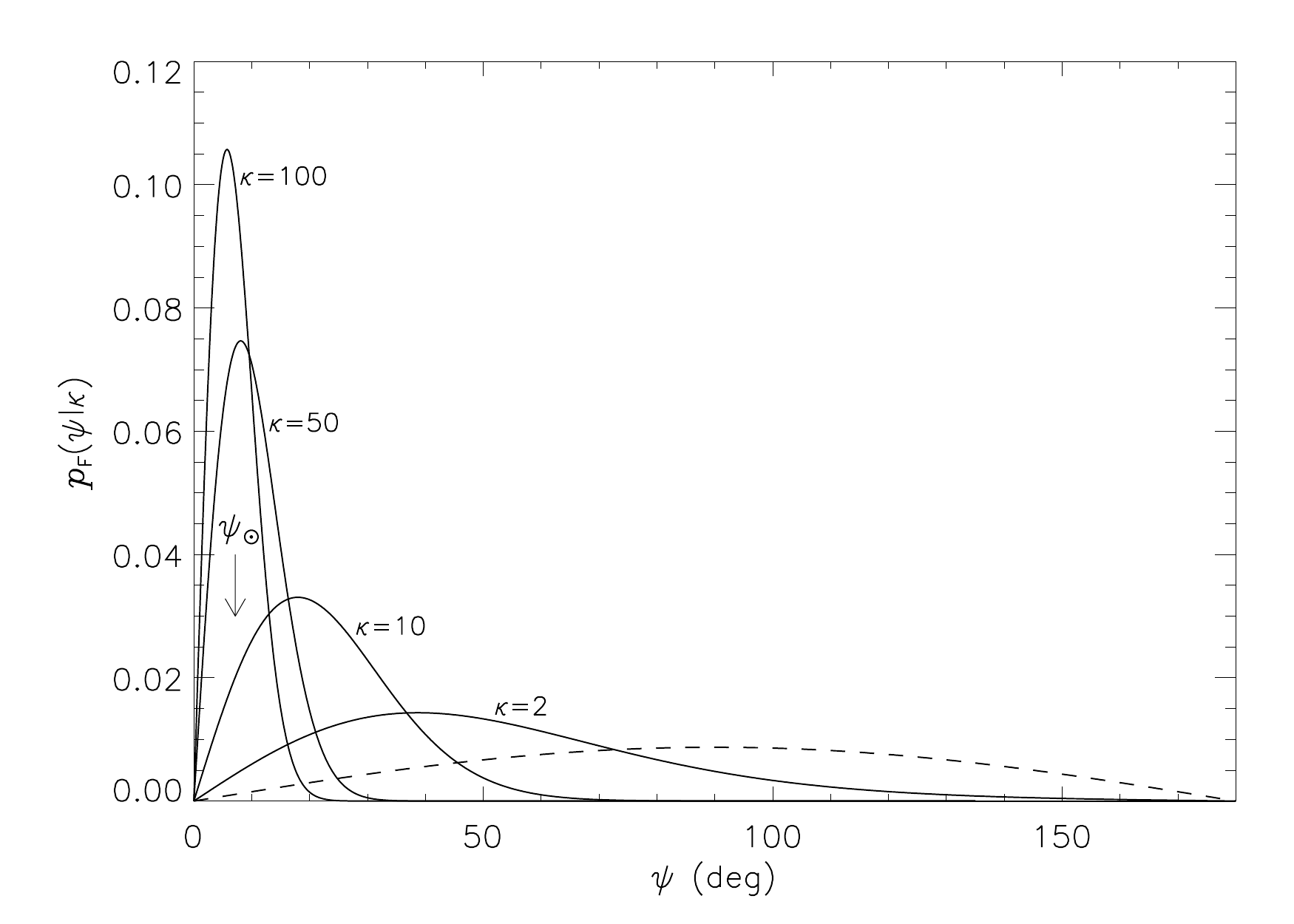}
\includegraphics*[scale=0.68]{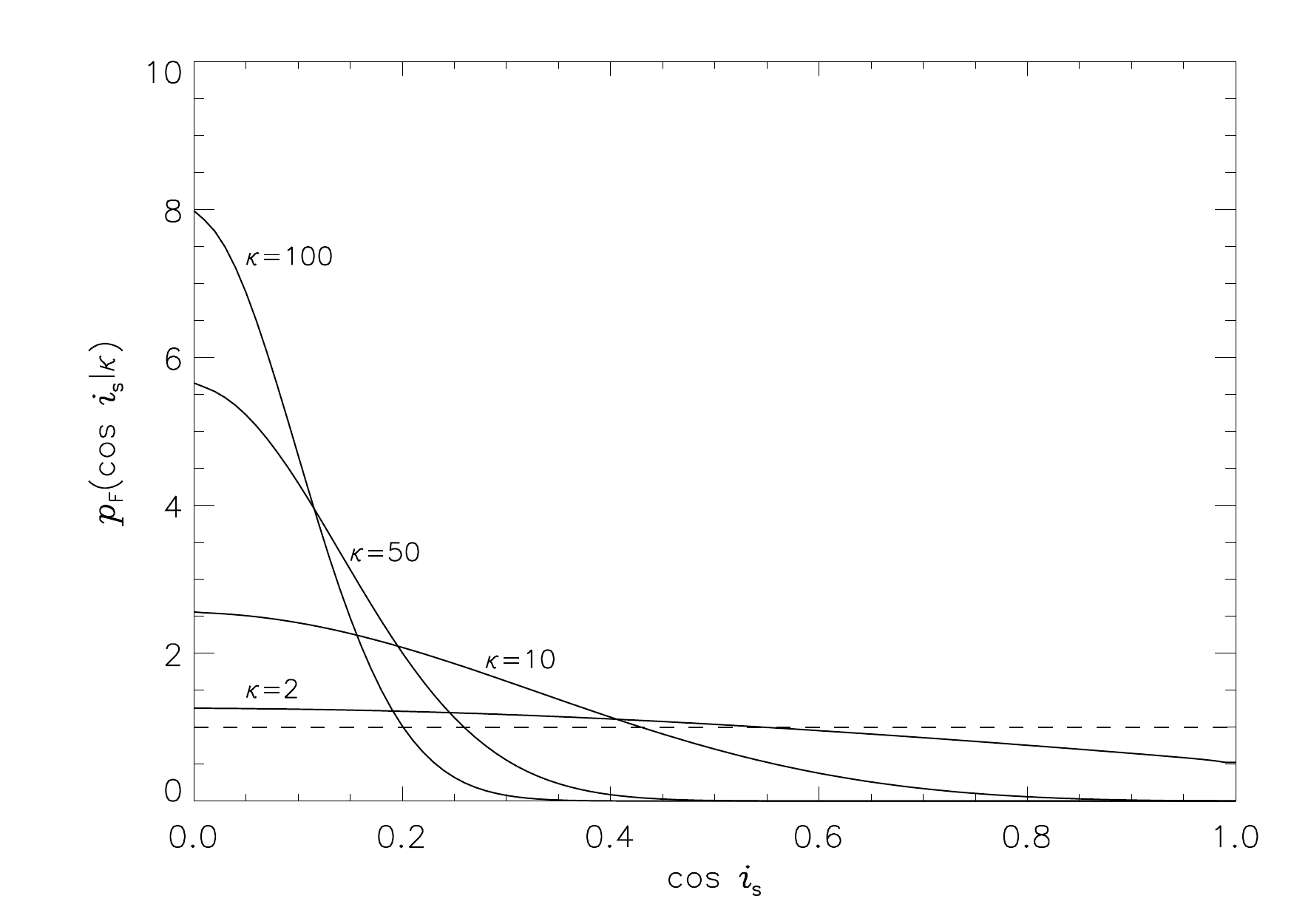}
\caption{\small Fisher probability distribution. Top panel: The individual Fisher distributions are characterized by different concentration parameters $\kappa$, namely, $\kappa\!=\!2,10,50$, and 100. The solar value $\psi_\sun\!=\!7.155\degr$ is shown for reference \citep{BeckGiles}. Bottom panel: The distribution $p_{\rm F}(\cos i_{\rm s}|\kappa)$ is shown for $\kappa\!=\!2,10,50$, and 100. The isotropic distribution is depicted by a dashed curve in both panels.\label{fig:Fisher}}
\end{figure}

{\it A Fisher distribution.} An isotropic distribution for $\psi$ is known to be inadequate \citep[e.g.,][]{Winn06}, which also follows from our realization above that $\mathbf{n}_{\rm s}$ and $\mathbf{n}_{\rm o}$ are apparently correlated. We choose to model the distribution function of $\psi$ as a Fisher distribution \citep{Fisher}, the analog of a zero-mean Gaussian distribution on a sphere. The Fisher distribution has been previously proposed by \citet{Fabrycky09} to model $f_{\boldsymbol{\alpha}}(\psi)$. Its probability density function is given by  
\begin{equation}
\label{eq:Fisher}
f_{\boldsymbol{\alpha}}(\psi) \equiv p_{\rm F}(\psi|\kappa) = \frac{\kappa}{2\sinh \kappa} \exp(\kappa \cos\psi) \sin\psi \, , 
\end{equation}
where $\kappa$ measures the probability concentration around $\psi\!=\!0$. As $\kappa\!\rightarrow\!0$, the distribution becomes the isotropic distribution, whereas for $\kappa\!\rightarrow\!\infty$ it becomes a Rayleigh distribution of width $\sigma\!=\!\kappa^{-1/2}$. However, our observable is $\cos i_{\rm s}$ and not $\psi$, and so we instead use the probability distribution for $\cos i_{\rm s}$ given $\kappa$ as derived by \citet{Morton14} (hereafter MW14):
\begin{equation}
\label{eq:Fisher2}
f_{\boldsymbol{\alpha}}(\cos i_{\rm s}) \equiv p_{\rm F}(\cos i_{\rm s}|\kappa) = \frac{2\kappa}{\pi \sinh \kappa} \int_{\cos i_{\rm s}}^1 \frac{\cosh(\kappa\sqrt{1-y^2})}{\sqrt{1-y^2}} \frac{1}{\sqrt{1-(\cos i_{\rm s}/y)^2}} \, {\rm d}y \, .
\end{equation}
We refer to this parameterized model as the `single-Fisher model' and adopt a uniform prior for $\kappa$ in the range [0,200]. Figure \ref{fig:Fisher} shows $p_{\rm F}(\psi|\kappa)$ and $p_{\rm F}(\cos i_{\rm s}|\kappa)$ for a set of concentration parameters $\kappa$. Finally, we sample the joint PPD of the model parameters $\boldsymbol{\alpha}$ (or simply $\kappa$ in the present case) following a hierarchical Bayesian scheme.

\begin{figure}[!t]
\centering
\includegraphics[width=0.75\linewidth]{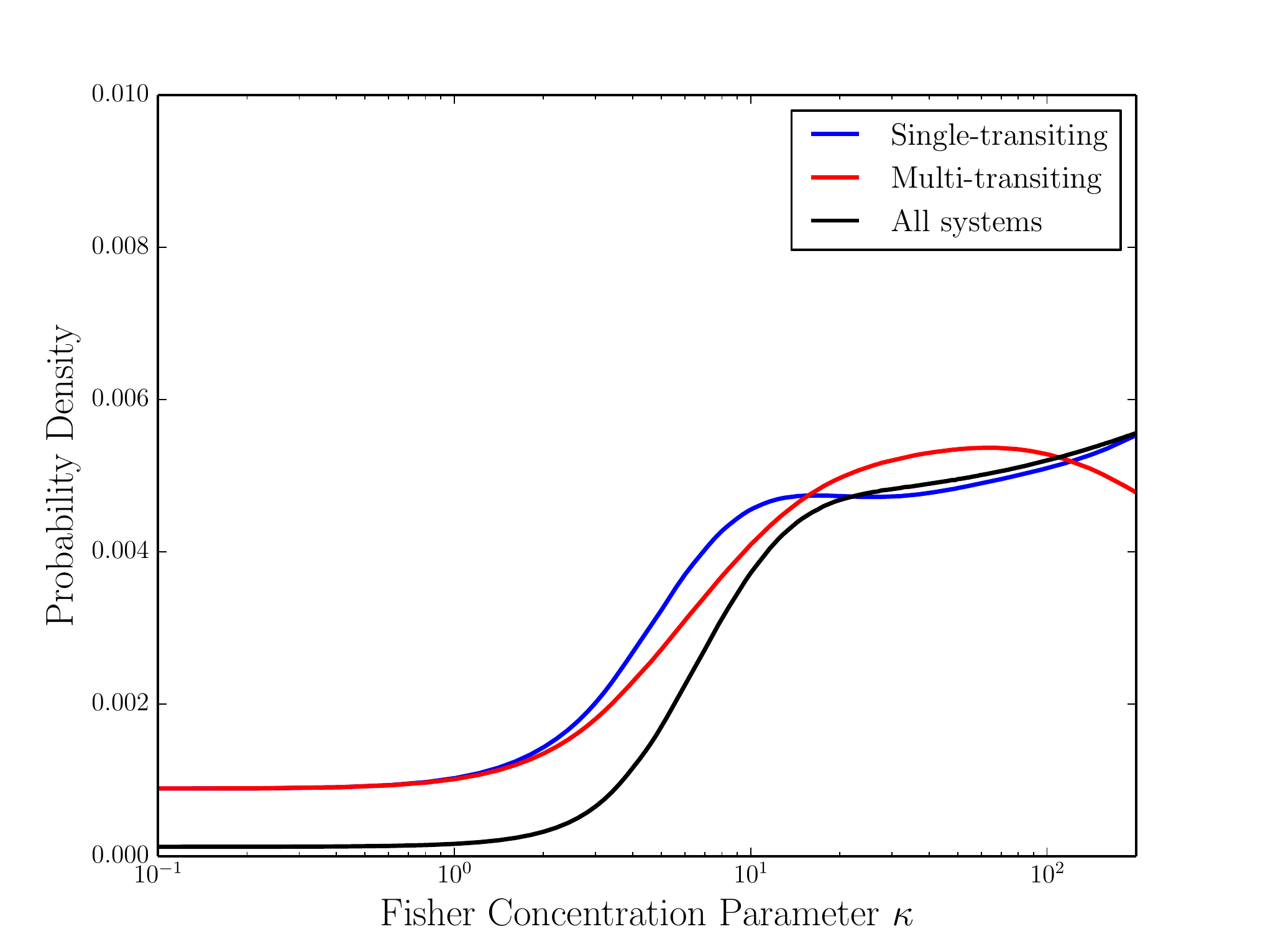}
\caption{\small Posterior probability distribution of the concentration parameter $\kappa$ in the single-Fisher model (asteroseismic sample). Distributions for the different cohorts of systems are color-coded. A logarithmic scale is used along the horizontal axis to emphasize the sharp fall-off at low $\kappa$.\label{fig:post_singleF_kappa_astero}}
\end{figure}

The PPD of the concentration parameter $\kappa$ is shown in Fig.~\ref{fig:post_singleF_kappa_astero}. Distributions for the different cohorts of systems are color-coded. Table \ref{tb:reshierarchical} presents a summary of the results. The posterior of $\kappa$ is dominated by the (uniform) prior over most of the range [0,200], the exception being at small $\kappa$, where the posterior falls sharply at $\kappa\!\approx\!10$ (the logarithmic scale along the horizontal axis in Fig.~\ref{fig:post_singleF_kappa_astero} emphasizes this point). Based on the lower bounds of the $95.4\,\%$ HPD credible regions (see Table \ref{tb:reshierarchical}), it can be stated that the concentration parameter $\kappa$ is significantly different from zero, thus ruling out isotropy.

The constraints on spin-orbit alignment made possible by an analysis of the asteroseismic sample are in qualitative agreement with the outcome of the analysis presented in \citet[][]{Hirano14}, who have estimated $i_{\rm s}$ for a sample of 25 (coincidentally) KOIs based on measurements of their rotation period, rotational line broadening, and stellar radius. Furthermore, our results cannot be used to support the recent finding by MW14 that the obliquities of systems with a single transiting planet are systematically larger than those with multiple transiting planets. MW14 have, however, considered a substantially larger sample, having performed an ensemble analysis of a compilation of 70 KOIs whose obliquity measurements had previously been published \citep[these included most of the KOIs analyzed by][]{Hirano14}.

The small size of the asteroseismic sample prevents us from placing more stringent constraints on spin-orbit alignment and more systems would be needed to draw firmer conclusions. In the next section we explore the possibility of extending the MW14 sample by combining it with our asteroseismic sample, thus forming the largest ensemble to date of measured $i_{\rm s}$ for exoplanet-host stars. We will be particularly interested in assessing the impact of adding our asteroseismic sample (representing just over $1/4$ of the combined sample) to the ensemble analysis.  

\subsection{Combined sample}\label{sec:comb}
\begin{figure}[!t]
\centering
\includegraphics[width=0.7\linewidth]{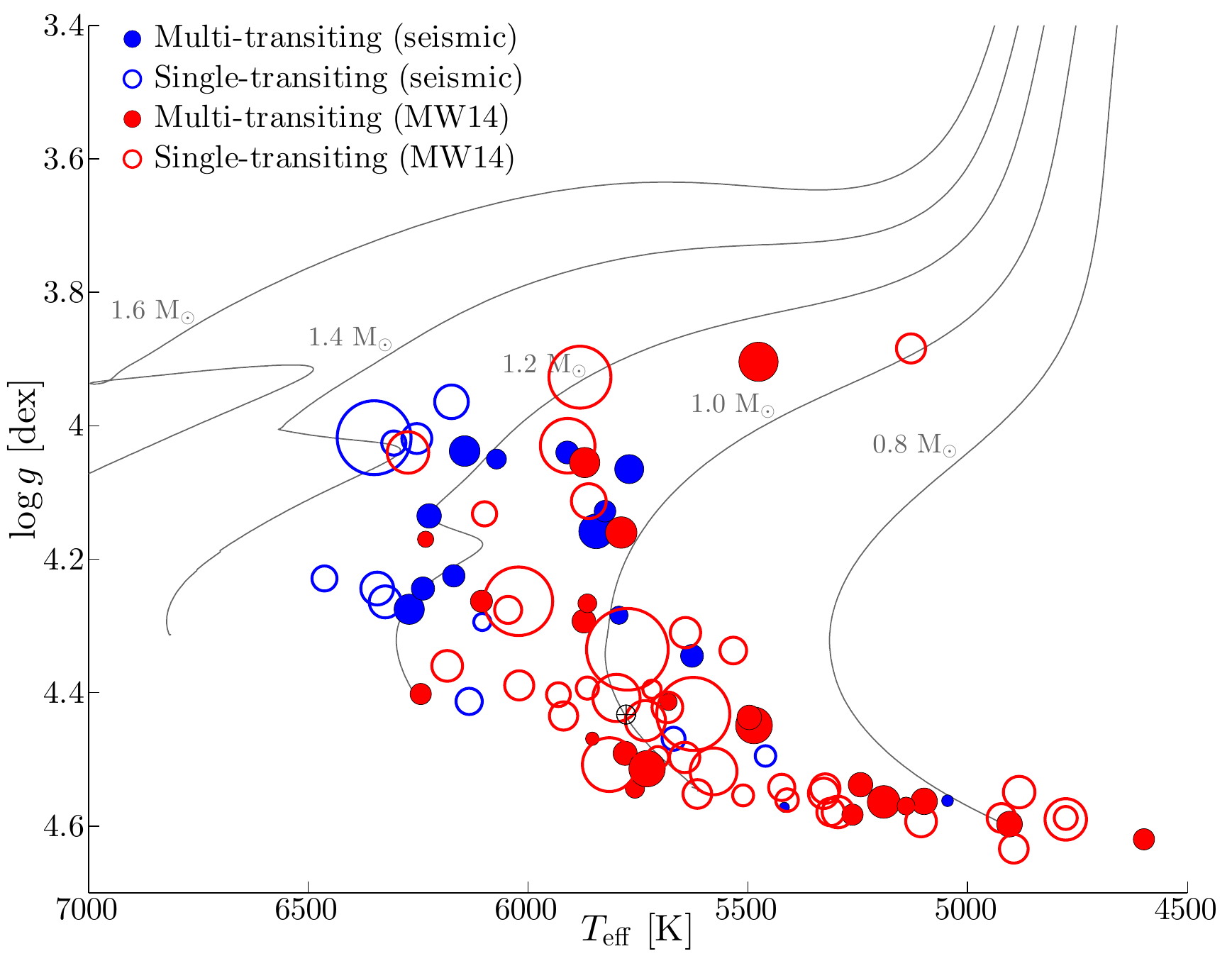}
\caption{\small Surface gravity vs.~effective temperature for the KOIs in the combined sample. Symbol size scales linearly with planetary size (for multiple-planet systems, the smallest planet is considered). For reference, a hypothetical solar twin harboring an Earth-size planet is represented by `$\earth$'. Solar-calibrated evolutionary tracks spanning the mass range $0.8$--$1.6\,{\rm M}_\sun$ (in steps of $0.2\,{\rm M}_\sun$) are shown as continuous lines.\label{fig:HR2}}
\end{figure}

We combined the asteroseismic sample with the compilation by MW14 in order to place statistical constraints on the spin-orbit alignment of exoplanet systems. There are 62 additional (non-overlapping) systems in the sample of MW14, of which 23 are multi-transiting and 39 are single-transiting (we note that KOI-2636 has in the meantime been reclassified as multi-transiting). We thus obtain a combined sample of 87 systems, of which 37 are multi-transiting and 50 are single-transiting. There are at least three reasons for considering an analysis that uses the combined sample: \begin{inparaenum}[(i)] \item as already mentioned, the combined sample is, to date, the largest ensemble of measured $i_{\rm s}$ for exoplanet-host stars; \item the two samples being combined are complementary in terms of $T_{\rm eff}$ coverage (see Fig.~\ref{fig:HR2}); \item the hierarchical Bayesian approach adopted in this work (cf.~Sect.~\ref{sec:statastero}) is able to handle the inhomogeneous nature of the combined sample in a straightforward manner\end{inparaenum}.

\begin{figure}[!t]
\centering
\includegraphics[width=0.8\linewidth]{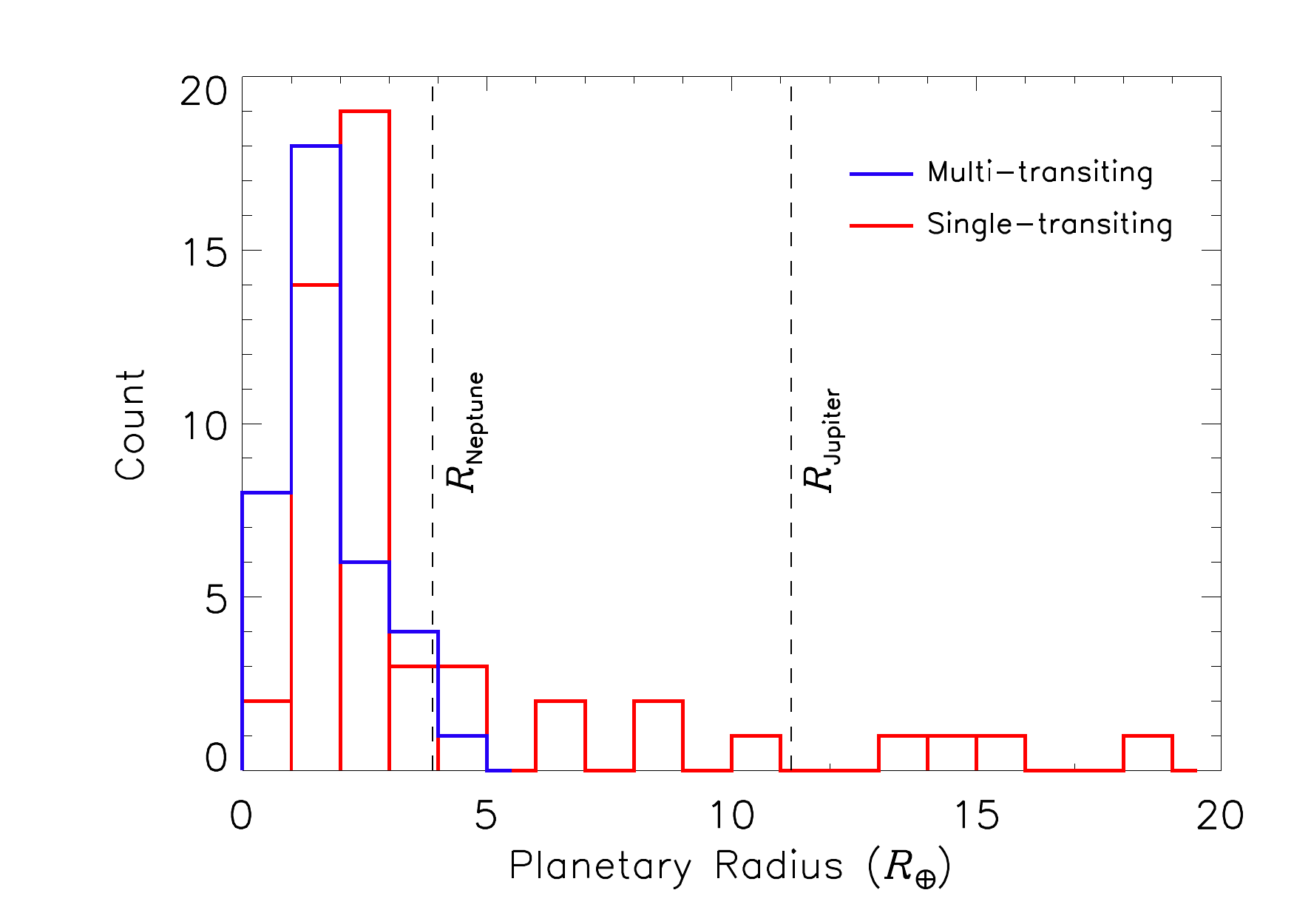}
\caption{\small Histogram of planetary radius for planets in the combined sample. For multiple-planet systems, we consider the radius of the smallest planet.\label{fig:rad_hist}}
\end{figure}

Figure \ref{fig:HR2} displays the stars in the combined sample in a $\log g$ vs.~$T_{\rm eff}$ diagram. Multi- and single-transiting systems are seen to span similar ranges of effective temperature ($4500\!\la\!T_{\rm eff}\!\la\!6500\:{\rm K}$) and surface gravity ($3.9\!\la\!\log g\!\la\!4.6\:{\rm dex}$). Stars belonging to the asteroseismic sample are responsible for populating the hot (and hence massive) end of the parameter space, where flux modulations due to starspots are hardly detectable and a measurement of the photometric $P_{\rm rot}$ is thus made difficult \citep[e.g.,][]{Radick82}. Conversely, the cool end of the parameter space is mainly populated by stars belonging to the MW14 sample, since their oscillation amplitudes are too small, and the stars in general too faint, to allow detecting the oscillations \cite[e.g.,][]{CampanteSurfGrav}. Furthermore, planets in multi-transiting systems tend to be systematically smaller than planets in single-transiting systems \citep[cf.][]{Latham11}, a tendency made clearer in Fig.~\ref{fig:rad_hist}. We also note that the fraction of multi-transiting systems is higher than in the full {\it Kepler} sample of planet-candidate host stars, where multi-transiting systems account for only $23\,\%$ of the total \citep{Burke14}. The fact that stars in the combined sample are relatively bright may partly explain the high observed fraction of multi-transiting systems, since the smaller planets in these systems would be preferentially detected around bright stars.

Given samplings of the posterior probability distributions of $\cos i_{\rm s}$ for the $N\!=\!87$ stars in the combined sample, we once more try to infer the distribution of the spin-orbit angle $\psi$ following a hierarchical Bayesian scheme (cf.~Sect.~\ref{sec:statastero}). Having now access to a larger sample, we decided to also employ a slightly more complex model to represent the distribution function of $\psi$ in addition to the single-Fisher model.

{\it A sum of an isotropic and a Fisher distribution.} An alternative to modeling the distribution function of $\psi$ is by assuming that all spin-orbit angles are drawn either from an isotropic distribution (with probability $f$) or from a Fisher distribution (with probability $1-f$). This so-called `mixture model' can be used to describe a scenario where two different channels exist by which planets migrate, ultimately giving rise to two different obliquity populations. One of these channels produces an isotropic distribution of spin-orbit angles and $f$ can then be interpreted as the fraction of systems that follow such a migration route. We thus have:
\begin{equation}
\label{eq:mixture}
f_{\boldsymbol{\alpha}}(\psi) \equiv \frac{f}{2}\sin\psi + (1-f)\,p_{\rm F}(\psi|\kappa) \, , 
\end{equation} 
or, in terms of the observable $\cos i_{\rm s}$,
\begin{equation}
\label{eq:mixture2}
f_{\boldsymbol{\alpha}}(\cos i_{\rm s}) \equiv f + (1-f)\,p_{\rm F}(\cos i_{\rm s}|\kappa) \, .
\end{equation}
The mixture model has two free parameters, namely, $f$ and $\kappa$. We assign uniform priors to both parameters.

\begin{figure}[!t]
\centering
\includegraphics[width=0.75\linewidth]{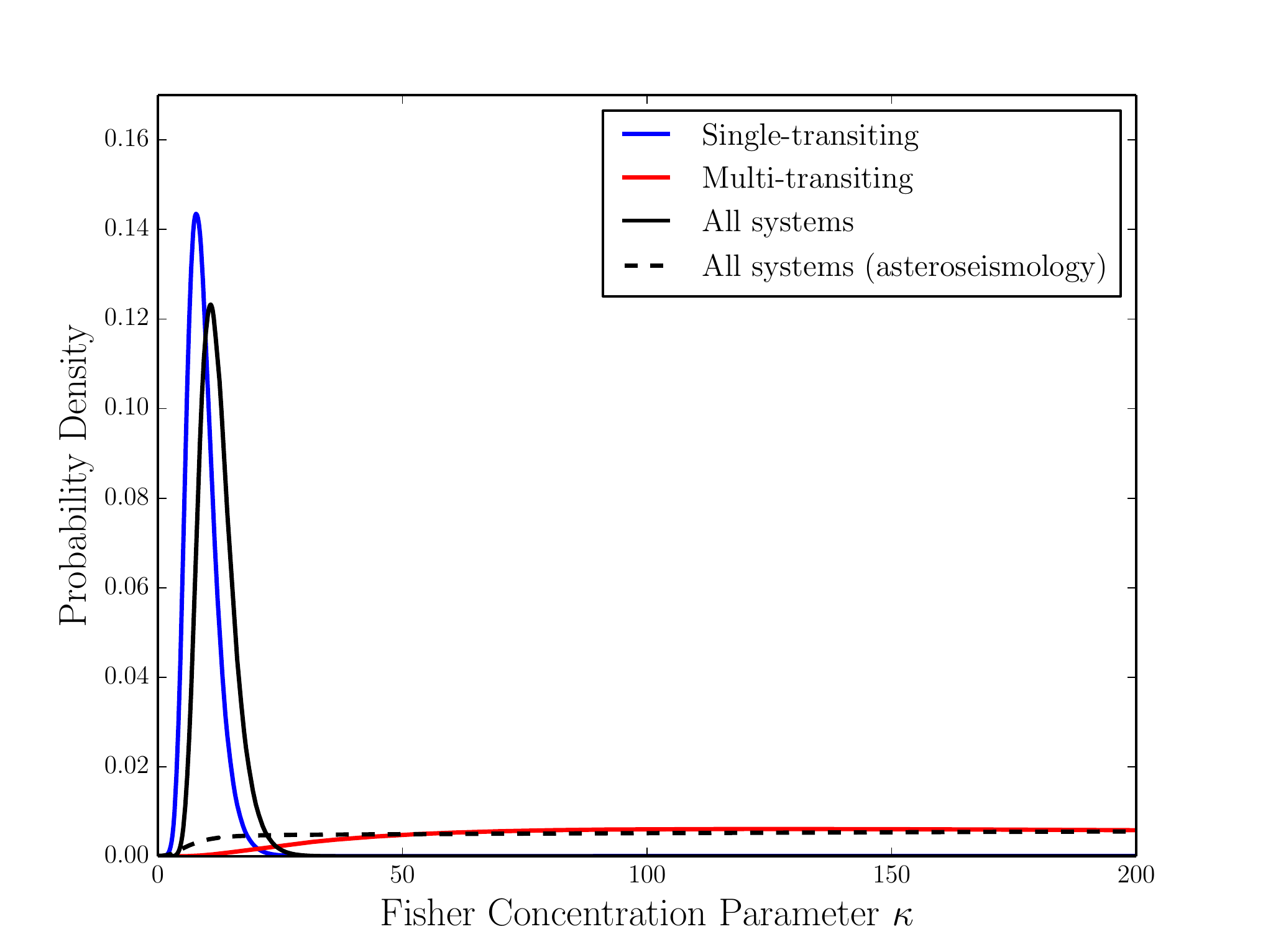}
\caption{\small Posterior probability distribution of the concentration parameter $\kappa$ in the single-Fisher model (combined sample). Distributions for the different cohorts of systems are color-coded. The dashed black line depicts the posterior obtained basing the analysis on the asteroseismic sample alone.\label{fig:post_singleF_kappa}}
\end{figure}

\begin{figure}[!t]
\centering
\includegraphics*[scale=0.56]{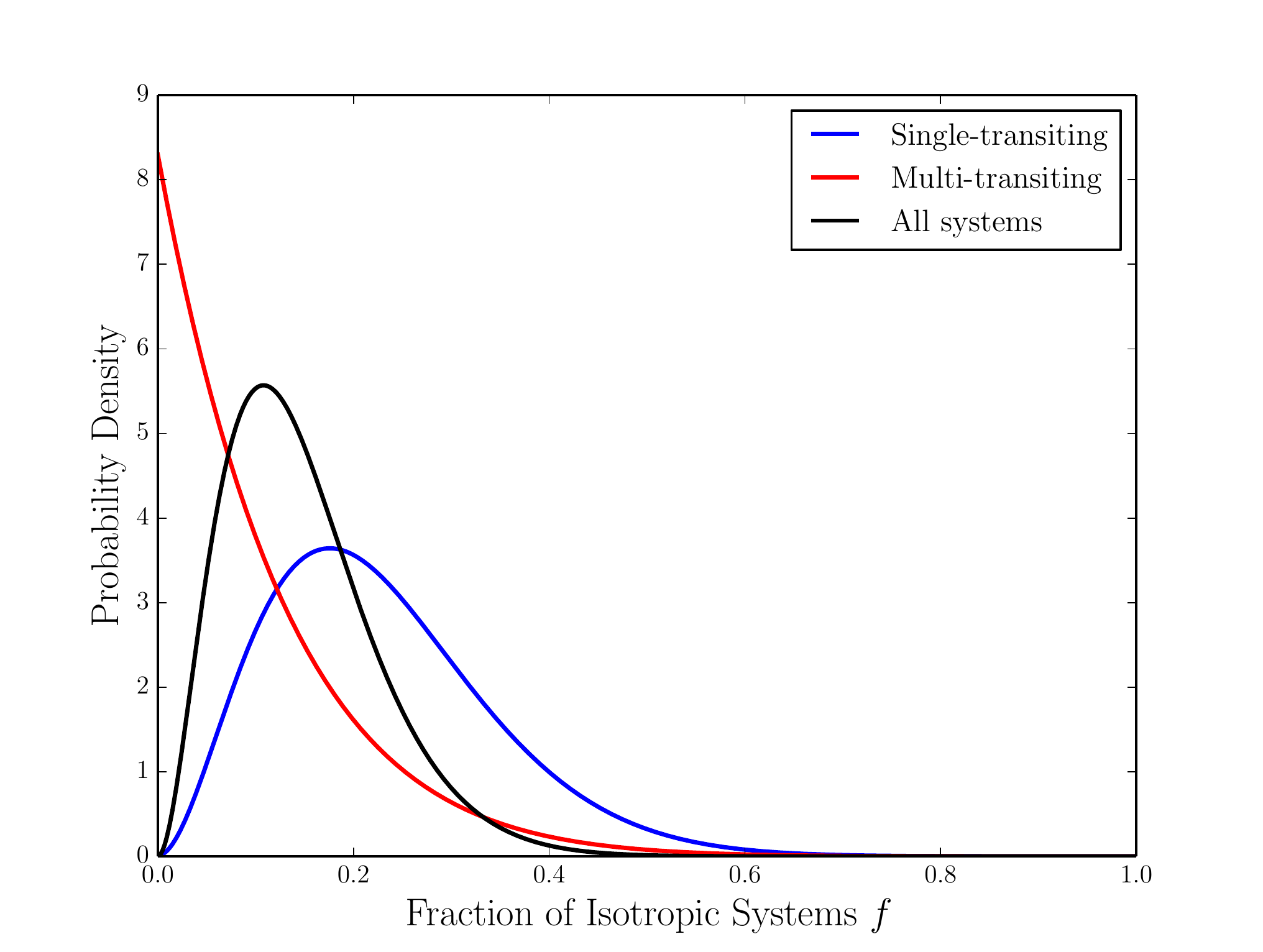}
\includegraphics*[scale=0.56]{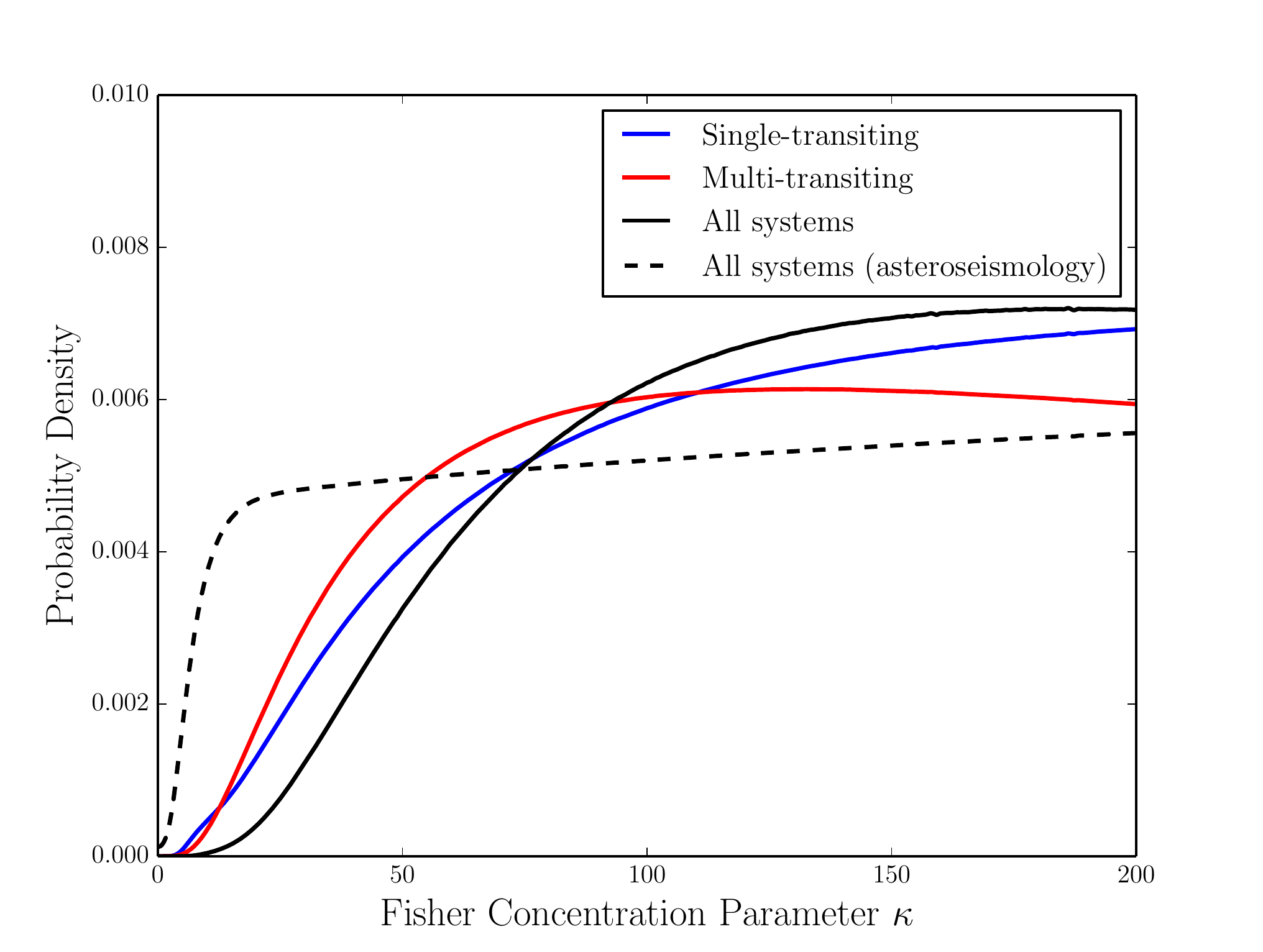}
\caption{\small Posterior probability distributions of the fraction $f$ of isotropic systems (top panel) and the concentration parameter $\kappa$ (bottom panel) in the mixture model. Distributions for the different cohorts of systems are color-coded. The dashed black line in the bottom panel depicts the posterior obtained basing the analysis on the asteroseismic sample alone (where the single-Fisher model has been used).\label{fig:post_mixture}}
\end{figure}

The PPD of the concentration parameter $\kappa$ in the single-Fisher model is shown in Fig.~\ref{fig:post_singleF_kappa}. Figure \ref{fig:post_mixture} shows the PPDs of the fraction $f$ of isotropic systems (top panel) and the concentration parameter $\kappa$ (bottom panel) in the mixture model. Distributions for the different cohorts of systems are color-coded in each panel. Table \ref{tb:reshierarchical} presents a summary of the results, where we also report the Bayesian model evidence, $E$, for each cohort of systems. The Bayesian evidence is computed by taking the integral of Eq.~(\ref{eq:postalpha}) over the parameter space spanned by the model parameters $\boldsymbol{\alpha}$.

MW14 have also considered the single-Fisher model, and the addition of the asteroseismic sample in the present analysis leads to consistent results within the quoted credible regions. For multi-transiting systems, however, the posteriors of $\kappa$ in both works differ in shape and appear to be dominated at large $\kappa$ by their respective priors. Our choice of a uniform prior on $\kappa$, as opposed to $p(\kappa)\!\propto\!(1+\kappa^2)^{-3/4}$ in MW14, was made to prevent the underestimation of $\kappa$ when $\kappa$ is large. We confirmed our suspicion that the prior may be dominating at large $\kappa$ by repeating the analysis having used the latter prior on $\kappa$ instead. We take this as an indication that the available data may not be able to fully constrain the posterior of $\kappa$ for multi-transiting systems, in particular at large $\kappa$.

Figures \ref{fig:post_singleF_kappa} and \ref{fig:post_mixture} possess a number of other interesting features: \begin{inparaenum}[(i)] \item the posteriors of $\kappa$ (in both models) and $f$ are not compatible with an isotropic distribution for $\psi$; \item multi-transiting systems tend to be characterized by a large $\kappa$ or, equivalently, by low obliquities; \item the posterior of $\kappa$ is similar for multi-transiting systems irrespective of the model being considered, but a difference exists in the case of single-transiting systems\end{inparaenum}. These points will be used in support of the discussion presented in the next section.

\section{Discussion and conclusions}\label{sec:conclusions}
The main outputs of this work are twofold, namely, the presentation of individual asteroseismic results on the stellar inclination angle for a sample of {\it Kepler} host stars (Sect.~\ref{sec:astero}) and the placement of statistical constraints on the spin-orbit alignment of exoplanet systems (Sect.~\ref{sec:stat}).

\subsection{Asteroseismic analysis}\label{sec:conclusions_asteroseismic}
The asteroseismic sample considered herein consists of 25 {\it Kepler} solar-type host stars. Two of the systems, HAT-P-7 and Kepler-25, are of particular interest, since the available measurements of $i_{\rm o}$, $i_{\rm s}$, and $\lambda$ allow constraining the spin-orbit angle $\psi$. We find that the orbit of the hot-Jupiter HAT-P-7b (Kepler-2b) is likely to be retrograde ($93.5\degr\!<\!\psi\!<\!138.4\degr$, as given by the $68.3\,\%$ HPD credible region), in good agreement with previous works by \citet{Benomar14} and \citet{Lund14}. The orbit of Kepler-25c, on the other hand, seems to be well aligned with the stellar spin axis ($1.6\degr\!<\!\psi\!<\!19.3\degr$, as given by the $68.3\,\%$ HPD credible region). Our results thus do not support the statement made by \citet{Benomar14} that this system is (mildly) misaligned. The $68.3\,\%$ credible regions for $\psi$ in both works overlap. This is the case despite the differences in the adopted data preparation and data analysis methodologies. Moreover, the $95.4\,\%$ credible region found by \citet{Benomar14} contains a nonnegligible probability of alignment. Hence, one cannot talk of a significant discrepancy between the results, i.e., the statement made by \citet{Benomar14} claiming misalignment may be regarded as an over-interpretation not strongly supported by their results.

We note that Kepler-25 is a late F-type star ($T_{\rm eff}\!\approx\!6270\:{\rm K}$), and consequently it exhibits broad mode profiles that hinder our ability to resolve and extract signatures of rotation in the power spectrum ($\nu_{\rm s}/\Gamma\!=\!0.46$; see Table \ref{tb:results}). Coupled to a low S/N in the p modes (${\rm S/N}\!=\!1.0$; see Table \ref{tb:results}), this means that the effect of the correlated background noise on the mode profiles may bias the outcome of the peak-bagging analysis. A reduced visibility of the multiplet components is an issue common to a substantial fraction of the stars in the asteroseismic sample (see Table \ref{tb:results}). To test our asteroseismic method, and in particular the robustness of the returned uncertainties on $i_{\rm s}$, we have produced artificial power spectra with varying input $i_{\rm s}$ for a representative set of stars in the asteroseismic sample (see Appendix \ref{append:artificial}). We managed to retrieve the input $i_{\rm s}$ at the 1-$\sigma$ level in $75\,\%$ of the simulated cases (21/28), and at the 2-$\sigma$ level in $\sim\!96\,\%$ of the cases (27/28), leading us to conclude that the returned uncertainties are robust.

As a further sanity check, detailed peak-bagging was conducted for all stars using an affine-invariant ensemble MCMC algorithm \citep{emcee,Lund14}. The main purpose of this exercise was to both investigate the impact of the sampling algorithm on the resulting joint PPD and the impact of using a different frequency range (or number of observed modes) in the peak-bagging analysis. The overall agreement between the Metropolis--Hastings (on which our results are based; see Appendix \ref{append:specfit}) and affine-invariant implementations is excellent.

Besides HAT-P-7, there are five other systems in the asteroseismic sample that could potentially be misaligned, namely, Kepler-50, Kepler-93, Kepler-145, Kepler-409, and KOI-280. Interestingly, Kepler-50, Kepler-93, and Kepler-145 are all multiple-planet systems (but see discussion on Kepler-93 below). This interpretation is based on the $68.3\,\%$ HPD credible regions. However, inclination angles close to $90\degr$ have nonnegligible probability, and so our results are consistent with alignment at the $95.4\,\%$ (or 2-$\sigma$) level. In light of these results, the system surrounding the red-giant star Kepler-56 \citep{Huber13} remains as the only unambiguous detection to date of a misaligned multiple-planet system \citep[we note that the reported misalignment of the super-Earth 55 Cnc e is controversial;][]{Bourrier,LopezMorales}.

Kepler-50 was found by \citet{Chaplin13} to have its spin axis nearly perpendicular to the line of sight, which matches our current findings only at the 2-$\sigma$ level. We have ruled out the effect of a different peak-bagging prescription as the cause of this discrepancy by analyzing the power spectrum in that work with our current approach. This discrepancy is then likely due to the different temporal coverages: in the present work we used an additional six quarters of {\it Kepler} observations compared to the 18 months of data already used by \citet{Chaplin13}. An increased temporal coverage is known to reduce existing biases on the fitted $i_{\rm s}$ \citep{Ballot08}. We demonstrate the importance of this effect in Appendix \ref{append:artificial} by performing tests with degraded Sun-as-a-star data with white noise added to levels comparable with Kepler-50. Kepler-93 is a single-transiting system with a long-period, non-transiting object (Kepler-93c) that was detected using the RV method, although its planetary nature is yet to be confirmed \citep{Marcy14,Dressing15}. It remains to be tested whether Kepler-93c may have been responsible for scattering Kepler-93b inward onto a short-period orbit. In an attempt to explain the potential misalignments for these multiple-planet systems by the presence of a bound stellar companion, we searched the high-resolution imaging surveys of \citet{Howell11}, \citet{Adams12}, \citet{Law14}, and \citet{LilloBox14}, as well as the high-resolution spectroscopy survey of \citet{Kolbl15}. Within the detectability limits of these surveys no companions have been detected for the three multiple-planet systems in question. A recent work combining broadband adaptive optics and non-redundant aperture masking \citetext{A.~L.~Kraus et al., submitted} has, however, led to the detection of companions in the Kepler-93 and Kepler-145 systems which, if gravitationally bound, must be late-type dwarfs at large ($\ga\!100\:{\rm AU}$) separations.

\subsection{Statistical analysis}\label{sec:conclusions_statistical}
Figure \ref{fig:averageposterior} and the analysis performed in Sect.~\ref{sec:statastero} based on the asteroseismic sample suggest that the directions of the stellar spin and planetary orbital axes are correlated. This was confirmed by the analysis performed in Sect.~\ref{sec:comb} based on the combined sample. But do the measured spin-orbit orientations reflect primordial conditions? The degree of spin-orbit alignment is correlated with the planet-to-star mass ratio, the orbital distance, and the stellar effective temperature \citep{Albrecht12}. Tidal effects are not expected to be significant for small planets and planets in long orbits \citep{Albrecht13}, but could play an important role in systems with close-in giants (i.e., $P_{\rm o}\!\leq\!10\:{\rm d}$ and $R_{\rm p}\!>\!6\,R_\earth$). For these systems, we calculated the alignment timescale $\tau_{\rm CE}$ presented by \citet{Albrecht12}, which is calibrated using binary-star data (see their eq.~2). The estimated value of $\tau_{\rm CE}\!\sim\!9.8\:{\rm Gyr}$ for the hot-Jupiter system HAT-P-7 far exceeds the age of the star as derived from asteroseismology, $t\!=\!2.12^{+0.27}_{-0.24}\:{\rm Gyr}$ \citep{VSA15}. The single-transiting systems Kepler-17, Kepler-43, and Kepler-63 also host a close-in giant, and all have $\tau_{\rm CE}$ exceeding the estimated age of the Universe ($16$, $49$, and $\sim\!10^5\:{\rm Gyr}$, respectively). Therefore, the planetary systems considered in this work should not have had their obliquities damped by tides, meaning that the spin-orbit orientations should in principle reflect nearly primordial conditions.

A related aspect that is of particular relevance in explaining the formation of hot-Jupiter systems has to do with the observation that multi-transiting systems tend to be characterized by a large concentration (i.e., around $\psi\!=\!0$) parameter $\kappa$ or, equivalently, by low obliquities (e.g., $\kappa\!>\!13$ based on the asteroseismic sample). A tendency for low obliquities in multi-transiting systems was first reported by \citet{Albrecht13}. Under the assumption that the planetary orbits in multi-transiting systems are nearly coplanar and trace the plane of the protoplanetary disk, this would imply that the high obliquities observed for hot-Jupiter systems are associated with planet migration. Our analysis thus favors migration mechanisms capable of exciting large obliquities in explaining hot-Jupiter formation --- planet-planet scattering \citep[e.g.,][]{RasioFord,Nagasawa08}, Kozai cycles \cite[e.g.,][]{Holman97,Kozai,FabryckyTremaine}, or secular chaos \citep[e.g.,][]{WuLithwick} --- over the paradigm of inward spiraling along the protoplanetary disk --- as in the case of Type II migration \citep[e.g.,][]{TypeII,LinPapaloizou}.

The results from the analysis of the asteroseismic sample cannot be used to support the finding by MW14 that the obliquities of systems with a single transiting planet are systematically larger than those with multiple transiting planets, which has been taken as evidence that a substantial fraction of {\it Kepler}'s single-transiting systems are in fact a separate population of compact multiple-planet systems characterized by large mutual inclinations. We should note, however, that the sensitivity of the statistical analysis performed on the asteroseismic sample is limited by the small number of systems (11 single- and 14 multi-transiting systems).

Finally, based on the analysis of the combined sample, we looked into the statistical merits of the two models for the distribution of $\psi$, namely, the single-Fisher model and the mixture model. This was done by computing the so-called Bayes' factor, $F_{\rm B}$, given by the ratio of the Bayesian evidences $E$ (see Table \ref{tb:reshierarchical}). The Bayes' factor is a summary of the evidence provided by the data in favor of one statistical model as opposed to a competing model. A scale for the interpretation of $F_{\rm B}$ was given by \citet{Jeffreys61} \citep[see also][]{KassRaftery}. For multi-transiting systems, there is `substantial' evidence (i.e., $0.5\!<\!\log_{10}F_{\rm B}\!=\!0.92\!<\!1$) favoring the single-Fisher model over the mixture model. The converse is true for single-transiting systems, with the mixture model being `decisively' favored (i.e., $\log_{10}F_{\rm B}\!=\!2.05\!>\!2$). This explains why the posterior of $\kappa$ is similar for multi-transiting systems irrespective of the model being considered, since the inclusion of an isotropic component via the mixture model is not supported by the data. It also suggests that the obliquity distribution of single-transiting systems may be multimodal and therefore better explained by the mixture model, pointing to two distinct channels by which planets migrate (with $\sim\!20\,\%$ of these systems belonging to a `dynamically hot' category). In both cases, however, the (statistically favored) posterior of $\kappa$ still appears to be dominated at large $\kappa$ by the prior.

\subsection{Outlook}\label{sec:conclusions_outlook}
Throughout the duration of the {\it Kepler} mission, asteroseismology has played an important role in the characterization of host stars and their planetary systems. This work presents the first analysis of an ensemble of asteroseismic obliquity measurements made for transiting systems, being another example of the enduring synergy between asteroseismology and exoplanetary science. The prospect of using asteroseismology to measure the obliquities of systems with evolved hosts will be addressed when data from the {\it TESS} mission \citep[Transiting Exoplanet Survey Satellite;][]{TESS} become available. The planned {\it PLATO} mission \citep[PLAnetary Transits and Oscillations of stars;][]{PLATO} will further offer the possibility of extending these asteroseismic measurements to bright solar-type hosts in wide fields with a coverage of 2--3 years.

\appendix
\section{An analytical expression for $p(\psi|i_{\rm s},i_{\rm o})$}\label{append:psipost}
Suppose an observer has measured $i_{\rm s}$ and $i_{\rm o}$ for a particular transiting system. What can then be inferred about $\psi$? Here we derive an analytical expression for $p(\psi|i_{\rm s},i_{\rm o})$, the posterior probability distribution for the spin-orbit angle $\psi$ conditioned on the stellar inclination angle $i_{\rm s}$ and on the orbital inclination angle $i_{\rm o}$. As far as the authors are aware, such derivation has not been presented in the literature.

We start by deriving an expression for $p(i_{\rm s}|\psi,i_{\rm o})$, i.e., the posterior probability distribution for $i_{\rm s}$ conditioned on $\psi$ and $i_{\rm o}$. We assume that for a given $\psi$, the probability distribution of the azimuthal angle $\phi$ is uniformly distributed between $-\pi$ and $+\pi$ \citep[see][]{Fabrycky09}. We then express $i_{\rm s}$ in terms of $\psi$, $i_{\rm o}$, and $\phi$ by eliminating $\lambda$ from Eqs.~(\ref{eq:geom1})--(\ref{eq:geom3}):
\begin{equation}
\label{eq:append1}
i_{\rm s}(\psi,i_{\rm o},\phi) = \arccos(\cos\psi\cos i_{\rm o} - \sin\psi\sin i_{\rm o}\cos\phi) \, . 
\end{equation}
Since $i_{\rm s}(\psi,i_{\rm o},-\phi)=i_{\rm s}(\psi,i_{\rm o},\phi)$, we need only consider $\phi$ in the interval $[0,\pi]$. Next, we transform variables \citep[e.g.,][chap.~5]{Meyer70} from $\phi$ to $i_{\rm s}$ in order to arrive at an expression for $p(i_{\rm s}|\psi,i_{\rm o})$:
\begin{align}
\label{eq:append2}
p(i_{\rm s}|\psi,i_{\rm o}) = p(\phi|\psi,i_{\rm o})&\left|\frac{\partial\phi}{\partial i_{\rm s}}\right| = \frac{1}{\pi}\left|\frac{\partial\phi}{\partial i_{\rm s}}\right| \nonumber \\
= \frac{1}{\pi} \frac{\sin i_{\rm s}}{\sqrt{\sin^2\psi\sin^2i_{\rm o}-(\cos\psi\cos i_{\rm o}-\cos i_{\rm s})^2}} \, &, \quad \text{for all } |i_{\rm o}-\psi| < i_{\rm s} < \pi-|\pi-i_{\rm o}-\psi| \, .
\end{align}
Note that the angles $i_{\rm s}$ and $\psi$ are allowed to vary in the interval $[0,\pi]$ in the previous equation, whereas we restrict $i_{\rm o}$ to the interval $[0,\pi/2]$.

We now resort to Bayes' theorem to derive an analytical expression for $p(\psi|i_{\rm s},i_{\rm o})$:
\begin{equation}
\label{eq:append3}
p(\psi|i_{\rm s},i_{\rm o}) \propto p(i_{\rm s}|\psi,i_{\rm o})\,p(\psi) \, ,
\end{equation}
where the prior probability $p(\psi)$ quantifies our assumptions on $\psi$. We will be adopting the uninformative prior probability $p(\psi)\!\propto\!\sin\psi$, which implies that $\mathbf{n}_{\rm s}$ and $\mathbf{n}_{\rm o}$ (see Fig.~\ref{fig:geometry}) are uncorrelated. Finally, by multiplying Eq.~(\ref{eq:append2}) by $\sin\psi$ we arrive at an (unnormalized) expression for $p(\psi|i_{\rm s},i_{\rm o})$:
\begin{equation}
\label{eq:append4}
p(\psi|i_{\rm s},i_{\rm o}) \propto \frac{\sin i_{\rm s}\sin\psi}{\sqrt{\sin^2\psi\sin^2i_{\rm o}-(\cos\psi\cos i_{\rm o}-\cos i_{\rm s})^2}} \, , \quad \text{for all } |i_{\rm o}-i_{\rm s}| < \psi < i_{\rm o}+i_{\rm s} \, .
\end{equation}
Note that the angle $\psi$ is allowed to vary in the interval $[0,\pi]$ in the previous equation, whereas we restrict $i_{\rm s}$ and $i_{\rm o}$ to the interval $[0,\pi/2]$.

\section{Modeling and fitting the power spectrum}\label{append:specfit}
Extracting signatures of rotation from the power spectrum is accomplished by a detailed fitting of the observed modes of oscillation, a procedure often referred to as peak-bagging. We are primarily interested in performing a global fit to the power spectrum \cite[e.g.,][]{Appourchaux08,Campante11,Appourchaux12}, whereby a selection of the observed modes are fitted simultaneously over a broad frequency range. We modeled the limit (noise-free) oscillation spectrum as a series of standard Lorentzian profiles, $\mathscr{O}(\nu;\mathbf{\Theta}_{\rm osc})$, which sit atop a background signal described by $\mathscr{B}(\nu;\mathbf{\Theta}_{\rm bg})$:
\begin{align}
\label{eq:specmodel}
\mathscr{P}(\nu;\mathbf{\Theta}) &= \mathscr{O}(\nu;\mathbf{\Theta}_{\rm osc}) + \mathscr{B}(\nu;\mathbf{\Theta}_{\rm bg}) \nonumber \\ 
&= \sum_{n',l} \sum_{m=-l}^{l} \frac{\mathscr{E}_{l m}(i_{\rm s}) H_{n'l}}{1+\left[\frac{2(\nu - \nu_{n'l0} - m\nu_{\rm s})}{\Gamma_{n'lm}}\right]^2} + \mathscr{B}(\nu;\mathbf{\Theta}_{\rm bg}) \, .
\end{align}
The sets of mode and background model parameters are denoted by $\mathbf{\Theta}_{\rm osc}$ and $\mathbf{\Theta}_{\rm bg}$. 

The inner sum in Eq.~(\ref{eq:specmodel}) runs over the azimuthal components of each multiplet, while the outer sum runs over the selection of observed modes. The dummy variable $n'$, which tags the observed order, takes values $n'\!=\!n$ for $l\!=\!1,2$ modes, $n'\!=\!n+1$ for $l\!=\!0$ modes, and $n'\!=\!n-1$ for $l\!=\!3$ modes. To reduce the number of parameters entering our model, the heights and linewidths of non-radial modes are defined based on the heights and linewidths of the neighboring radial mode(s), the latter being allowed to vary with frequency. The set of mode parameters is ultimately given by $\mathbf{\Theta}_{\rm osc}\!=\!\{\nu_{n'l0},H_{n'0},\Gamma_{n'0},i_{\rm s},\nu_{\rm s}\}$.

The parameter $H_{n'l}$ describes the height of a given mode of degree $l$, with the height of a multiplet component, $H_{n'lm}$, given by $\mathscr{E}_{l m}(i_{\rm s}) H_{n'l}$. $H_{n'l}$ is defined relative to the height of the neighboring radial mode(s) according to $H_{n'l}\!=\!V_l^2 H_{n'0}$ for $l\!=\!2$ modes or $H_{n'l}\!=\!V_l^2 [H_{n'0}+H_{(n'-1) 0}]/2$ for $l\!=\!1,3$ modes, where the $V_l^2$ describe the visibilities (in power) of modes of different $l$ relative to those with $l\!=\!0$ \citep[we adopted fixed values of $V_1\!=\!1.22$, $V_2\!=\!0.71$, and $V_3\!=\!0.14$, obtained taking into account the {\it Kepler} bandpass and the effect of limb darkening;][]{HandCamp}. The parameter $\Gamma_{n'lm}$ describes the mode linewidth, taken to be the same regardless of $m$ for a given mode of degree $l$, i.e., $\Gamma_{n'lm}\!\equiv\!\Gamma_{n'l}$. $\Gamma_{n'l}$ is defined relative to the linewidth of the neighboring radial mode(s) according to $\Gamma_{n'l}\!=\!\Gamma_{n'0}$ for $l\!=\!2$ modes or $\Gamma_{n'l}\!=\![\Gamma_{n'0}+\Gamma_{(n'-1) 0}]/2$ for $l\!=\!1,3$ modes.

The background signal was modeled as the superposition of three components:
\begin{equation}
\label{eq:backmodel}
\mathscr{B}(\nu;\mathbf{\Theta}_{\rm bg}) = B_0 + \eta^2(\nu) \left[\frac{B_{\rm gran}}{1+(2\pi\nu\,\tau_{\rm gran})^a} + \frac{B_{\rm act}}{\nu^2}\right] \, .
\end{equation}
A flat component $B_0$ is used to model photon shot-noise. The contribution from granulation is described by a Harvey-like profile \citep{Harvey85,Harvey93}, where $B_{\rm gran}$ is the height at $\nu\!=\!0$ of the granulation component, $\tau_{\rm gran}$ is the characteristic turnover timescale for granulation, and $a$ calibrates the amount of memory in the process. Finally, a power-law component describes the contribution from activity (characterized by the scale factor $B_{\rm act}$). This functional form results from considering a Harvey-like profile in the limit $2\pi\nu\,\tau_{\rm act}\!\gg\!1$ with the exponent set to 2 as in the original work by \citet{Harvey85}. Such a power law is well-suited to describe the decaying slope of the activity component. The attenuation factor $\eta^2$ is given by ${\rm sinc}^2\left[\pi/2\left(\frac{\nu}{\nu_{\rm Nyq}}\right)\right]$ for an integration duty cycle of $100\,\%$ \citep[e.g.,][]{Kallinger14}, where $\nu_{\rm Nyq}$ is the Nyquist frequency.

\begin{figure}[!t]
\figurenum{B1}
\centering
\includegraphics*[scale=0.65]{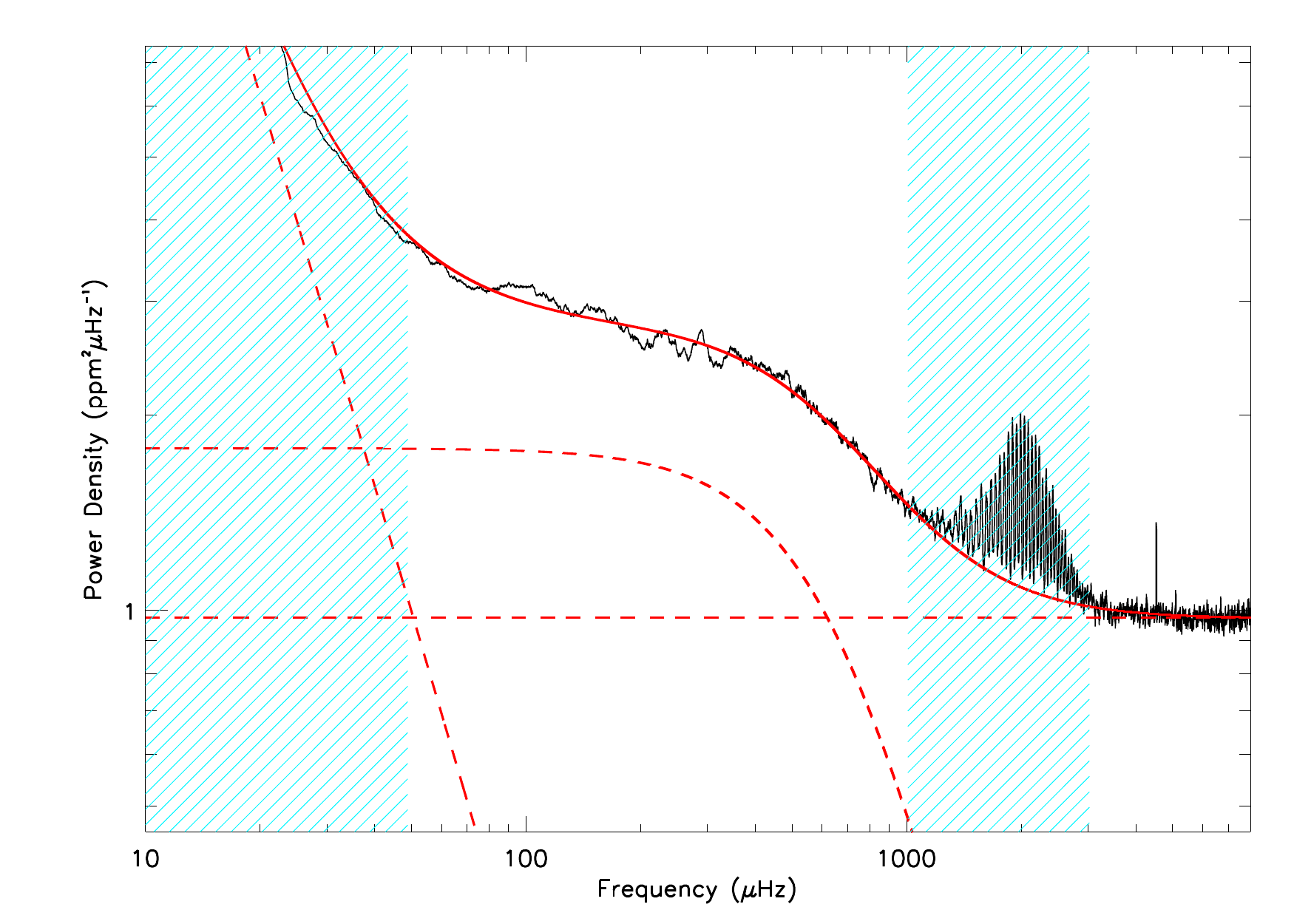}
\includegraphics*[scale=0.65]{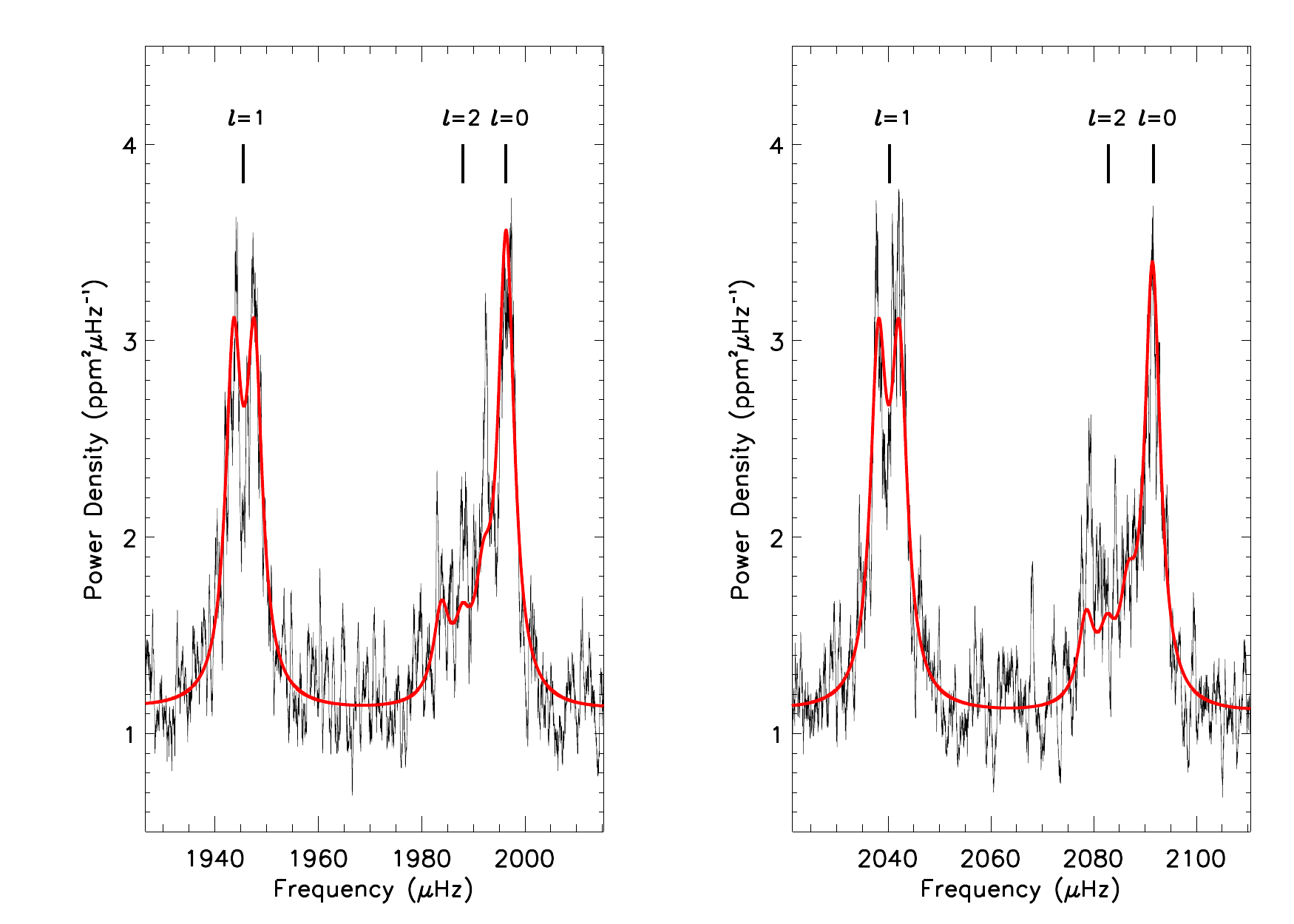}
\caption{\small Fit to the power spectrum of KIC~8866102 (Kepler-410 A, KOI-42). Top panel: Fit (solid red line) to the background signal using a maximum likelihood estimator. The background signal is modeled as the superposition of three components (dashed curves). The power law describing the contribution from activity dominates at low frequencies, whereas shot noise is the only nonnegligible component at high frequencies. The Harvey-like profile describing the contribution from granulation is conspicuous at intermediate frequencies. Shaded areas indicate the frequency ranges excluded from the fit, including the oscillation power envelope at $\nu_{\rm max}\!\sim\!2000\:{\rm \mu Hz}$ ($\nu_{\rm max}$ represents the frequency of maximum oscillation amplitude). Bottom panels: Fit (solid red line) to the oscillation spectrum across two contiguous orders located around $\nu_{\rm max}$. The best-fitting model is the output of an MCMC analysis. Modes are tagged according to their angular degree.\label{fig:peakbag}}
\end{figure}

We started by fitting the background signal using a maximum likelihood estimator prior to the extraction of the mode parameters \citep[e.g.,][and references therein]{Karoff13}. The spectral range considered in this preliminary fit excluded the oscillation modes and the very low frequencies (Fig.~\ref{fig:peakbag}, top panel). Exclusion of the very low frequencies results from the power law being only able to describe the decaying slope of the activity component. The background model parameters, i.e., $\mathbf{\Theta}_{\rm bg}\!=\!\{B_0,B_{\rm act},B_{\rm gran},\tau_{\rm gran},a\}$, determined in this way are subsequently held fixed in Eq.~(\ref{eq:specmodel}) at their maximum likelihood estimates. 

Mode parameters were then optimized in a Bayesian manner using an MCMC (Markov chain Monte Carlo) sampler that employs the Metropolis--Hastings algorithm \citep[Fig.~\ref{fig:peakbag}, bottom panels;][]{HandCamp,CampantePhD}. In this way, we were able to map the joint posterior probability distribution (PPD) of $\mathbf{\Theta}_{\rm osc}$:
\begin{equation}
\label{eq:ppd}
p(\mathbf{\Theta}_{\rm osc}|D,I) \propto p(\mathbf{\Theta}_{\rm osc}|I)\,p(D|\mathbf{\Theta}_{\rm osc},I) \, ,
\end{equation}
where $p(\mathbf{\Theta}_{\rm osc}|I)$ is the prior probability of $\mathbf{\Theta}_{\rm osc}$ based on any relevant prior information $I$ and $p(D|\mathbf{\Theta}_{\rm osc},I)$ is the likelihood of the observed data $D$ \citep[a description of the likelihood function can be found in, e.g.,][]{Toutain94}. The PPD of a given mode parameter can then be simply arrived at through marginalization of the joint PPD. 

A series of comprehensive fits to the power spectra of the stars in our sample have been conducted in a separate work \citep{DaviesKages}. These fits considered the frequency range containing all observed modes and their output is used in the present analysis to define initial guesses for the mode parameters. Since we are mainly interested in estimating the stellar inclination angle, we focus here not on the full set of observed modes but instead on a selection of those modes, which necessarily span a somewhat narrower frequency range. The selection of the (continuous) frequency range used in the present analysis was done on a star-by-star basis, having taken into account the linewidths of the modes and their S/N as obtained from the above preliminary fits. This meant selecting those observed orders (a minimum of five) whose modes have the highest S/N and for which the intrinsic $\nu_{\rm s}/\Gamma$ ratio is more favorable when it comes to resolving the azimuthal components \citep[cf.][]{Ballot06}. It should be noted that choosing the number of observed orders is a trade-off. The inclusion of more radial orders improves the precision in the final estimates. However, if orders are included where the modes show only modest or low S/N, improved precision may come at the cost of reduced accuracy (increased bias), rendering the outputs less robust. A measure of the S/N and $\nu_{\rm s}/\Gamma$ at the frequency of maximum oscillation amplitude $\nu_{\rm max}$, as obtained from the present analysis, is given for each star in Table \ref{tb:results}.

The main advantage of a Bayesian approach when compared to a frequentist approach is the ability to incorporate relevant prior information through Bayes' theorem (Eq.~\ref{eq:ppd}). We imposed uniform priors on all mode parameters with the exception of $i_{\rm s}$, for which we assumed an isotropic orientation in the sphere \citep{Abt01}, i.e., $p(i_{\rm s}|I)\!=\!\sin i_{\rm s}$ or, equivalently, a uniform prior on $\cos i_{\rm s}$. The stellar inclination angle was allowed to vary between $-\pi/2$ and $\pi$ and all samples lying outside the interval $[0,\pi/2]$ were then reflected about the $i_{\rm s}\!=\!0$ and $\pi/2$ boundaries. This was done to avoid potential boundary effects in the MCMC sampling.

\section{Tests with artificial data}\label{append:artificial}
To validate our asteroseismic method, we have performed a series of tests with artificial data. The reduced visibility of the multiplet components in many of the stars in the asteroseismic sample (see Table \ref{tb:results}) may raise concerns as to how robust the returned uncertainties on $i_{\rm s}$ are and how prone the MCMC fitting is to biases caused by the moderate-to-low S/N in the p modes. Fits to the power spectrum as described in Appendix \ref{append:specfit} are computationally expensive. We have thus produced artificial power spectra for a representative set of stars in the asteroseismic sample, which include all six potentially misaligned systems (based on our results for $i_{\rm s}$) plus Kepler-25. For each star, we have considered two input $i_{\rm s}$ with two noise realizations per input $i_{\rm s}$, thus leading to a total of 28 fitted artificial spectra. For HAT-P-7 and Kepler-145 the input $i_{\rm s}$ was taken to be either $90\degr$ or $30\degr$, whereas for the remainder of the stars we considered either $90\degr$ or $60\degr$. Having fixed the input $i_{\rm s}$, we then used the tabulated $v\sin i_{\rm s}$ and asteroseismic radius to determine the input $\nu_{\rm s}$. The power spectra were directly generated in the frequency domain, having preserved the same frequency resolution as in the real data and properly modeled the correlation between the background noise and the excitation function \citep[cf.][]{Chaplin08}. The adopted mode linewidths and S/N were based on the observed linewidths and S/N of radial modes (not split by rotation). We managed to retrieve the input $i_{\rm s}$ at the 1-$\sigma$ level in 21 of the fits ($75\,\%$), at the 1.5-$\sigma$ level in 26 fits ($\sim\!93\,\%$), and at the 2-$\sigma$ level in 27 fits ($\sim\!96\,\%$). The returned uncertainties are based on the Bayesian HPD credible regions. Drawing a parallel between Bayesian credible regions and the more familiar frequentist confidence intervals, we are led to conclude that the returned uncertainties are robust.

\begin{figure}[!t]
\figurenum{C1}
\centering
\includegraphics[width=0.9\linewidth]{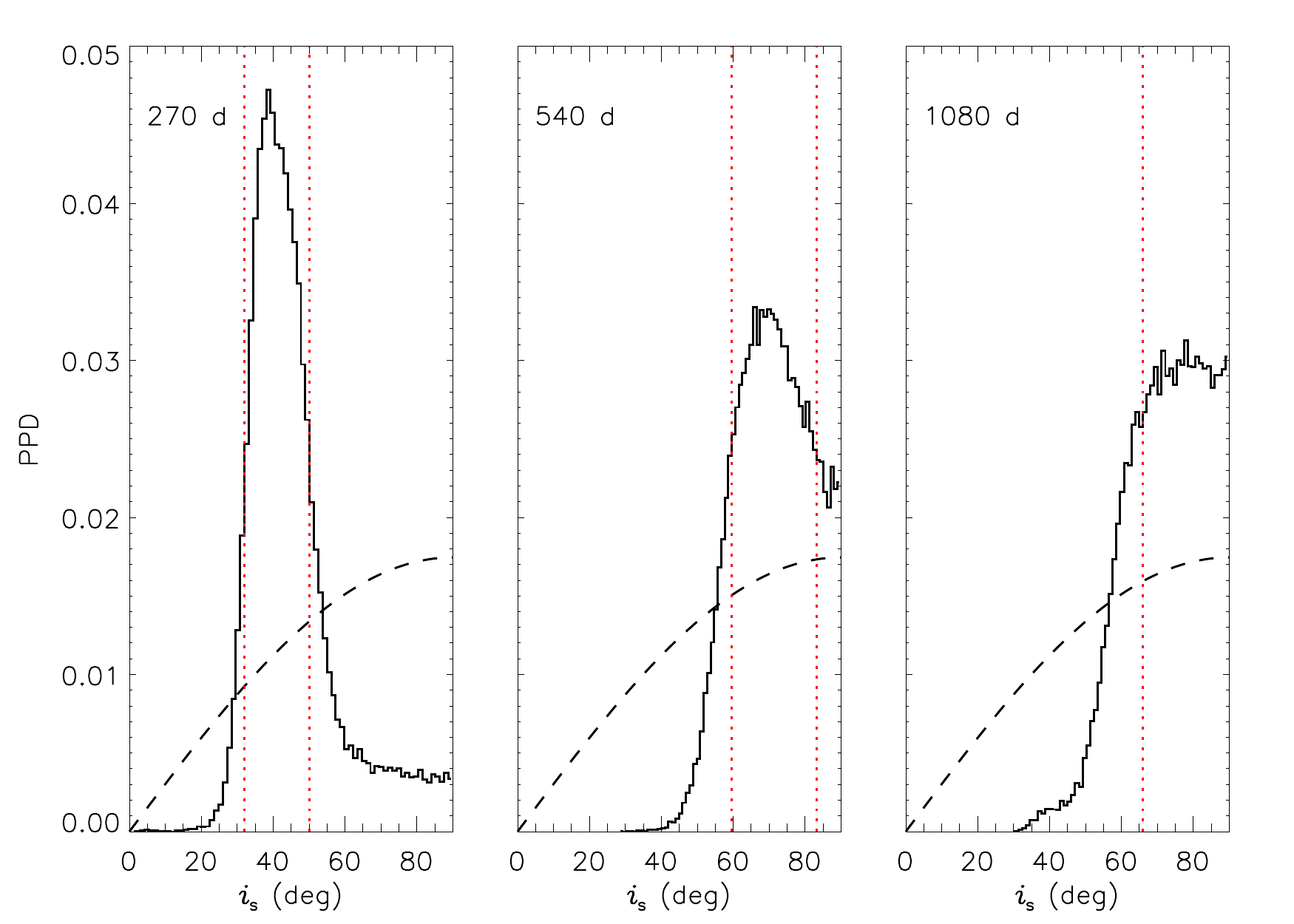}
\caption{\small Output of tests with degraded Sun-as-a-star data. We used data provided by the green channel of the SPM/VIRGO instrument on board {\it SoHO} with white noise added to levels comparable with Kepler-50. The temporal coverage of the time series increases from left to right: 270 days (left-hand panel), 540 days (middle panel), and 1080 days (right-hand panel). Marginalized PPDs of $i_{\rm s}$ are shown as histograms. Dotted lines enclose the $68.3\,\%$ HPD credible regions. For reference, the dashed curves represent the (uninformative) isotropic prior on $i_{\rm s}$ adopted in the asteroseismic analysis.\label{fig:Sun_anglepost}}
\end{figure}

\begin{figure}[!p]
\figurenum{D1}
\centering
\includegraphics[width=0.8\linewidth]{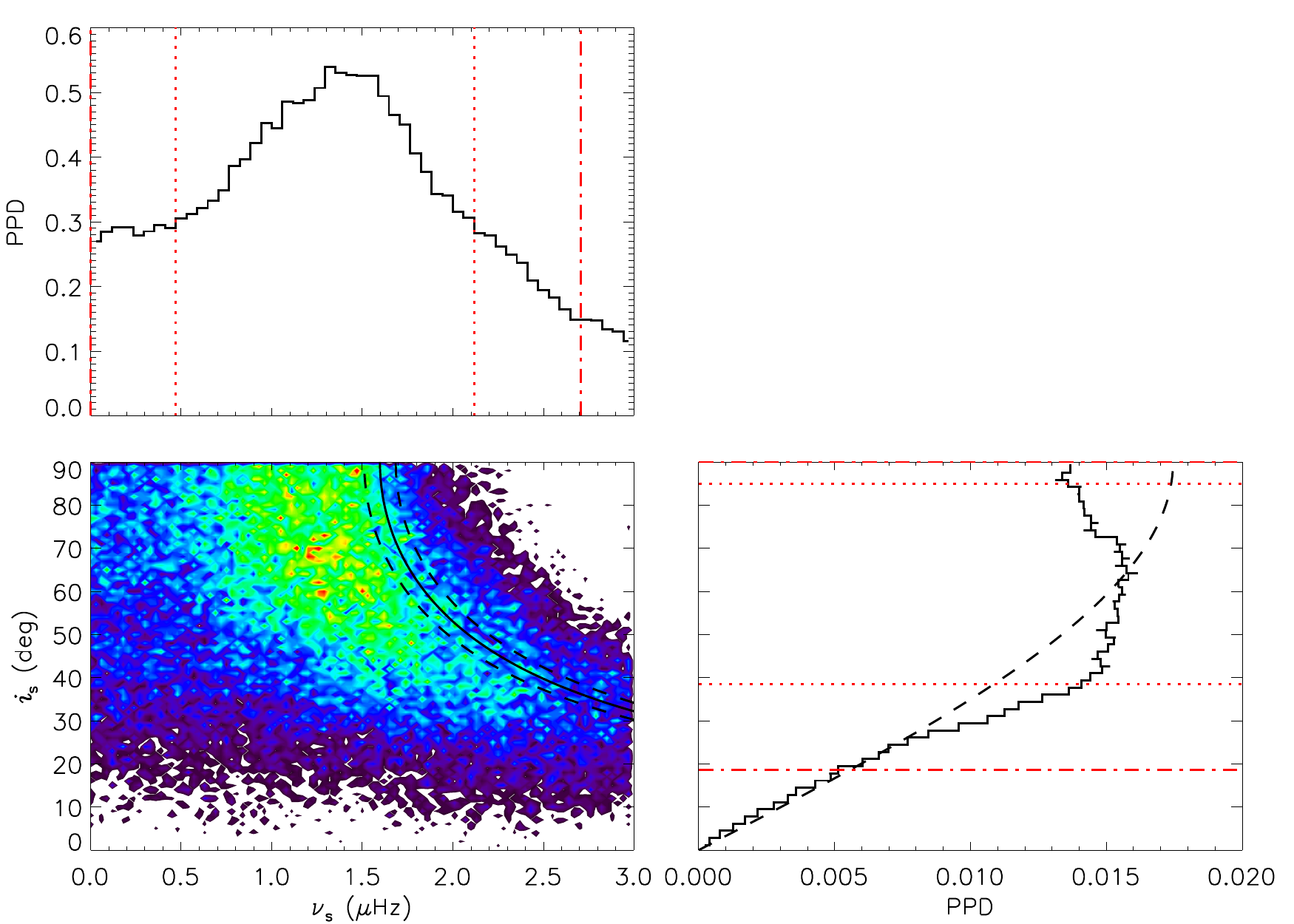}
\caption{\small Asteroseismic results on KIC~3425851 (KOI-268). Similar to Fig.~\ref{fig:ppd008866102}.\label{fig:ppd003425851}}
\end{figure}

We have also validated our asteroseismic method by performing a series of tests with degraded Sun-as-a-star data. Here, we are mostly interested in showing the effect of an increased temporal coverage in reducing the bias affecting the retrieved $i_{\rm s}$. Using data provided by the green channel of the triple Sun PhotoMeter/Variability of solar IRradiance and Gravity Oscillations (SPM/VIRGO) instrument \citep{VIRGO} on board the {\it SoHO} spacecraft, with white noise added to levels comparable with Kepler-50, we have performed our asteroseismic analysis. This was done by treating the Sun as a {\it Kepler} target of magnitude $m_{\rm Kep}\!=\!10$ (and 10.5) observed during 270, 540, and 1080 days (the equivalent to 3, 6, and 12 {\it Kepler} quarters). The pristine time series were taken from around solar minimum and share the same starting time stamp. Figure \ref{fig:Sun_anglepost} shows the output from the fits to the case with $m_{\rm Kep}\!=\!10$. The solar spin axis is inclined with respect to the normal of the ecliptic by $\sim\!7\degr$. The actual inclination for a particular block of SPM/VIRGO data will depend on the ephemerides of the Sun and is assumed to lie in the reference interval $[83\degr,90\degr]$. For reference, $\nu_{\rm s}/\Gamma\!\approx\!0.4$ at $\nu_{\rm max}$ for the Sun. The bias affecting the retrieved $i_{\rm s}$ is reduced as we increase the temporal coverage \citep[cf.][]{Ballot08}. The solar inclination is retrieved at the 1-$\sigma$ level for the longest of the time series (or the equivalent to 12 {\it Kepler} quarters). We note that there are only three stars in the asteroseismic sample with a temporal coverage shorter than ten quarters (see Table \ref{tb:sample}): the posteriors of $i_{\rm s}$ for KOI-268 (Q6.1--Q8.3; Fig.~\ref{fig:ppd003425851}) and Kepler-129 (Q6.1--Q7.3; Fig.~\ref{fig:ppd010586004}) are dominated by the prior, whereas that of the bright ($m_{\rm Kep}\!=\!8.72$) star Kepler-444 (Q15.1--Q17.2; Fig.~\ref{fig:ppd006278762}) shows signs of bimodality.

\section{Additional asteroseismic results}\label{append:seismicres}
The joint PPD of $i_{\rm s}$ and $\nu_{\rm s}$, as well as the corresponding marginalized PPDs are shown below for the remainder of the stars in the asteroseismic sample.

\clearpage

\begin{figure}[!p]
\figurenum{D2}
\centering
\includegraphics[width=0.8\linewidth]{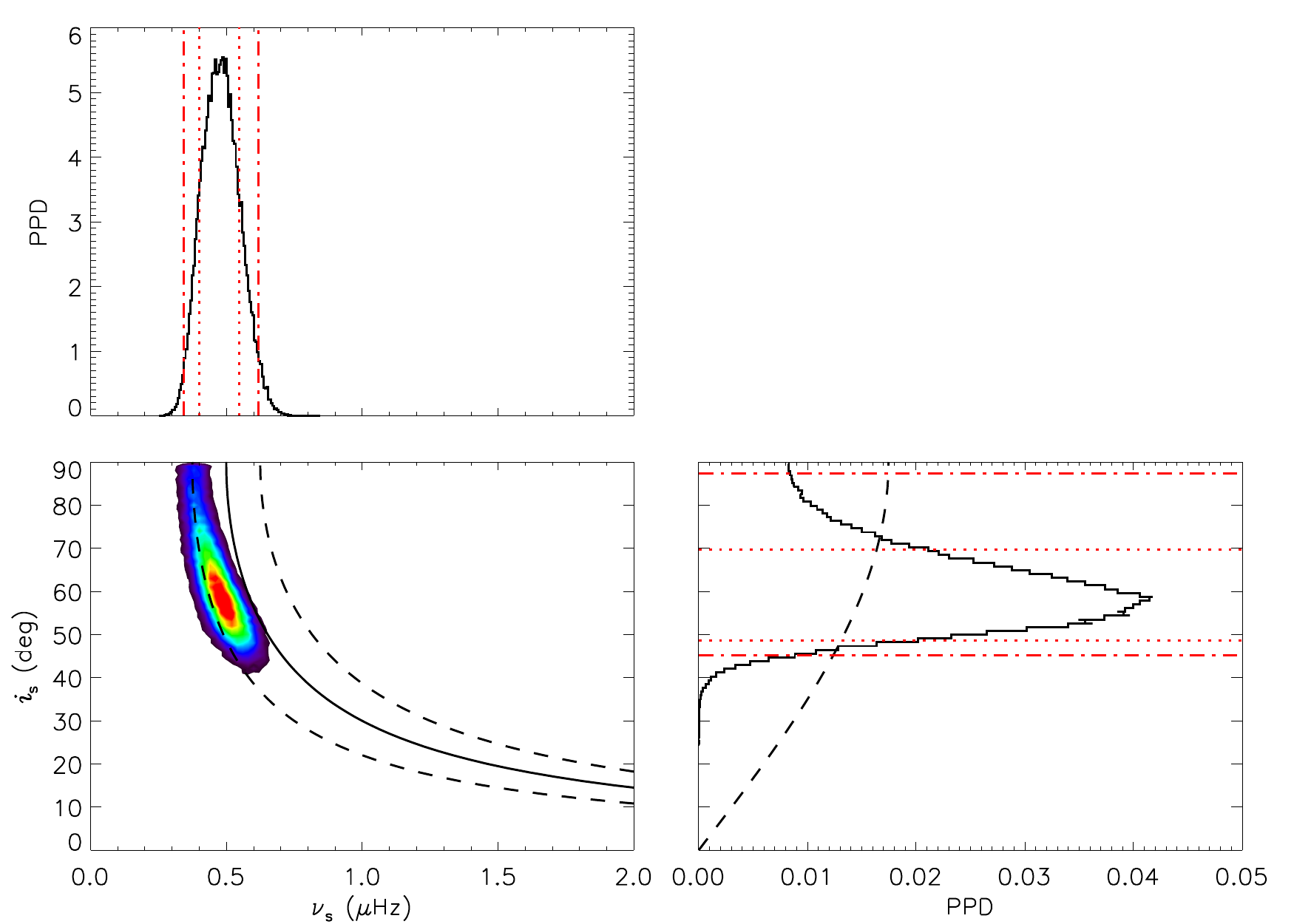}
\caption{\small Asteroseismic results on KIC~3544595 (Kepler-93, KOI-69). Similar to Fig.~\ref{fig:ppd008866102}.\label{fig:ppd003544595}}
\end{figure}

\clearpage

\begin{figure}[!p]
\figurenum{D3}
\centering
\includegraphics[width=0.8\linewidth]{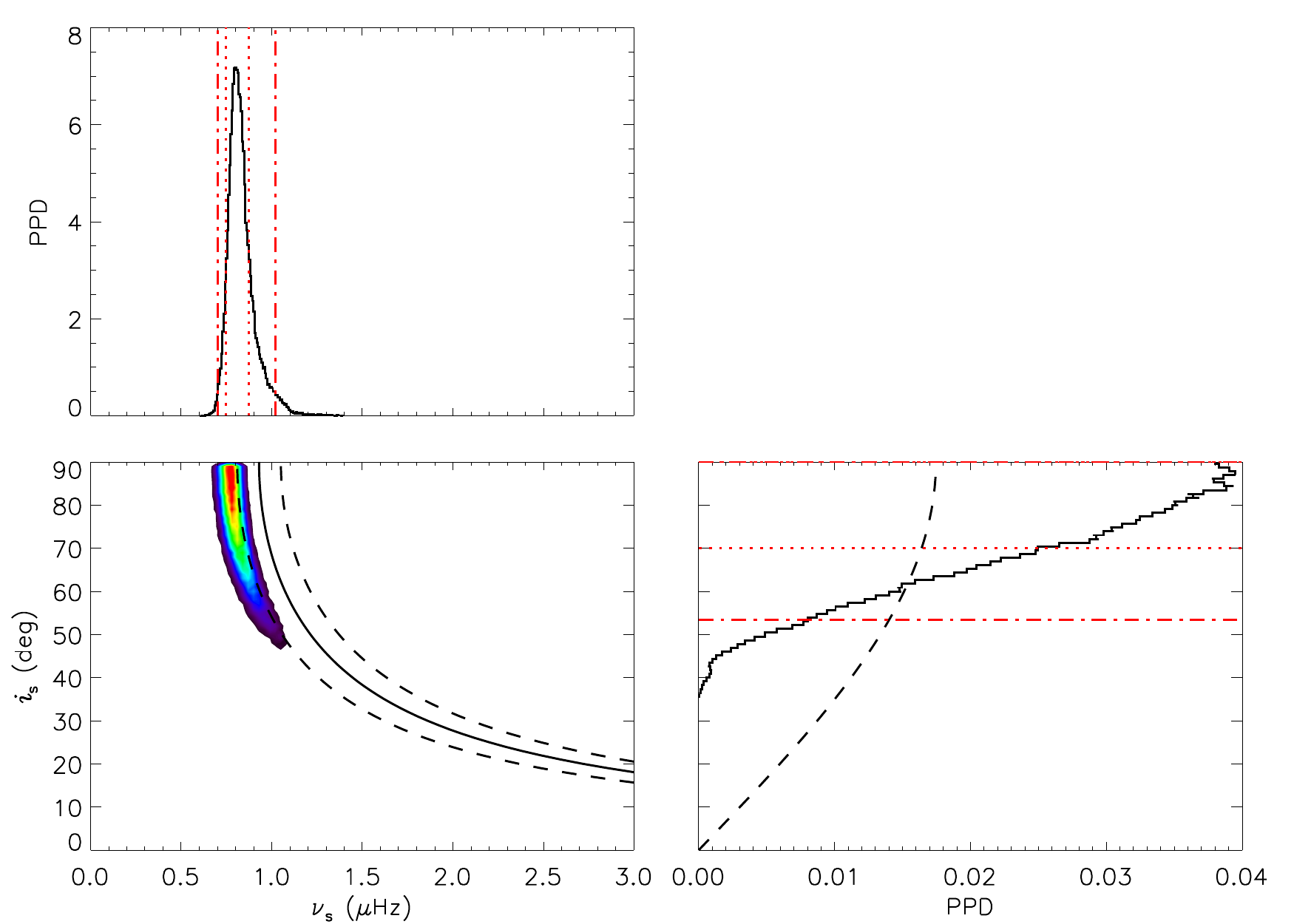}
\caption{\small Asteroseismic results on KIC~3632418 (Kepler-21, KOI-975). Similar to Fig.~\ref{fig:ppd008866102}.\label{fig:ppd003632418}}
\end{figure}

\begin{figure}[!p]
\figurenum{D4}
\centering
\includegraphics[width=0.8\linewidth]{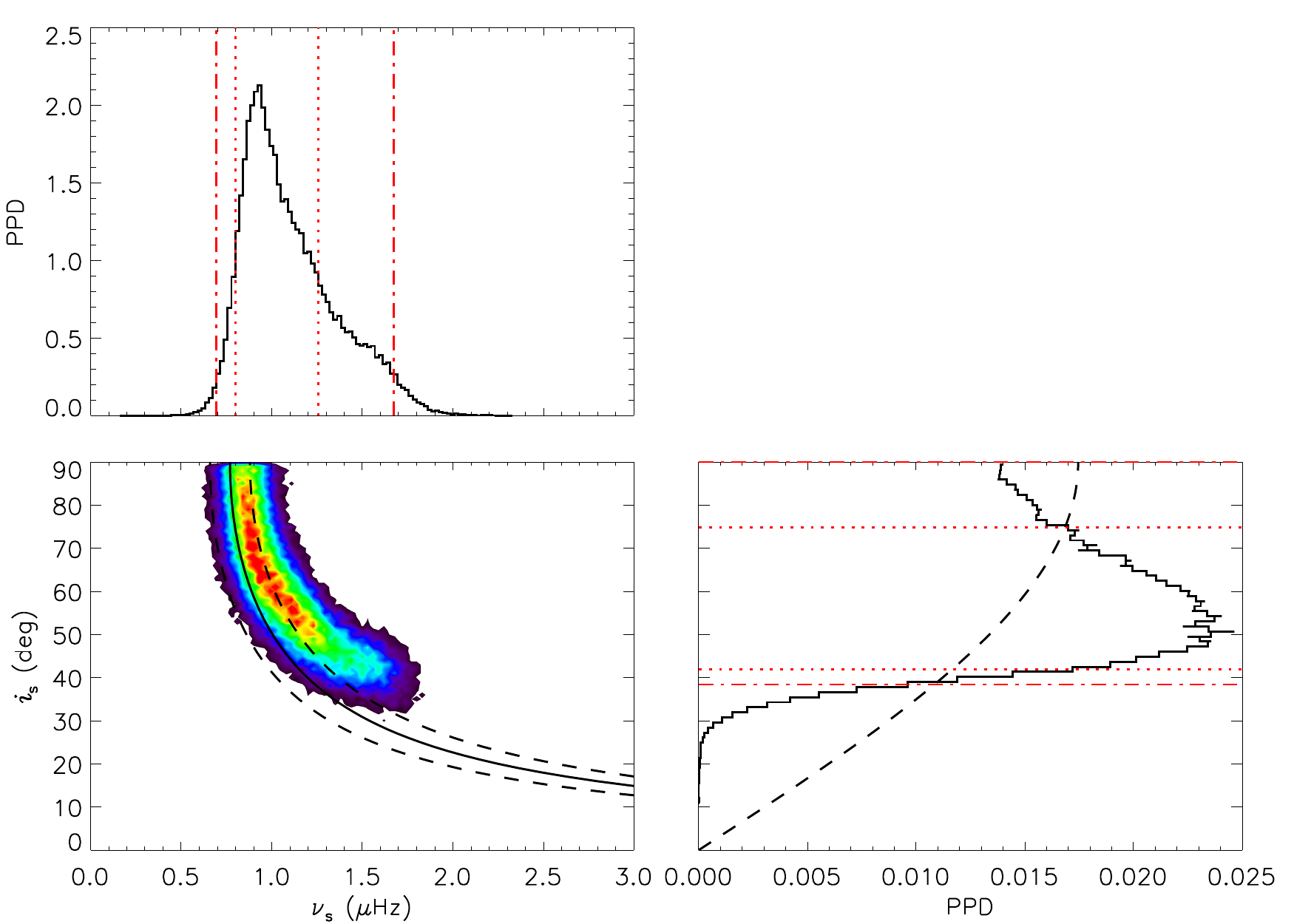}
\caption{\small Asteroseismic results on KIC~4141376 (KOI-280). Similar to Fig.~\ref{fig:ppd008866102}.\label{fig:ppd004141376}}
\end{figure}

\clearpage

\begin{figure}[!p]
\figurenum{D5}
\centering
\includegraphics[width=0.8\linewidth]{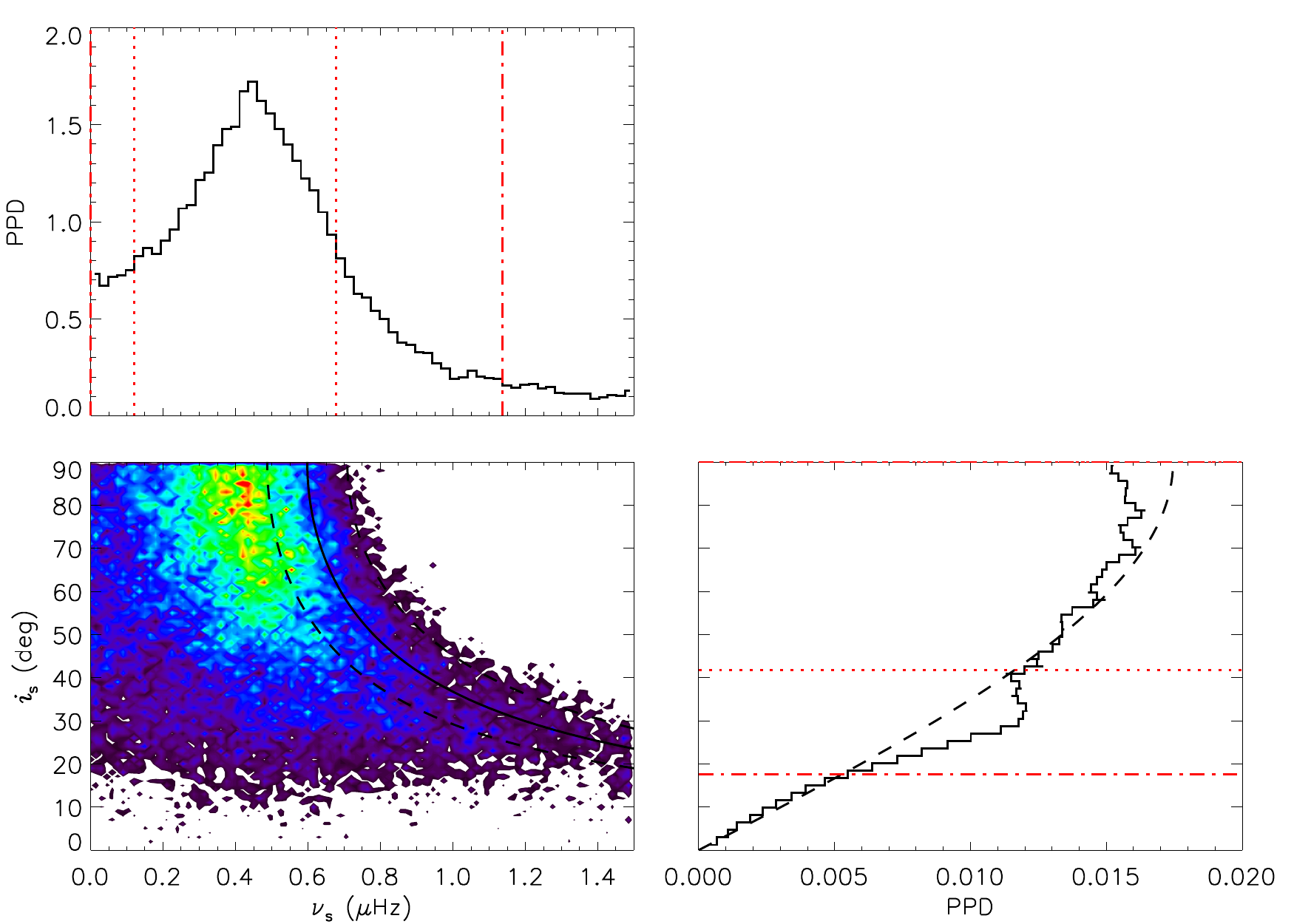}
\caption{\small Asteroseismic results on KIC~4914423 (Kepler-103, KOI-108). Similar to Fig.~\ref{fig:ppd008866102}.\label{fig:ppd004914423}}
\end{figure}

\begin{figure}[!p]
\figurenum{D6}
\centering
\includegraphics[width=0.8\linewidth]{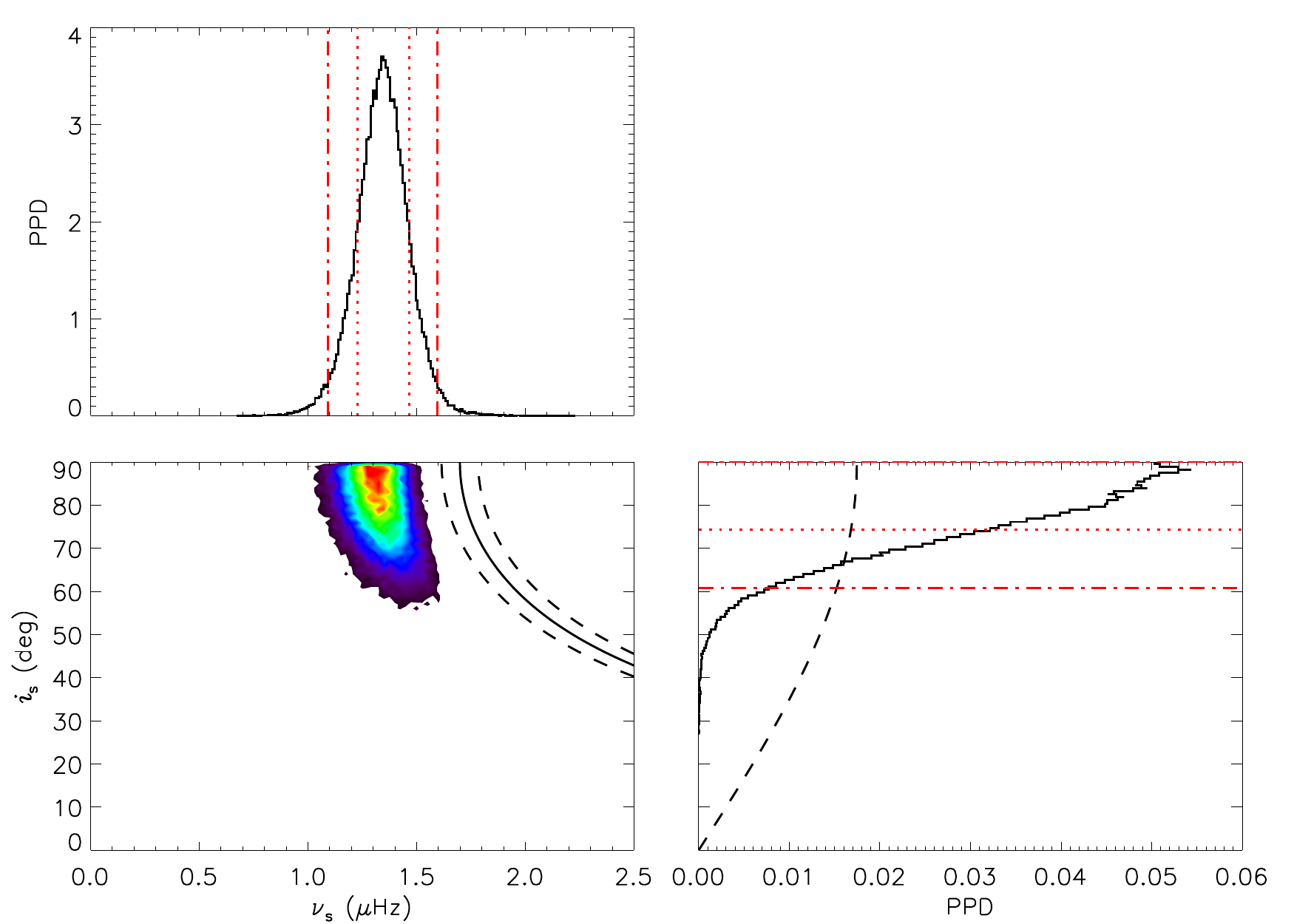}
\caption{\small Asteroseismic results on KIC~5866724 (Kepler-65, KOI-85). Similar to Fig.~\ref{fig:ppd008866102}.\label{fig:ppd005866724}}
\end{figure}

\clearpage

\begin{figure}[!p]
\figurenum{D7}
\centering
\includegraphics[width=0.8\linewidth]{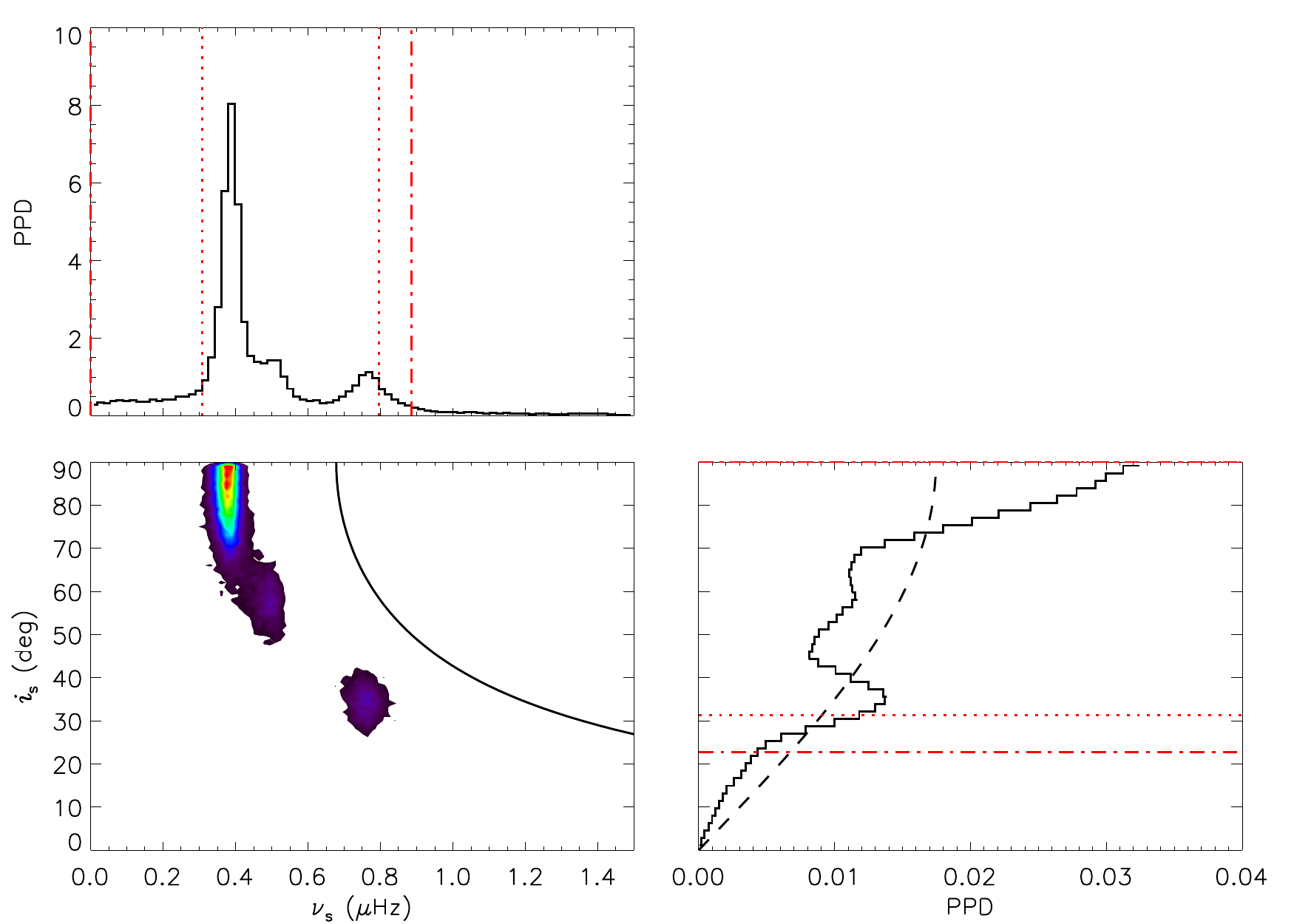}
\caption{\small Asteroseismic results on KIC~6278762 (Kepler-444, KOI-3158). Similar to Fig.~\ref{fig:ppd008866102}.\label{fig:ppd006278762}}
\end{figure}

\begin{figure}[!p]
\figurenum{D8}
\centering
\includegraphics[width=0.8\linewidth]{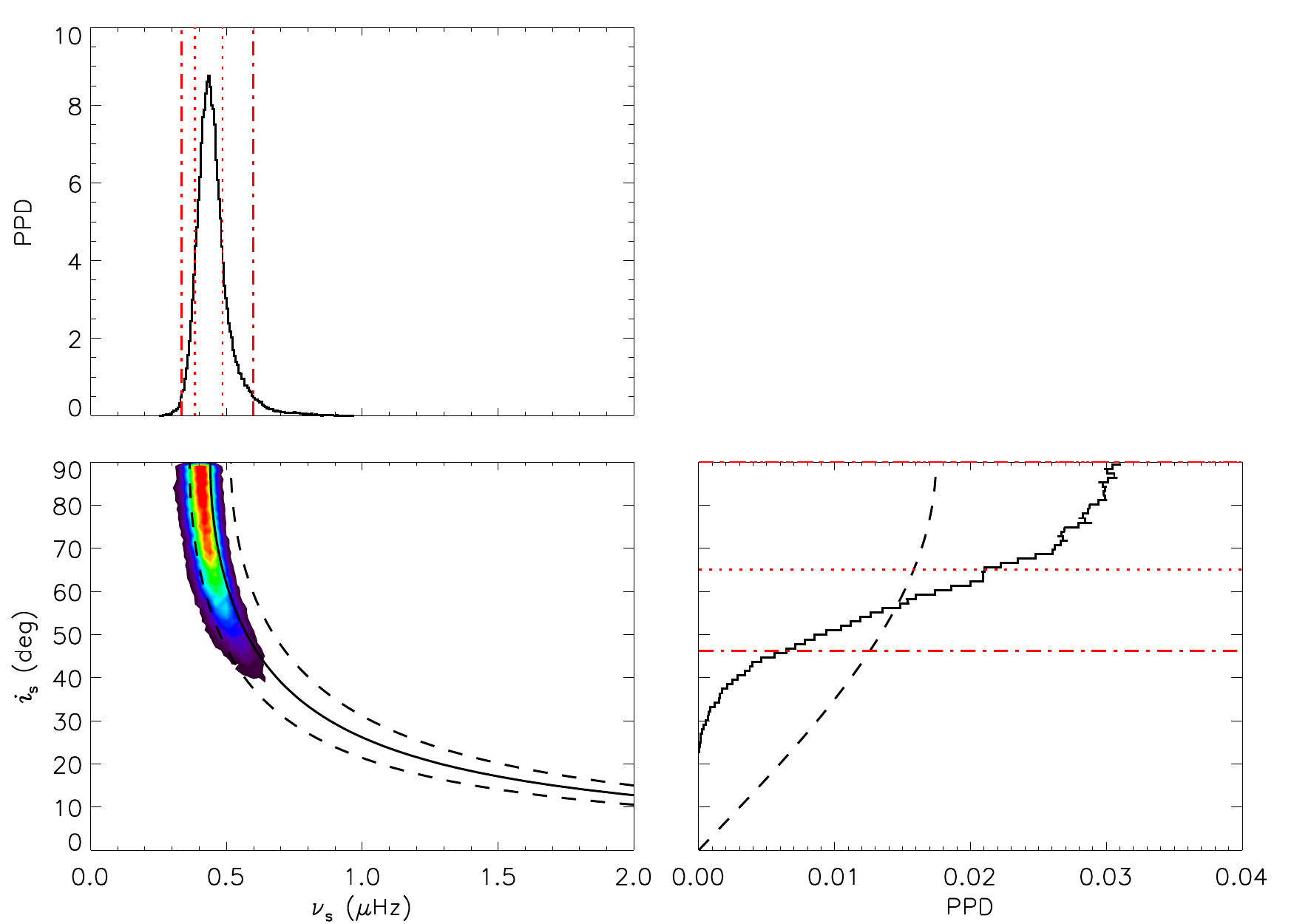}
\caption{\small Asteroseismic results on KIC~6521045 (Kepler-100, KOI-41). Similar to Fig.~\ref{fig:ppd008866102}.\label{fig:ppd006521045}}
\end{figure}

\clearpage

\begin{figure}[!p]
\figurenum{D9}
\centering
\includegraphics[width=0.8\linewidth]{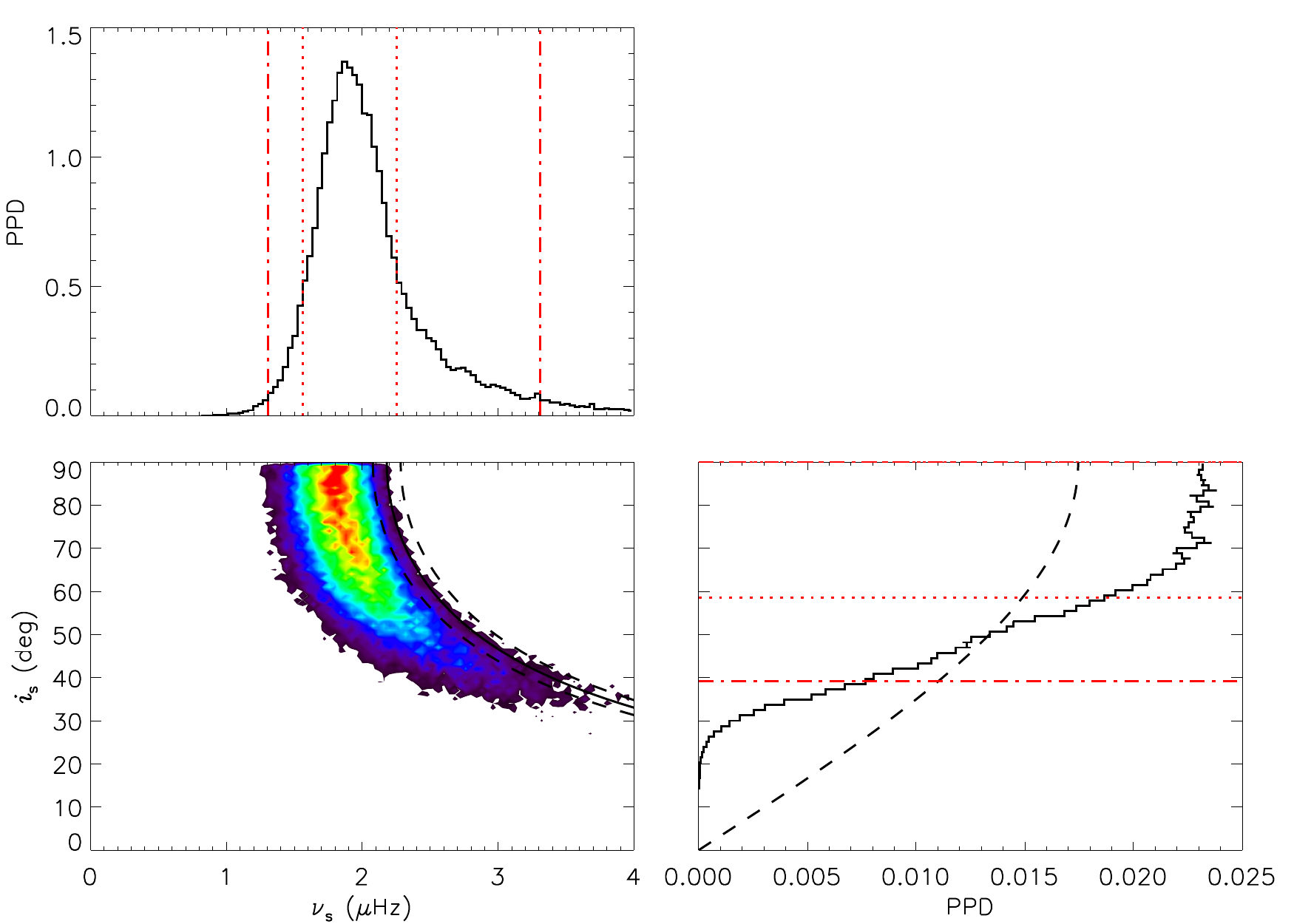}
\caption{\small Asteroseismic results on KIC~7670943 (KOI-269). Similar to Fig.~\ref{fig:ppd008866102}.\label{fig:ppd007670943}}
\end{figure}

\begin{figure}[!p]
\figurenum{D10}
\centering
\includegraphics[width=0.8\linewidth]{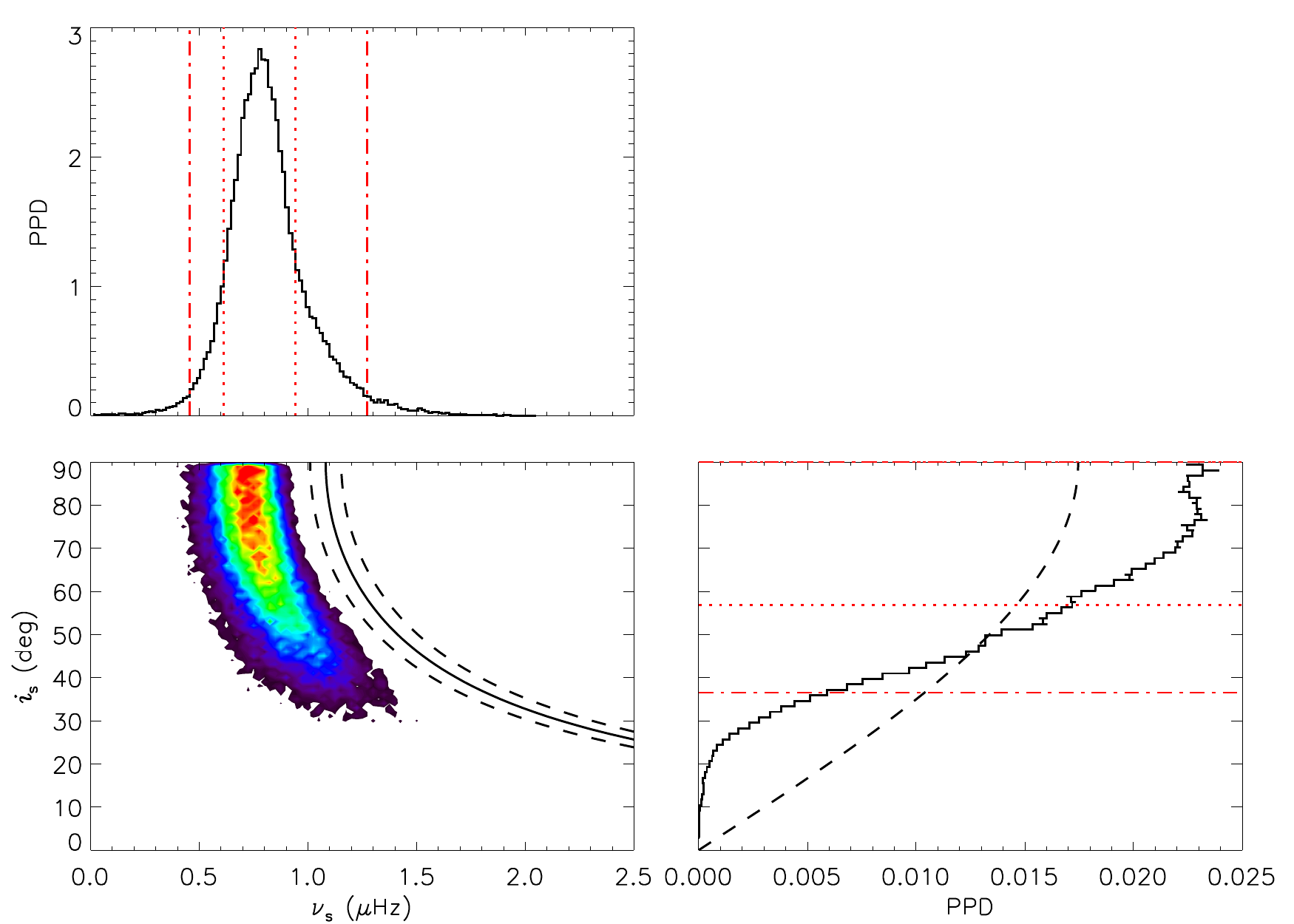}
\caption{\small Asteroseismic results on KIC~8077137 (Kepler-128, KOI-274). Similar to Fig.~\ref{fig:ppd008866102}.\label{fig:ppd008077137}}
\end{figure}

\clearpage

\begin{figure}[!p]
\figurenum{D11}
\centering
\includegraphics[width=0.8\linewidth]{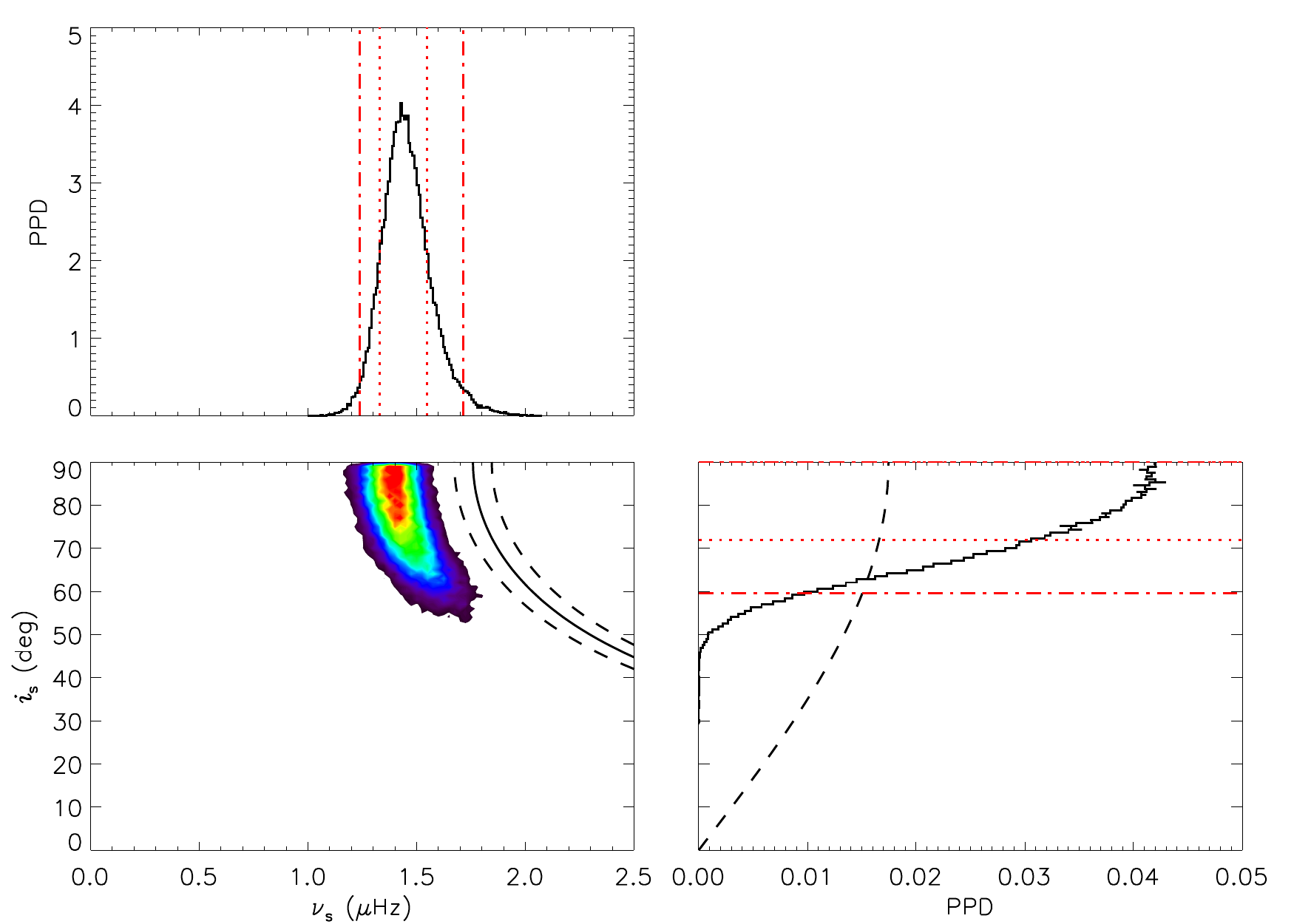}
\caption{\small Asteroseismic results on KIC~8292840 (Kepler-126, KOI-260). Similar to Fig.~\ref{fig:ppd008866102}.\label{fig:ppd008292840}}
\end{figure}

\begin{figure}[!p]
\figurenum{D12}
\centering
\includegraphics[width=0.8\linewidth]{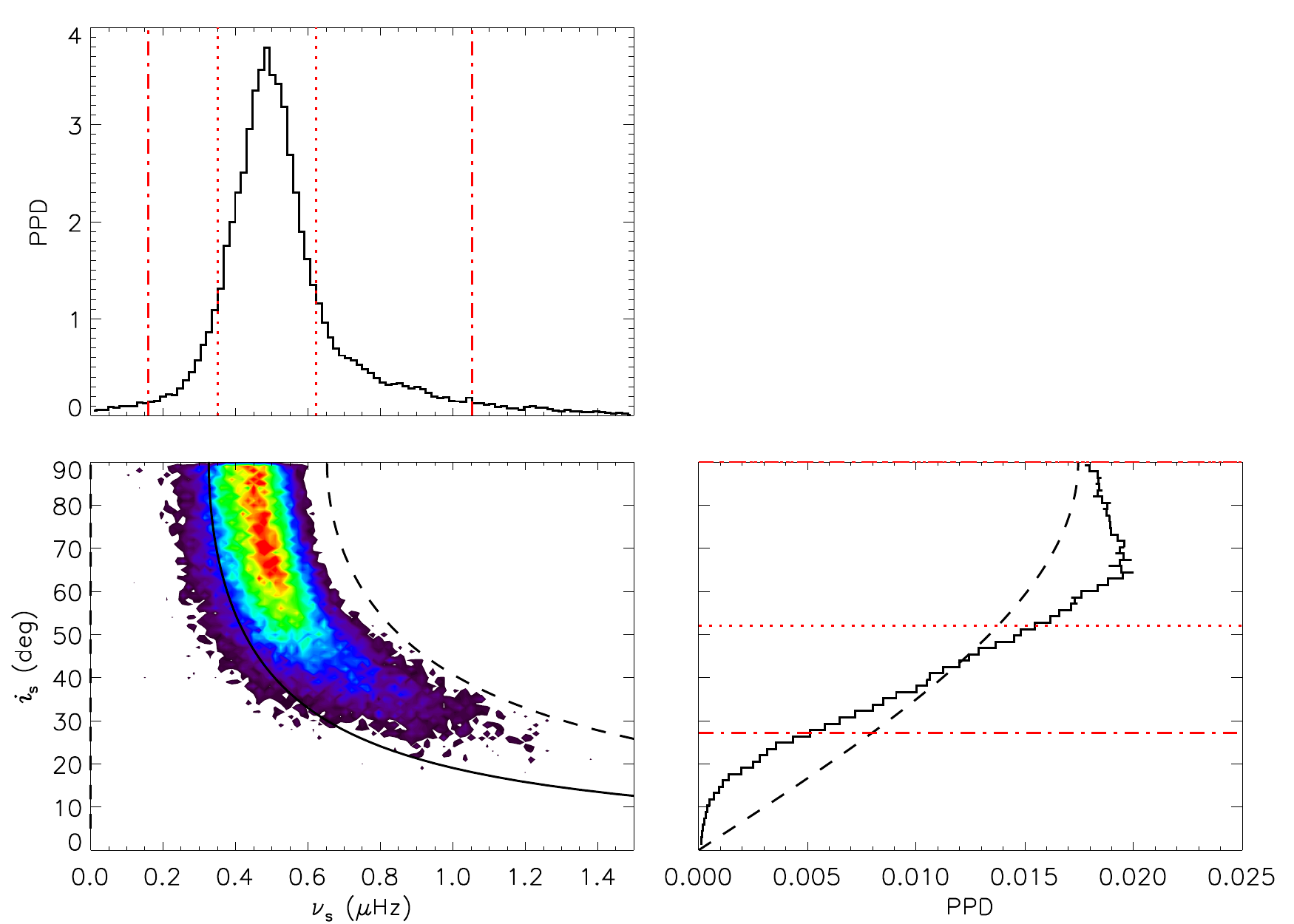}
\caption{\small Asteroseismic results on KIC~8478994 (Kepler-37, KOI-245). Similar to Fig.~\ref{fig:ppd008866102}.\label{fig:ppd008478994}}
\end{figure}

\clearpage

\begin{figure}[!p]
\figurenum{D13}
\centering
\includegraphics[width=0.8\linewidth]{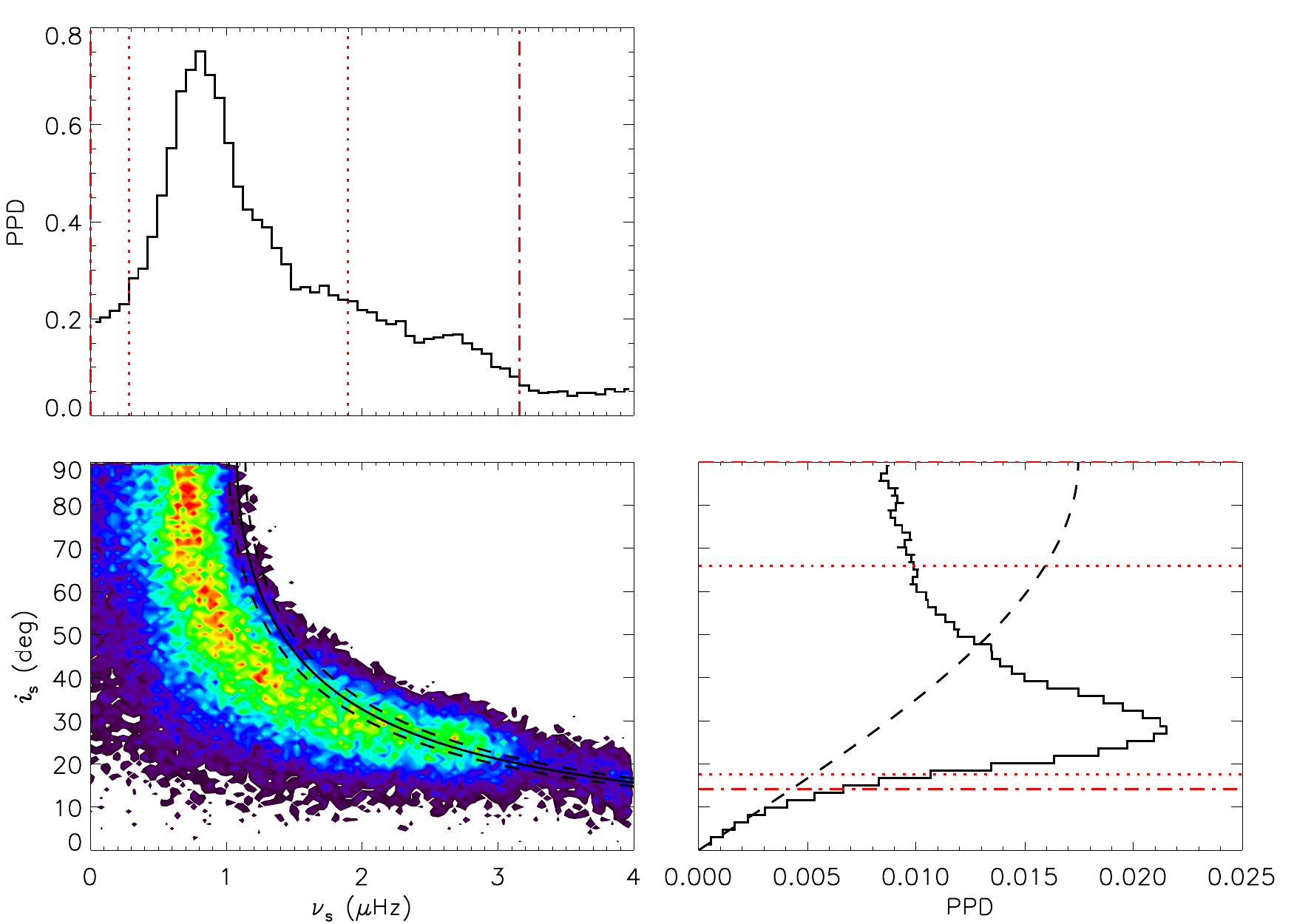}
\caption{\small Asteroseismic results on KIC~8494142 (Kepler-145, KOI-370). Similar to Fig.~\ref{fig:ppd008866102}.\label{fig:ppd008494142}}
\end{figure}

\begin{figure}[!p]
\figurenum{D14}
\centering
\includegraphics[width=0.8\linewidth]{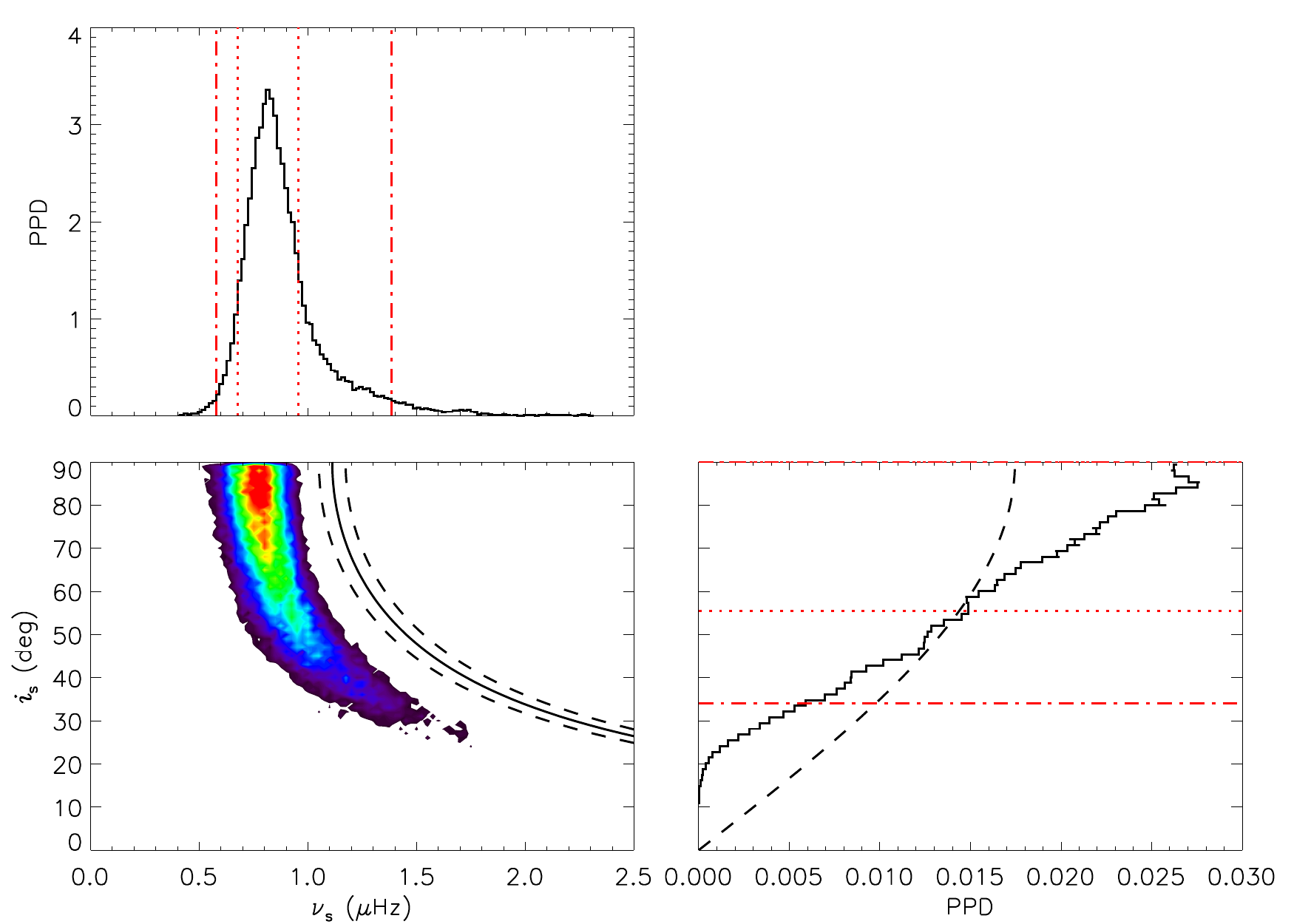}
\caption{\small Asteroseismic results on KIC~9414417 (KOI-974). Similar to Fig.~\ref{fig:ppd008866102}.\label{fig:ppd009414417}}
\end{figure}

\clearpage

\begin{figure}[!p]
\figurenum{D15}
\centering
\includegraphics[width=0.8\linewidth]{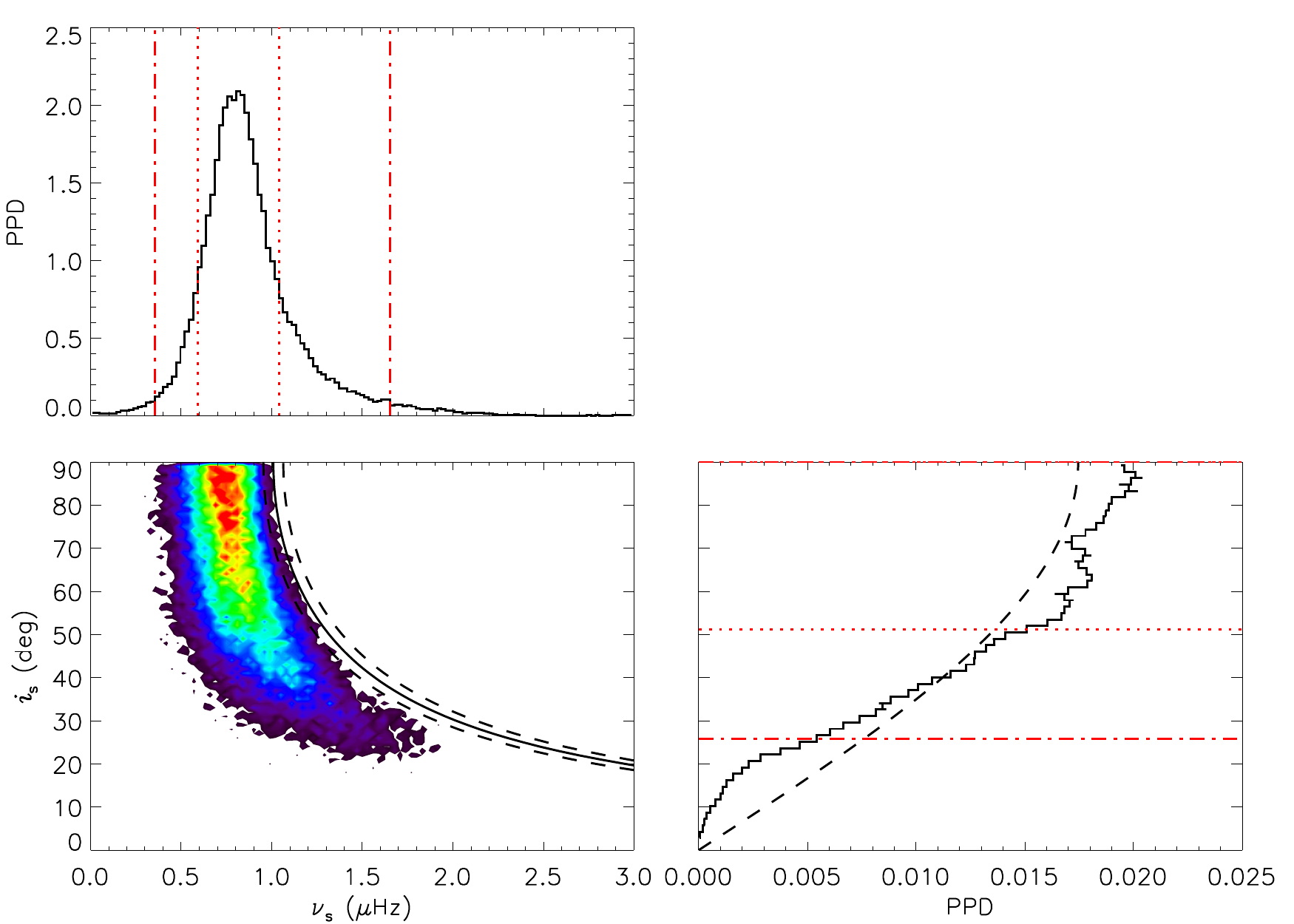}
\caption{\small Asteroseismic results on KIC~9592705 (KOI-288). Similar to Fig.~\ref{fig:ppd008866102}.\label{fig:ppd009592705}}
\end{figure}

\begin{figure}[!p]
\figurenum{D16}
\centering
\includegraphics[width=0.8\linewidth]{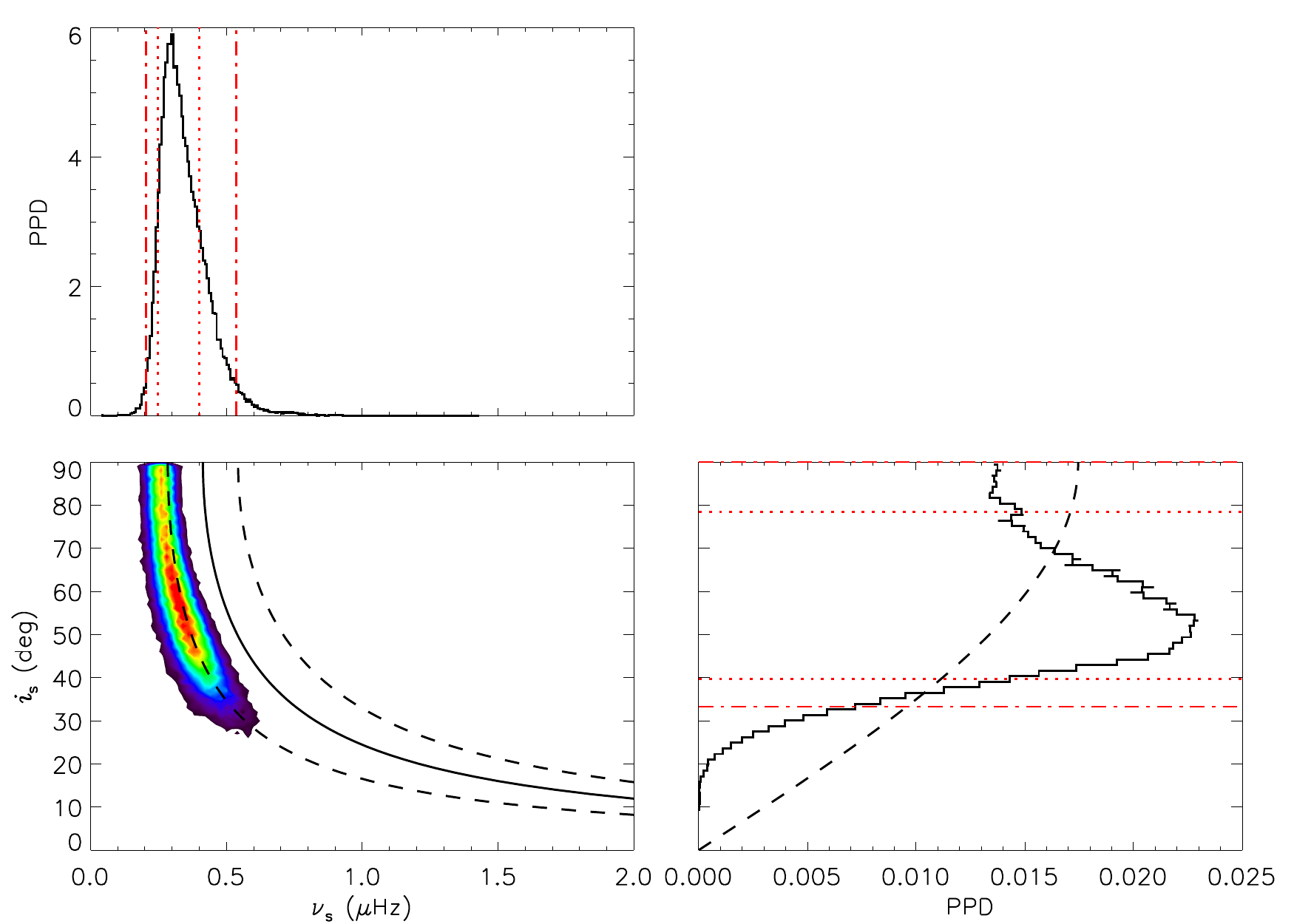}
\caption{\small Asteroseismic results on KIC~9955598 (Kepler-409, KOI-1925). Similar to Fig.~\ref{fig:ppd008866102}.\label{fig:ppd009955598}}
\end{figure}

\clearpage

\begin{figure}[!p]
\figurenum{D17}
\centering
\includegraphics[width=0.8\linewidth]{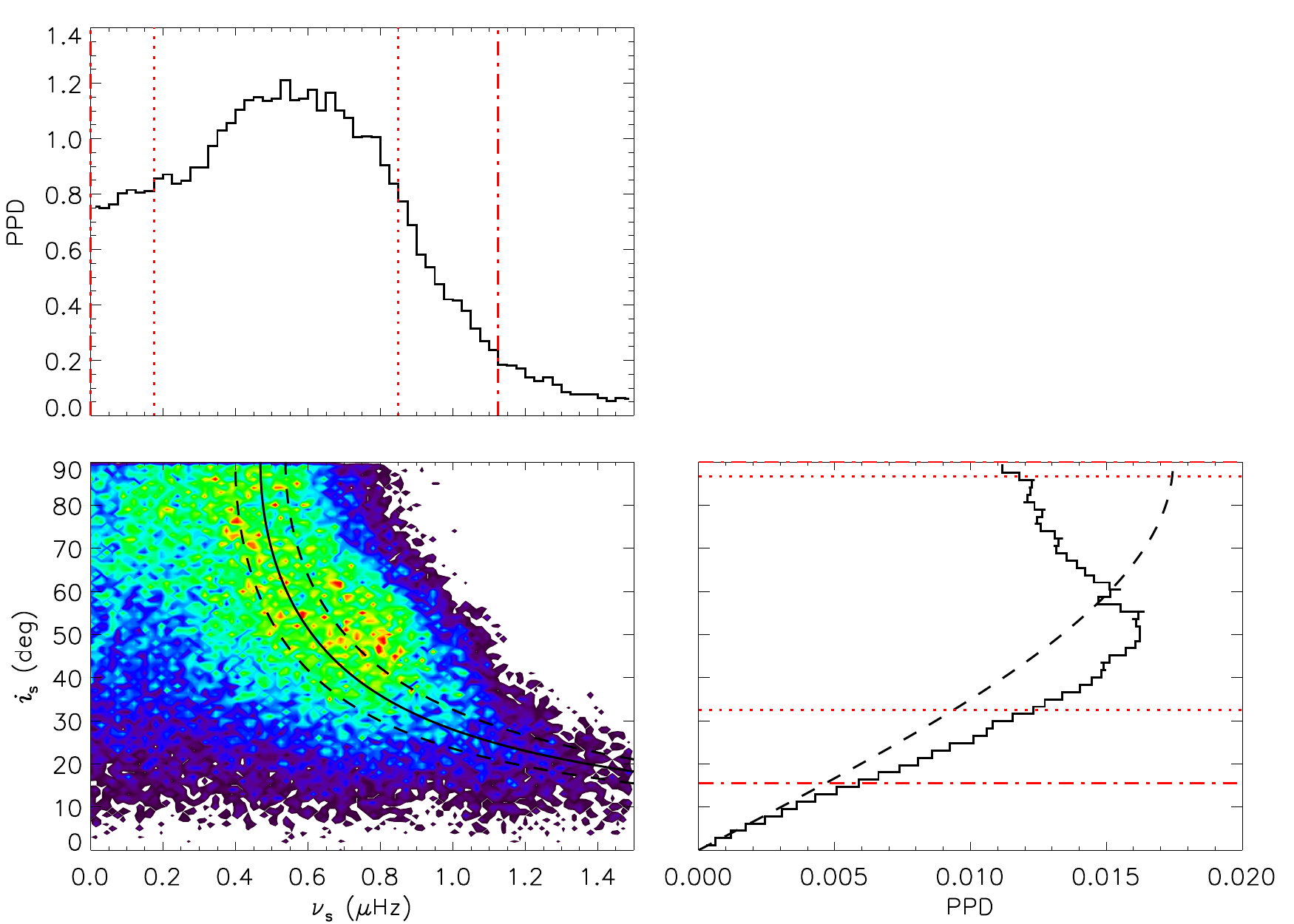}
\caption{\small Asteroseismic results on KIC~10586004 (Kepler-129, KOI-275). Similar to Fig.~\ref{fig:ppd008866102}.\label{fig:ppd010586004}}
\end{figure}

\begin{figure}[!p]
\figurenum{D18}
\centering
\includegraphics[width=0.8\linewidth]{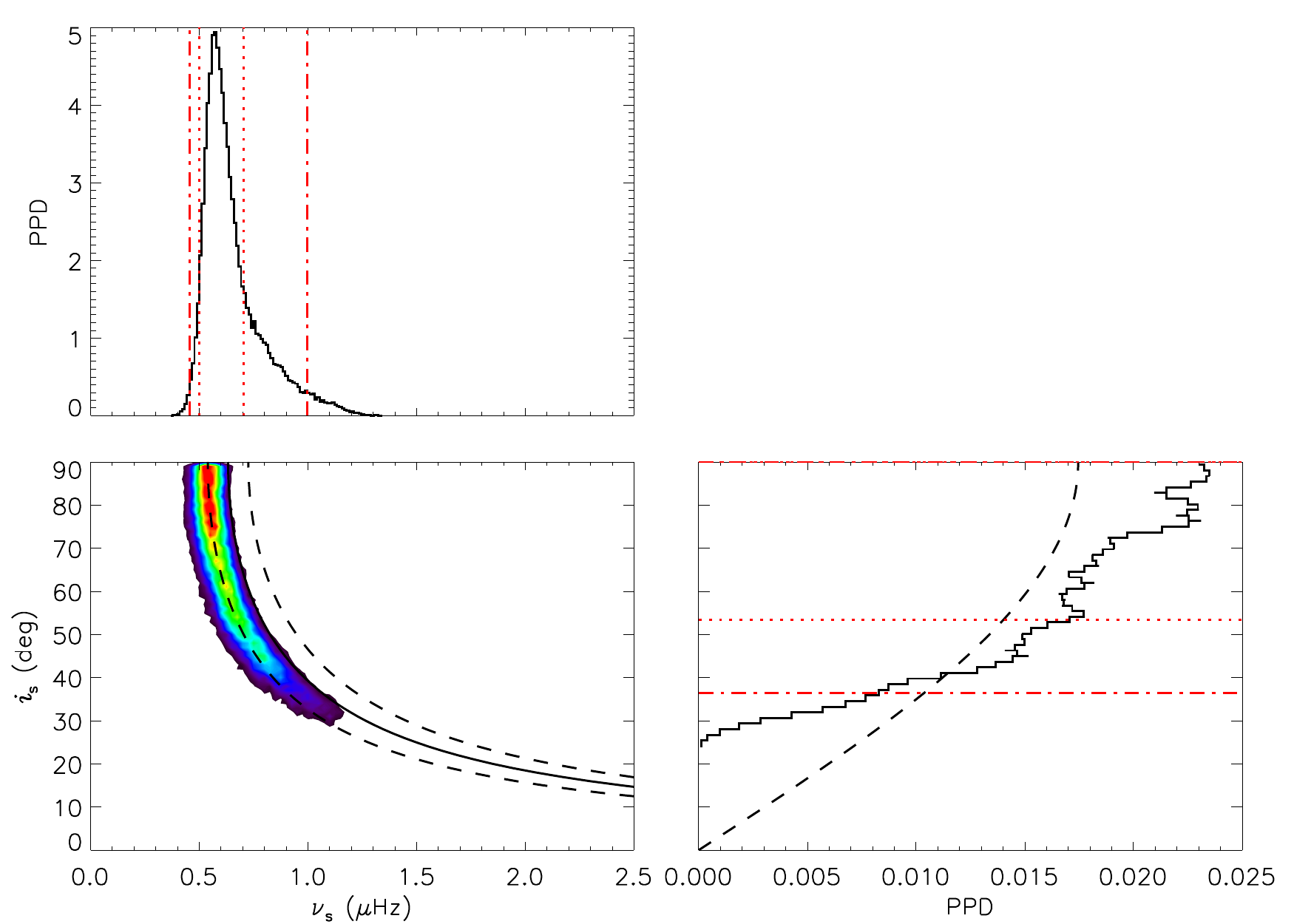}
\caption{\small Asteroseismic results on KIC~10963065 (Kepler-408, KOI-1612). Similar to Fig.~\ref{fig:ppd008866102}.\label{fig:ppd010963065}}
\end{figure}

\clearpage

\begin{figure}[!p]
\figurenum{D19}
\centering
\includegraphics[width=0.8\linewidth]{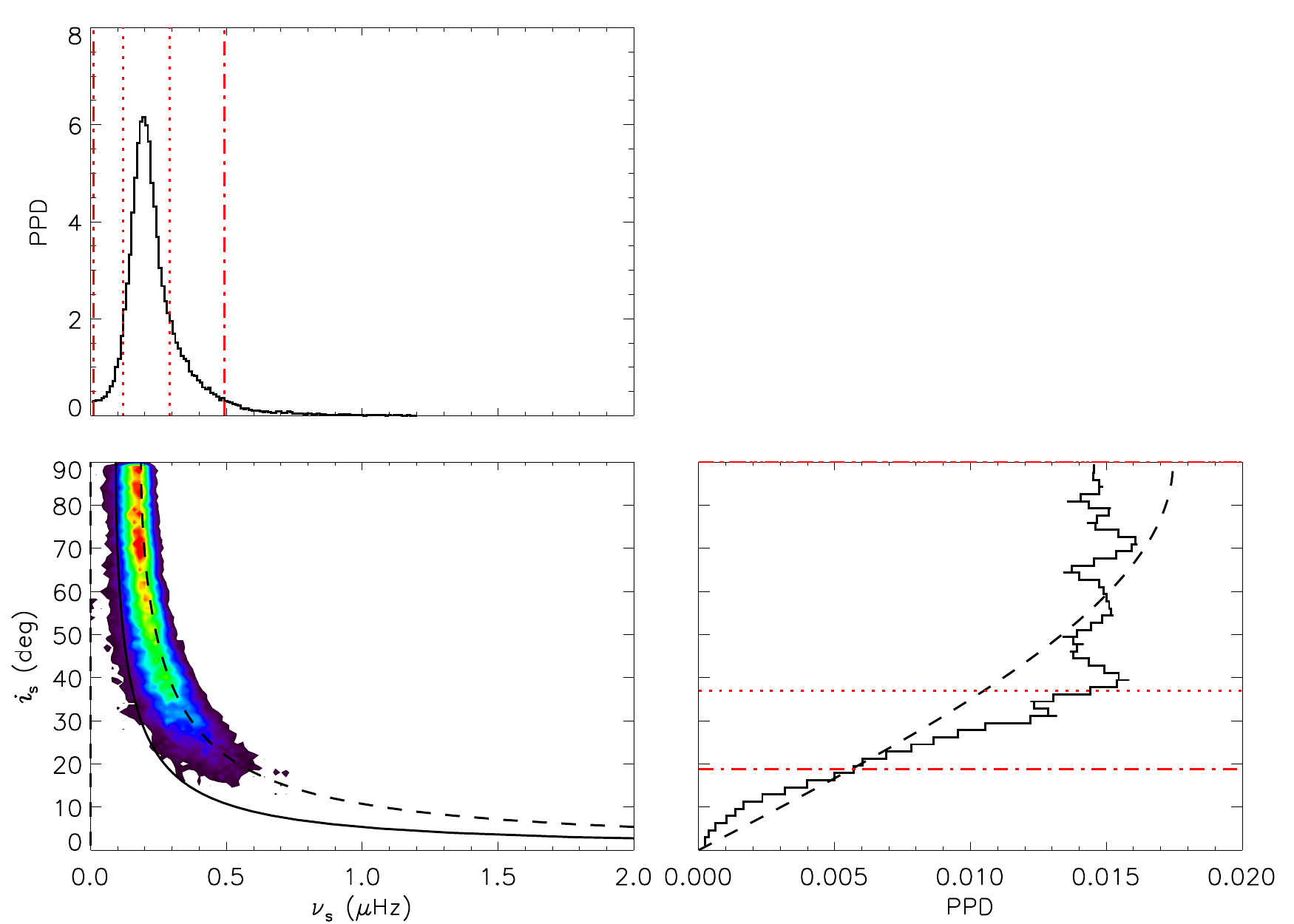}
\caption{\small Asteroseismic results on KIC~11295426 (Kepler-68, KOI-246). Similar to Fig.~\ref{fig:ppd008866102}.\label{fig:ppd011295426}}
\end{figure}

\begin{figure}[!p]
\figurenum{D20}
\centering
\includegraphics[width=0.8\linewidth]{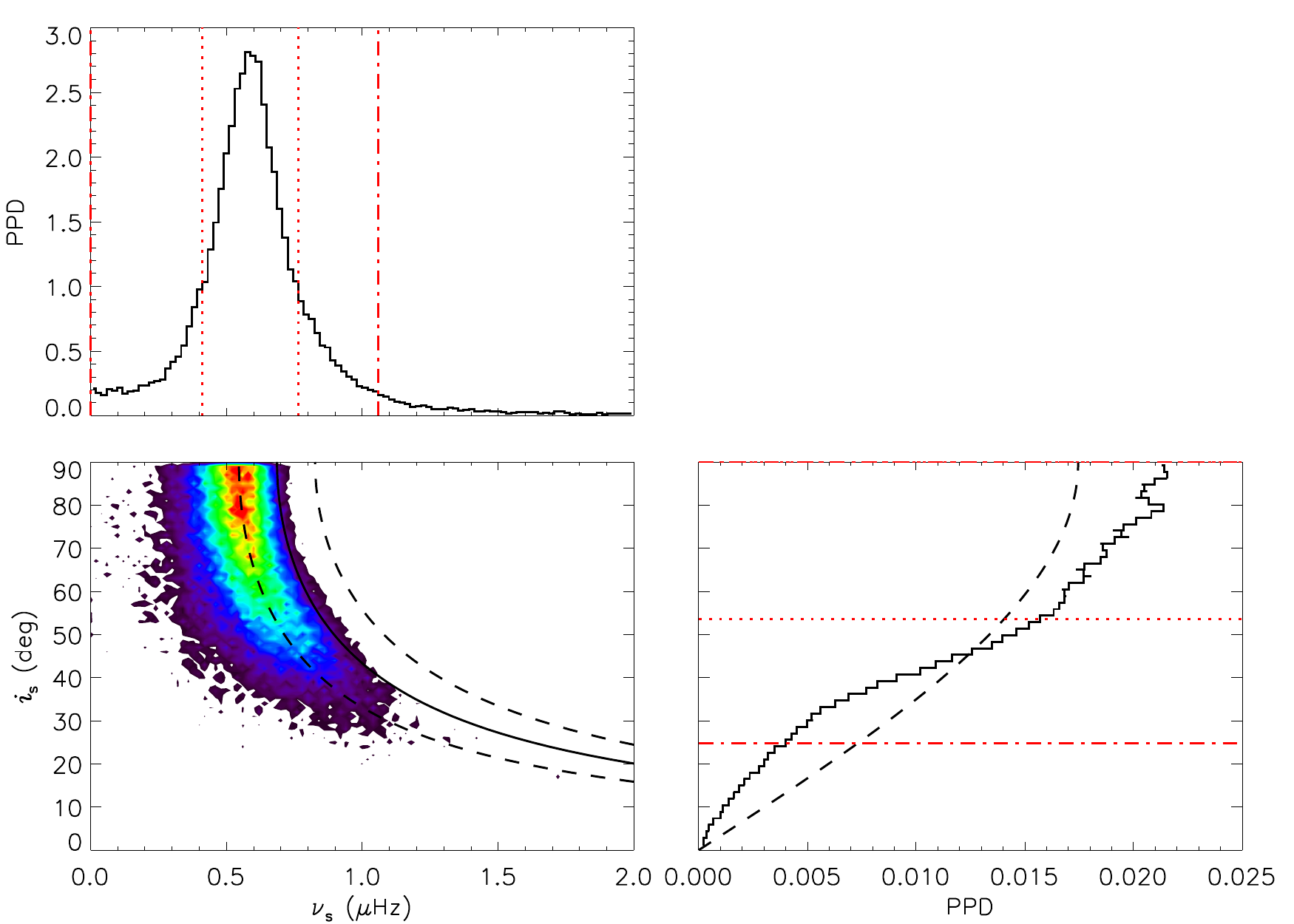}
\caption{\small Asteroseismic results on KIC~11401755 (Kepler-36, KOI-277). Similar to Fig.~\ref{fig:ppd008866102}.\label{fig:ppd011401755}}
\end{figure}

\clearpage

\begin{figure}[!p]
\figurenum{D21}
\centering
\includegraphics[width=0.8\linewidth]{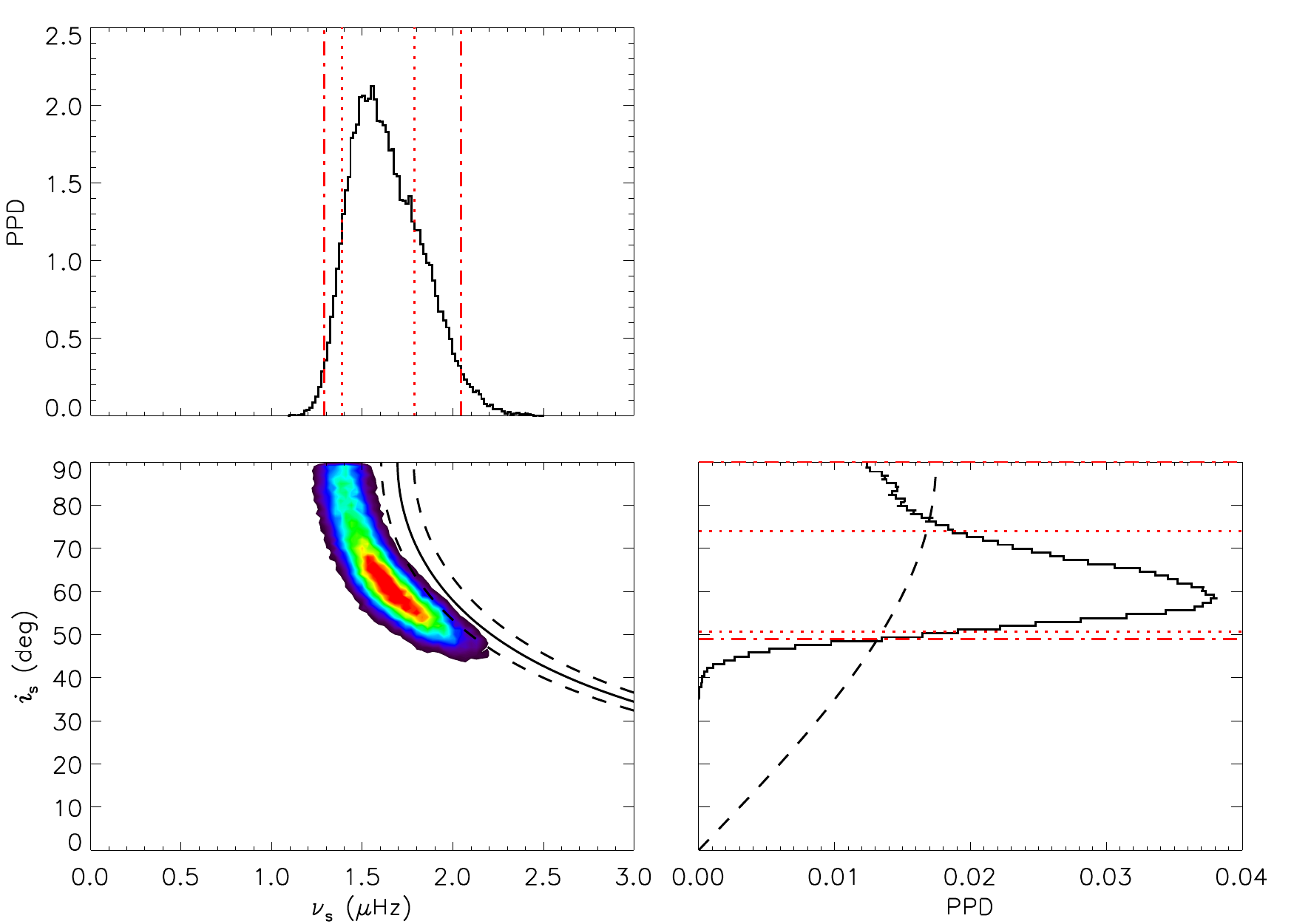}
\caption{\small Asteroseismic results on KIC~11807274 (Kepler-50, KOI-262). Similar to Fig.~\ref{fig:ppd008866102}.\label{fig:ppd011807274}}
\end{figure}

\begin{figure}[!p]
\figurenum{D22}
\centering
\includegraphics[width=0.8\linewidth]{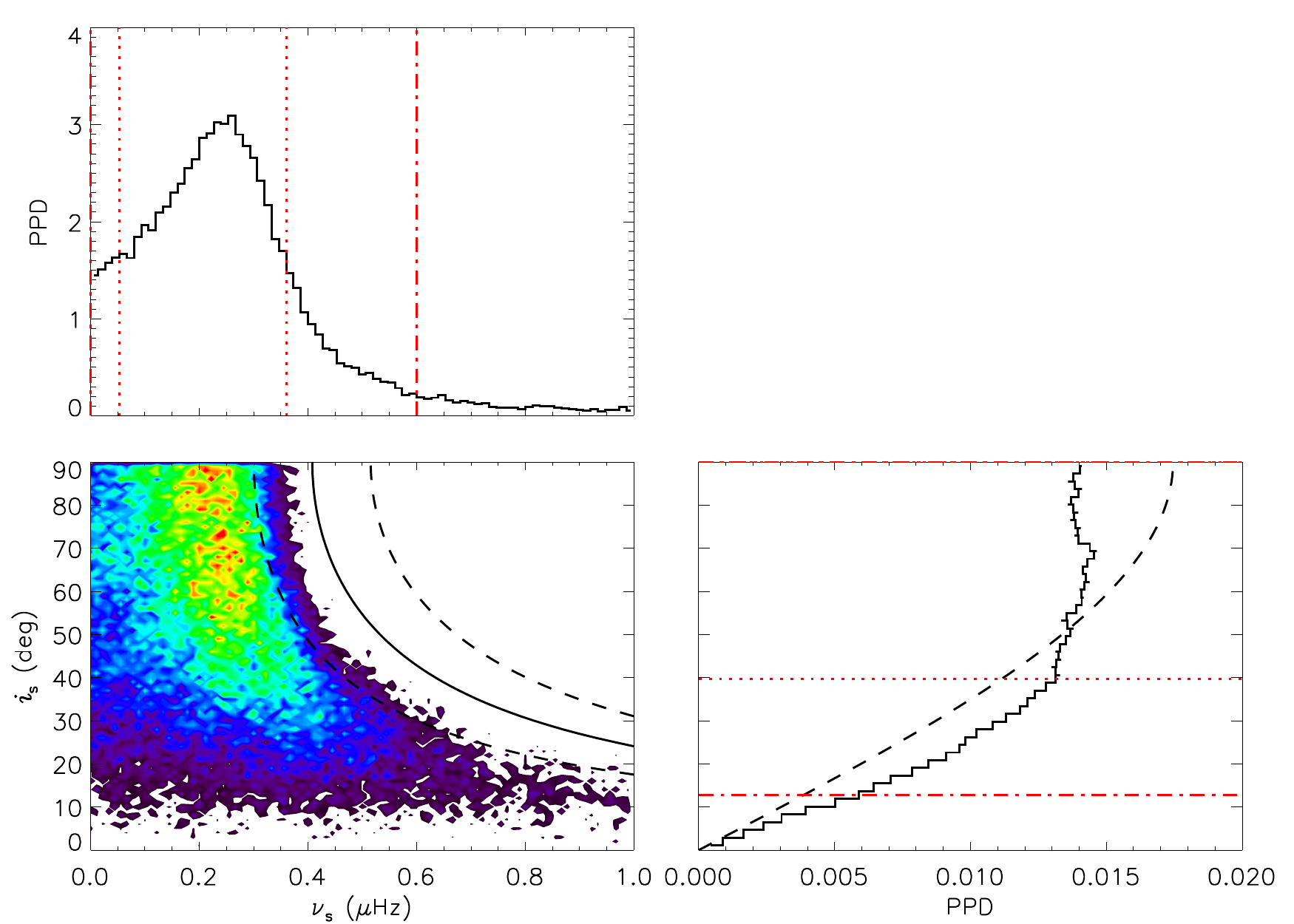}
\caption{\small Asteroseismic results on KIC~11904151 (Kepler-10, KOI-72). Similar to Fig.~\ref{fig:ppd008866102}.\label{fig:ppd011904151}}
\end{figure}

\clearpage

\section{Hierarchical Bayesian analysis}\label{append:hierarchical}
We wish to sample the joint PPD of the model parameters $\boldsymbol{\alpha}$,
\begin{equation}
\label{eq:postalpha}
p(\boldsymbol{\alpha}|\{\cos i_{\rm s}\}_{n=1}^N) \propto p(\boldsymbol{\alpha}) \, \mathcal{L}_{\boldsymbol{\alpha}} \, ,
\end{equation}
where $p(\boldsymbol{\alpha})$ is the prior on $\boldsymbol{\alpha}$ and $\mathcal{L}_{\boldsymbol{\alpha}}$ is the marginalized likelihood for the model parameters. Here, we employ the hierarchical probabilistic method developed by \citet{Hogg10}, whereby the true distribution of a given quantity is inferred by taking as input a set of posterior samplings of that same quantity (obtained using an uninformative prior). According to this formalism, a sampling approximation to the marginalized likelihood for $\boldsymbol{\alpha}$ is given by
\begin{equation}
\label{eq:marlike}
\mathcal{L}_{\boldsymbol{\alpha}} \approx \prod_{n=1}^N \frac{1}{K} \sum_{k=1}^K \frac{f_{\boldsymbol{\alpha}}(\cos i_{{\rm s}\,(nk)})}{p_0(\cos i_{{\rm s}\,(nk)})} \, ,
\end{equation}
where $\cos i_{{\rm s}\,(nk)}$ is the $k$th sample of the $n$th posterior of $\cos i_{\rm s}$. The sum is over $K\!=\!1000$ posterior samples of each individual system, a number of samples large enough to guarantee that our analysis is not prone to sampling variance. The ratio inside the sum is between the model distribution and the uninformative prior on which the sampling was based (i.e., a uniform prior on $\cos i_{\rm s}$). Simply put, we are conducting a forward modeling of the observed data by defining a model distribution $f_{\boldsymbol{\alpha}}(\psi)$ for the true $\psi$ and finding the model parameters $\boldsymbol{\alpha}$ that best explain those data. The use of input from several different observers/fitters, each of whom may have sampled their posteriors with different uninformative priors $p_0$, does not constitute a problem and can be straightforwardly accounted for in Eq.~(\ref{eq:marlike}).

\acknowledgments
This paper includes data collected by the {\it Kepler} mission. Funding for the {\it Kepler} mission is provided by the NASA Science Mission Directorate. Some of the data presented in this paper were obtained from the {\it Kepler} Asteroseismic Science Operations Center (KASOC) database. T.L.C., J.S.K., G.R.D., W.J.C., D.B., Y.P.E., and T.S.H.N.~acknowledge the support of the UK Science and Technology Facilities Council (STFC). Funding for the Stellar Astrophysics Centre is provided by The Danish National Research Foundation (Grant DNRF106). The research is supported by the ASTERISK project (ASTERoseismic Investigations with SONG and {\it Kepler}) funded by the European Research Council (Grant agreement no.: 267864). The research leading to the presented results has received funding from the European Research Council under the European Community's Seventh Framework Programme (FP7/2007-2013) / ERC grant agreement no.~338251 (StellarAges). Work by J.N.W.~was supported by the NASA Origins program (NNX11AG85G). S.B.~acknowledges support from NASA grant NNX13AE70G. C.K.~and V.S.A.~acknowledge the support of the Villum Foundation. This research has made use of the NASA Exoplanet Archive, which is operated by the California Institute of Technology, under contract with the National Aeronautics and Space Administration under the Exoplanet Exploration Program. This study has made use of Ren\'e Heller's Holt--Rossiter--McLaughlin Encyclopaedia. This research made use of the \texttt{Python} package \texttt{obliquity} developed by Timothy Morton and made available at \url{https://github.com/timothydmorton/obliquity}.

{\it Facilities:} \facility{{\it Kepler}}

\bibliographystyle{apj}
\bibliography{biblio}

\begin{deluxetable}{rlccccccccc}
\tablecolumns{11}
\tabletypesize{\scriptsize}
\tablewidth{0pc}
\tablecaption{Asteroseismic sample of {\it Kepler} Objects of Interest.\label{tb:sample}}
\tablehead{
\colhead{KIC} & \colhead{KOI} & \colhead{Kepler ID} & \colhead{$m_{\rm Kep}$} & \colhead{Quarters} & \colhead{No.~of} & \colhead{$R_{\rm p}$} & \colhead{$P_{\rm o}$} & \colhead{$v\sin i_{\rm s}$} & \colhead{$P_{\rm rot}$} & \colhead{Refs.} \\
\colhead{} & \colhead{} & \colhead{} & \colhead{} & \colhead{} & \colhead{Planets} & \colhead{($R_\earth$)} & \colhead{(d)} & \colhead{(${\rm km\,s^{-1}}$)} & \colhead{(d)} & \colhead{}
}
\startdata
3425851 & 268 & \nodata & 10.56 & Q6.1--Q8.3 & 1 & 3.03 & 110.4 & $9.5\pm0.5$ & $7.873\pm0.001$ & 1,9 \\
3544595 & 69 & Kepler-93 & 9.93 & Q2.3--Q17.2 & 2 & 1.48 & $>$3650 & $2.0\pm0.5$ & \nodata & 1 \\
3632418 & 975 & Kepler-21 & 8.22 & Q5.1--Q17.2 & 1 & 1.64 & 2.8 & $7.75\pm1.0$ & $12.591\pm0.036$ & 2,9 \\
4141376 & 280 & \nodata & 11.07 & Q6.1--Q17.2 & 1 & 1.94 & 11.9 & $3.5\pm0.5$ & $15.78\pm2.12$ & 1,9 \\
4349452 & 244 & Kepler-25 & 10.73 & Q5.1--Q17.2 & 3 & 2.60 & 123 & $8.2\pm0.2$ & $23.147\pm0.039$ & 3,9 \\
4914423 & 108 & Kepler-103 & 12.29 & Q3.1--Q12.3 & 2 & 3.37 & 179.6 & $3.8\pm0.7$ & \nodata & 1 \\
5866724 & 85 & Kepler-65 & 11.02 & Q3.1--Q17.2 & 3 & 1.42 & 8.1 & $10.4\pm0.5$ & $7.911\pm0.155$ & 1,10 \\
6278762\tablenotemark{a} & 3158 & Kepler-444 & 8.72 & Q15.1--Q17.2 & 5 & 0.40 & 9.7 & $\sim\!2.2$ & \nodata & 4 \\
6521045 & 41 & Kepler-100 & 11.20 & Q3.1--Q17.2 & 3 & 1.32 & 35.3 & $2.9\pm0.5$ & $24.988\pm2.192$ & 1,9 \\
7670943 & 269 & \nodata & 10.93 & Q6.1--Q17.2 & 1 & 1.75 & 18.0 & $13.5\pm0.6$ & $5.274\pm0.033$ & 1,9 \\
8077137 & 274 & Kepler-128 & 11.39 & Q6.1--Q17.2 & 2 & 1.13 & 22.8 & $7.7\pm0.5$ & \nodata & 1 \\
8292840 & 260 & Kepler-126 & 10.50 & Q5.1--Q17.2 & 3 & 1.52 & 100.3 & $10.4\pm0.5$ & \nodata & 1 \\
8478994 & 245 & Kepler-37 & 9.71 & Q5.1--Q17.2 & 4 & 0.30 & 51.2 & $1.1\pm1.1$ & $28.79\pm3.29$ & 5,11 \\
8494142 & 370 & Kepler-145 & 11.93 & Q7.1--Q17.2 & 2 & 2.65 & 42.9 & $8.9\pm0.5$ & \nodata & 1 \\
8866102\tablenotemark{b} & 42 & Kepler-410 A & 9.36 & Q3.1--Q17.2 & $>$1 & 2.84 & 17.8 & $15.0\pm0.5$ & $20.850\pm0.007$ & 1,9 \\
9414417 & 974 & \nodata & 9.58 & Q6.1--Q17.2 & 1 & 2.49 & 53.5 & $9.3\pm0.5$ & $10.847\pm0.002$ & 1,9 \\
9592705 & 288 & \nodata & 11.02 & Q6.1--Q17.2 & 1 & 3.17 & 10.3 & $9.4\pm0.5$ & $13.380\pm0.099$ & 1,9 \\
9955598 & 1925 & Kepler-409 & 9.44 & Q5.1--Q17.2 & 1 & 1.19 & 69.0 & $1.6\pm0.5$ & \nodata & 1 \\
10586004 & 275 & Kepler-129 & 11.70 & Q6.1--Q7.3 & 2 & 2.37 & 82.2 & $3.4\pm0.5$ & \nodata & 1 \\
10666592 & 2 & Kepler-2 & 10.46 & Q0--Q17.2 & 1 & 15.04 & 2.2 & $3.8\pm0.5$ & \nodata & 6 \\
10963065 & 1612 & Kepler-408 & 8.77 & Q2.3--Q17.2 & 1 & 0.82 & 2.5 & $3.4\pm0.5$ & $12.444\pm0.172$ & 1,9 \\
11295426 & 246 & Kepler-68 & 10.00 & Q5.1--Q17.2 & 3 & 0.95 & 580 & $0.5\pm0.5$ & \nodata & 7 \\
11401755 & 277 & Kepler-36 & 11.87 & Q6.1--Q17.2 & 2 & 1.49 & 16.2 & $4.9\pm1.0$ & \nodata & 8 \\
11807274 & 262 & Kepler-50 & 10.42 & Q6.1--Q17.2 & 2 & 1.71 & 9.4 & $11.7\pm0.6$ & $7.553\pm0.755$ & 1,9 \\
11904151 & 72 & Kepler-10 & 10.96 & Q2.2--Q17.2 & 2 & 1.47 & 45.3 & $1.9\pm0.5$ & \nodata & 1 \\
\enddata
\tablecomments{\scriptsize The number of planets refers to the total number of confirmed or else candidate transiting planets. $R_{\rm p}$ is the planetary radius (source: NASA Exoplanet Archive). For multiple-planet systems, the smallest planet in the system is considered. $P_{\rm o}$ is the orbital period (source: NASA Exoplanet Archive). For multiple-planet systems, the longest-period planet in the system is considered. See references below for sources of $v\sin i_{\rm s}$ and $P_{\rm rot}$.}
\tablenotetext{a}{\scriptsize A recent spectroscopic follow-up of this star with the Subaru High Dispersion Spectrograph (HDS) returned an upper limit for $v\sin i_{\rm s}$ (at the 2-$\sigma$ level) of $0.56\:{\rm km\,s^{-1}}$ \citetext{T.~Hirano, private communication}.}
\tablenotetext{b}{\scriptsize The $P_{\rm o}$ value is for the transiting planet (the TTV signal has a period of 957 days). The $P_{\rm rot}$ value is associated with a blended star \citep[][]{VanEylen14}.}
\tablerefs{\scriptsize (1) \citet{HuberKOIs}, (2) \citet{Kepler-21}, (3) \citet{Albrecht13}, (4) \citet{Kepler-444}, (5) \citet{Kepler-37}, (6) \citet{HAT-P-7b}, (7) \citet{Kepler-68}, (8) \citet{Kepler-36}, (9) \citet{McQuillan}, (10) \citet{Hirano12,Hirano14}, (11) \citet{Walkowicz13}.}
\end{deluxetable}

\begin{deluxetable}{ccccccc}
\tablecolumns{7}
\tabletypesize{\scriptsize}
\tablewidth{0pc}
\rotate
\tablecaption{Results on the spin-orbit angle $\psi$ for HAT-P-7 and Kepler-25.\label{tb:obliquities}}
\tablehead{
\colhead{} & \colhead{} & \colhead{} & \multicolumn{3}{c}{$\psi$} & \colhead{} \\
\cline{4-6} \\
\colhead{Kepler ID} & \colhead{$\lambda$} & \colhead{$i_{\rm o}$} & \colhead{Median} & \colhead{$68.3\,\%$ HPD Lower Bound} & \colhead{$68.3\,\%$ HPD Upper Bound} & \colhead{Refs.} \\
\colhead{} & \colhead{(deg)} & \colhead{(deg)} & \colhead{(deg)} & \colhead{(deg)} & \colhead{(deg)}
}
\startdata
HAT-P-7b & $182.5\pm9.4$\tablenotemark{a}/$-132.6^{+10.5}_{-16.3}$/$\mathbf{155\pm37}$ & $80.8^{+2.8}_{-1.2}$/$\mathbf{83.111\pm0.030}$ & $116.4$ & $93.5$ & $138.4$ & 1,2,3,4 \\
Kepler-25c & $-0.5\pm5.7$ & $87.27\pm0.05$ & $12.6$ & $1.6$ & $19.3$ & 5 \\
\enddata
\tablecomments{\scriptsize When several measurements of $\lambda$ or $i_{\rm o}$ are available (see references below), the adopted value is given in boldface. Values for $\lambda$ all come from measurements of the RM effect.}
\tablenotetext{a}{\scriptsize Equivalent to $\lambda\!=\!-177.5\degr\pm9.4\degr$ if brought into the $[-\pi,\pi]$ interval.}
\tablerefs{\scriptsize (1) \citet{Winn09}, (2) \citet{Narita09}, (3) \citet{Albrecht12}, (4) \citet{Morris13}, (5) \citet{Albrecht13}.}
\end{deluxetable}

\begin{deluxetable}{lcccccccc}
\tablecolumns{9}
\tabletypesize{\scriptsize}
\rotate
\tablewidth{0pc}
\tablecaption{Results on the stellar inclination angle $i_{\rm s}$ for the stars in the asteroseismic sample.\label{tb:results}}
\tablehead{
\colhead{} & \colhead{} & \colhead{} & \colhead{} & \colhead{} & \colhead{} & \multicolumn{3}{c}{$i_{\rm s}$} \\
\cline{7-9} \\
\colhead{KOI} & \colhead{Kepler ID} & \colhead{No.~of} & \colhead{Frequency Range} & \colhead{S/N} & \colhead{$\nu_{\rm s}/\Gamma$} & \colhead{$68.3\,\%$ HPD Credible Region} & \colhead{$95.4\,\%$ HPD Credible Region} & \colhead{MW14 Upper Bound} \\
\colhead{} & \colhead{} & \colhead{Orders} & \colhead{(${\rm \mu Hz}$)} & \colhead{} & \colhead{} & \colhead{(deg)} & \colhead{(deg)} & \colhead{(deg)}
}
\startdata
268 & \nodata & 6 & 1615--2170 & 0.30 & 0.21 & [38.5,85.0] & [18.6,90.0] & \nodata \\
69 & Kepler-93 & 6 & 2735--3610 & 2.7 & 0.49 & [48.6,69.8] & [45.1,87.4] & \nodata \\
975 & Kepler-21 & 8 & 810--1300 & 12.4 & 0.31 & [70.0,90.0] & [53.4,90.0] & 89.4 \\
280 & \nodata & 5 & 2520--3160 & 1.3 & 0.73 & [41.9,74.8] & [38.4,90.0] & 88.8 \\
244 & Kepler-25 & 6 & 1715--2300 & 1.0 & 0.46 & [68.7,90.0] & [54.3,90.0] & \nodata \\
108 & Kepler-103 & 5 & 1355--1765 & 1.0 & 0.28 & [41.7,90.0] & [17.6,90.0] & \nodata \\
85 & Kepler-65 & 6 & 1475--2010 & 1.4 & 0.56 & [74.3,90.0] & [60.8,90.0] & 90.0 \\
3158 & Kepler-444 & 6 & 3730--4800 & 2.2 & 0.31 & [31.3,90.0] & [22.7,90.0] & \nodata \\
41 & Kepler-100 & 7 & 1120--1660 & 4.3 & 0.32 & [65.0,90.0] & [46.2,90.0] & 89.4 \\
269 & \nodata & 6 & 1540--2075 & 1.0 & 0.54 & [58.5,90.0] & [39.1,90.0] & 63.6 \\
274 & Kepler-128 & 6 & 1065--1480 & 1.4 & 0.27 & [56.9,90.0] & [36.5,90.0] & \nodata \\
260 & Kepler-126 & 6 & 1615--2175 & 1.9 & 0.44 & [71.8,90.0] & [59.5,90.0] & \nodata \\
245 & Kepler-37 & 6 & 3710--4780 & 1.0 & 0.35 & [52.0,90.0] & [27.1,90.0] & \nodata \\
370 & Kepler-145 & 5 & 960--1270 & 1.3 & 0.36 & [17.6,65.9] & [14.1,90.0] & \nodata \\
42 & Kepler-410 A & 8 & 1550--2305 & 3.3 & 0.55 & [82.8,90.0] & [76.8,90.0] & \nodata \\
974 & \nodata & 8 & 800--1280 & 5.2 & 0.27 & [55.4,90.0] & [34.1,90.0] & 68.1 \\
288 & \nodata & 7 & 770--1150 & 1.8 & 0.27 & [51.2,90.0] & [25.9,90.0] & 90.0 \\
1925 & Kepler-409 & 7 & 2880--3950 & 5.1 & 0.42 & [39.7,78.4] & [33.2,90.0] & \nodata \\
275 & Kepler-129 & 5 & 1150--1505 & 2.1 & 0.40 & [32.4,86.6] & [15.5,90.0] & \nodata \\
2 & Kepler-2 & 7 & 840--1255 & 2.2 & 0.17 & [19.1,69.8] & [15.8,90.0] & \nodata \\
1612 & Kepler-408 & 8 & 1705--2530 & 10.1 & 0.28 & [53.5,90.0] & [36.5,90.0] & \nodata \\
246 & Kepler-68 & 8 & 1690--2500 & 10.0 & 0.21 & [36.9,90.0] & [18.7,90.0] & \nodata \\
277 & Kepler-36 & 6 & 980--1390 & 1.8 & 0.38 & [53.6,90.0] & [24.8,90.0] & \nodata \\
262 & Kepler-50 & 7 & 1170--1700 & 2.7 & 0.53 & [50.7,73.9] & [48.9,90.0] & 89.4 \\
72 & Kepler-10 & 5 & 2200--2790 & 1.9 & 0.37 & [39.7,90.0] & [12.7,90.0] & \nodata \\
\enddata
\tablecomments{\scriptsize The number of orders refers to the number of observed orders entering the fit (cf.~Appendix \ref{append:specfit}). Frequency ranges are approximate. The height-to-background ratio of the dipole mode in the vicinity of the frequency of maximum oscillation amplitude ($\nu_{\rm max}$) is taken as a measure of the S/N in the p modes. The ratio $\nu_{\rm s}/\Gamma$ refers to the order around $\nu_{\rm max}$ (for reference, $\Gamma\!\approx\!1.0\:{\rm \mu Hz}$ and $\nu_{\rm s}/\Gamma\!\approx\!0.4$ for the Sun). The rightmost column gives the $95\,\%$ confidence upper bound on $i_{\rm s}$ obtained by \citet{Morton14}.}
\end{deluxetable}

\begin{deluxetable}{cccccccccccc}
\tablecolumns{12}
\tabletypesize{\small}
\tablewidth{0pc}
\rotate
\tablecaption{Results from the hierarchical Bayesian analysis.\label{tb:reshierarchical}}
\tablehead{
\colhead{Model} & \multicolumn{3}{c}{Single-transiting} & & \multicolumn{3}{c}{Multi-transiting} & & \multicolumn{3}{c}{All} \\
\cline{2-4} \cline{6-8} \cline{10-12} \\
\colhead{} & \colhead{$\kappa$} & \colhead{$f$} & \colhead{$E$} & \colhead{} & \colhead{$\kappa$} & \colhead{$f$} & \colhead{$E$} & \colhead{} & \colhead{$\kappa$} & \colhead{$f$} & \colhead{$E$}
}
\startdata
\tableline
\multicolumn{12}{c}{Asteroseismic Sample} \\
\tableline
Single-Fisher & $106^{+92}_{-94}$ & \nodata & \nodata & & $101^{+96}_{-88}$ & \nodata & \nodata & & $107^{+93}_{-89}$ & \nodata & \nodata \\[0.2in]
\tableline
\multicolumn{12}{c}{Combined Sample} \\
\tableline
Single-Fisher & $8.7^{+6.7}_{-5.4}$ & \nodata & $1.10\times10^5$ & & $118^{+82}_{-83}$ & \nodata & $5.11\times10^7$ & & $11.5^{+7.5}_{-5.7}$ & \nodata & $1.61\times10^9$ \\[0.1in]
Mixture & $126^{+74}_{-86}$ & $0.21^{+0.24}_{-0.18}$ & $1.24\times10^7$ & & $116^{+84}_{-82}$ & $0.08^{+0.24}_{-0.08}$ & $6.16\times10^6$ & & $129^{+71}_{-80}$ & $0.13^{+0.16}_{-0.12}$ & $7.97\times10^{11}$ \\
\enddata
\tablecomments{\footnotesize We quote the median and $95.4\,\%$ HPD credible region for both $\kappa$ and $f$. The Bayesian model evidence, $E$, is reported for each cohort of systems in the combined sample.}
\end{deluxetable}

\end{document}